\documentclass[pdf,12pt]{article}
\usepackage{bm}
\usepackage{url}
\usepackage{amsfonts}
\usepackage{textgreek}
\usepackage{graphicx}
\usepackage{comment}
\usepackage{amsmath}
\usepackage{amssymb}
\usepackage{bm}
\usepackage{multicol}
\usepackage{epsfig}
\usepackage{float}
\usepackage{rotating}
\usepackage{colordvi}
\usepackage{booktabs}
\usepackage{multirow}
\usepackage{caption}
\usepackage{mathrsfs}
\usepackage{subfigure}
\DeclareMathAlphabet{\mathpzc}{OT1}{pzc}{m}{it}
\usepackage[normalem]{ulem}
\usepackage{color}
\usepackage{setspace}
\usepackage{enumitem}
\usepackage{setspace}

\newtheorem{theorem}{Theorem}

\newtheorem{corollary}[theorem]{Corollary}
\newtheorem{definition}[theorem]{Definition}

\newtheorem{proposition}{Proposition}

\newtheorem{assumption}[theorem]{Assumption}

\usepackage{titlesec}
\titleformat*{\section}{\large\bfseries}
\titleformat*{\subsection}{\normalsize\bfseries}

\titlespacing*{\section}
{0pt}{1.5ex plus 1ex minus .2ex}{1.3ex plus .2ex}
\titlespacing*{\subsection}
{0pt}{1.5ex plus 1ex minus .2ex}{1.3ex plus .2ex}

\usepackage{tikz}
\usepackage{tikz-3dplot}

\tikzset{>=latex}
\tdplotsetmaincoords{75}{110}

\begin{document}

\setlength{\abovedisplayskip}{6pt} 
\setlength{\belowdisplayskip}{6pt}
\setlength{\abovedisplayshortskip}{-0pt}
\setlength{\belowdisplayshortskip}{6pt}

\title{\Large Linear fractional relative risk aversion\thanks{Earlier versions of this paper have been presented at the ETSG 2025 Milan, the 39th Annual Meeting of the Applied Regional Science Conference, the CAE Research Workshop in International Economics (ETH Z\"{u}rich), and UQ\`{A}M. We thank Ikuto Aiba, Swati Dhingra, Peter Egger, Ruobing Huang, and Mathieu Parenti, as well as the participants at these conferences and workshops, for valuable comments and suggestions. Behrens gratefully acknowledges financial support from SSHRC Insight Grant \#435-2021-1017. Murata gratefully acknowledges financial support from JSPS KAKENHI Grant Number 24K04845.}}

\author{Kristian Behrens\thanks{Department of Economics, Universit\'e du Qu\'ebec \`a Montr\'eal ({\sc esg-uqam}), Canada; and {\sc cepr}, UK. E-mail: \texttt{behrens.kristian@uqam.ca}}
\and
Yasusada Murata\thanks{College of Economics, Nihon University, Japan.
E-mail: \texttt{murata.yasusada@nihon-u.ac.jp}}}
\date{April 23, 2026}
\maketitle
\begin{abstract}
We characterize the family of utility functions satisfying linear fractional relative risk aversion (LFRRA) in terms of the Gauss hypergeometric functions. We apply this  family, which nests various utility functions used in different strands of literature, to monopolistic competition and obtain the profit-maximizing price by generalizing the Lambert $W$ function. We let firm-level data decide whether the RRA in each sector or in the aggregate economy is increasing, decreasing, or constant, which in turn \mbox{determines} whether markups are decreasing, increasing, or constant with respect to marginal costs. 

\end{abstract}

\newpage
\clearpage

\section{Introduction}
\label{sec:intro}

The relative risk aversion (RRA) of a utility function $u$ of quantity $q$, defined as $-\frac{q u''(q)}{u'(q)}$, dates back to Arrow (1963, 1971) and Pratt (1964). It appears repeatedly in different contexts such as the Euler equation, which comes from the utility maximization problem of intertemporally optimizing consumers, and the Lerner (1934) index, which is derived from the profit maximization problem of imperfectly competitive firms.\footnote{See, e.g., Blanchard and Fisher (1989) for the former and Krugman (1979) for the latter.}
Therefore, the RRA has been widely used in various fields including economic growth, international trade, urban economics, as well as the economics of uncertainty.

In these fields of economics, different parametric specifications of $u$ have been used for deriving qualitative results. Examples include the utility function with hyperbolic absolute risk aversion (HARA)---which nests both constant relative risk aversion (CRRA) and constant absolute risk aversion (CARA) as special cases---and the constant revenue elasticity of marginal revenue (CREMR) utility function, which also nests CRRA but does not belong to HARA.\footnote{See, e.g., Merton (1971) for HARA and Mr\'{a}zov\'{a} et al.~(2021) for CREMR.}
These specifications have spurred the development of quantitative models and contributed to our understanding of the relationship between theory and numbers.\footnote{See, e.g., Redding and Rossi-Hansberg (2017) for the CRRA; Behrens et al.~(2014, 2017) for the CARA;  Simonovska (2015) for the (translated) log; and Mr\'{a}zov\'{a} and Neary (2017) and Arkolakis et al.~(2019) for the Pollak (1971) family. As shown below, the demand function derived from the Pollak family used in Mr\'{a}zov\'{a} and Neary (2017) and Arkolakis et al.~(2019) is isomorphic to that from the HARA.}

However, it is known that the choice of a utility function may significantly affect quantitative results (see, e.g., Behrens et al.~2020 for the CRRA and CARA cases). Furthermore, researchers do not necessarily know \emph{a priori} which specification provides the best fit to their data. To address these problems, we first propose a parametric family of utility functions that not only unifies the above-mentioned specifications but also yields a flexible RRA that is a second-order approximation of essentially any real analytic function of quantity. We then illustrate how to make a data-driven choice of a utility function from that family.

To this end, we start with \emph{linear fractional relative risk aversion} (LFRRA), defined as $-\frac{q u''(q)}{u'(q)} = \frac{\alpha q+\beta}{\alpha \sigma q+1}$, where $\alpha$, $\beta$, and $\sigma$ are parameters. It encompasses the HARA ($\beta=0$) and CREMR ($\beta=1$) cases, as well as the CRRA ($\alpha=0$ or $\beta=\frac{1}{\sigma}$), CARA ($\beta=\sigma=0$), quadratic ($\beta=0$ and $\sigma=-1$), and log ($\beta=0$ and $\sigma=1$) cases.\footnote{The CRRA can also be obtained from the HARA ($\beta=0$) and CREMR ($\beta=1$) by letting $\alpha \to \infty$.}
Furthermore, it is the $[1/1]$--Pad\'{e} approximant (Pad\'{e}, 1892) to essentially any real analytic function $\varphi(q)= \sum_{j=0}^\infty a_j q^j$ with coefficients $\{a_j\}$ and satisfies $\varphi(q) - \frac{\alpha q+\beta}{\alpha \sigma q+1} = O(q^3)$. We then show that the foregoing second-order differential equation can be rewritten as a hypergeometric differential equation, $z (1-z)v''(z) + (\beta-\frac{z}{\sigma})v'(z)=0$, using the change of variables $z=-\alpha \sigma q$ with $u(q) = v(z)$. Solving this differential equation allows us to characterize the family of utility functions satisfying the LFRRA in terms of the Gauss hypergeometric functions (Gauss, 1812).

We apply the LFRRA family of utility functions thus obtained to a model of monopolistic competition with heterogeneous firms, where the profit-maximizing price $p$, given marginal cost $m$, is defined as the solution to $\frac{p-m}{p} = -\frac{q u''(q)}{u'(q)}$. We derive the solution, $p=\frac{m}{\mu}$, where $\mu$ is the inverse markup function parametrized by $\{\alpha, \beta, \sigma\}$. It turns out that $\mu$ generalizes the Lambert $W$ function (Lambert, 1758; Euler, 1783; Corless et al., 1996), which has been used for the inverse markup for the CARA utility function in Behrens et al.~(2014, 2017, 2020). Our generalized Lambert $W$ function reduces to the original when $\alpha>0$ and $\beta=\sigma=0$.

The LFRRA family also generalizes the CRRA or the constant elasticity of substitution (CES) utility function in Melitz (2003), and constitutes a parametric version of the relative love for variety (RLV) in Zhelobodko et al.~(2012). Since the sign of the derivative of the RRA (or RLV) is given parametrically by the sign of $\alpha (1-\beta \sigma)$ in our model, we can take their comparative statics results to data. To illustrate this, we present a parsimonious way to estimate the key parameters $\{\alpha, \beta, \sigma\}$ using the  product-level markup and quantity data for a sample of firms from eleven sectors in De Loecker et al.~(2016). Our estimates tell us which sectors display increasing, decreasing, or constant RRA, which in turn  determines whether markups are decreasing, increasing, or constant with respect to marginal costs, respectively.

We believe that our framework is useful for the further development of qualitative and quantitative analyses for at least three reasons.

First, existing parametric approaches to the RRA typically focus on either CRRA, HARA or CREMR, or rely on a subcase of the HARA, such as the CARA, quadratic,  or log case. The LFRRA family nests all of these specifications and allows for new ones. Since it remains amenable to theoretical analysis by virtue of the generalized Lambert $W$ function, our approach may serve to enrich the qualitative analysis in economic growth, international trade, urban economics or the economics of uncertainty, \mbox{where the RRA has played a central role.}

Second, it is often the case that researchers do not have prior knowledge of whether a dataset satisfies the parameter restrictions on $\{\alpha, \beta, \sigma\}$ associated with the specification they choose. We address this question by letting the data decide the most appropriate specification from the LFRRA family of utility functions. Our estimates of $\beta$ are significantly different from $0$ and $1$ in ten out of eleven sectors and in the aggregate economy. The numbers of sectors that display increasing, decreasing, or constant RRA are four, four, and three, respectively, thus showing that having a flexible model accommodating these different cases is meaningful. We also find that the estimates for all sectors pooled are consistent with increasing~RRA.

Last, we show that our application to monopolistic competition based on directly additive preferences can be extended to the case of implicitly additive preferences. The LFRRA family of utility functions allows us to treat these two types of preferences in a unified manner and to explore various parametric specifications that are not covered by Zhelobodko et al.~(2012). We further extend the LFRRA family to include non-linear fractional forms and show that the associated utility functions can also be expressed in terms of the Gauss hypergeometric functions.

The remainder of the paper is organized as follows. Section~\ref{sec:LFRRA} introduces the LFRRA, characterizes the family of utility functions satisfying the LFRRA using the Gauss hypergeometric functions, and discusses a wide range of special cases. Section~\ref{sec:mc} embeds the LFRRA family of utility functions into a monopolistic competition model with heterogeneous firms, derives the inverse markup function in terms of the generalized Lambert $W$ function, and provides properties of markups, prices, and quantities. Section~\ref{sec:fit} takes the LFRRA family to data, provides estimates of our key parameters $\{\alpha, \beta, \sigma\}$, and detects which sectors display increasing, decreasing, or constant RRA. In Section~\ref{sec:extensions}, we apply our monopolistic competition analysis to the case of implicitly additive preferences and extend the LFRRA family to include non-linear fractional forms. Finally, Section~\ref{sec:conclusion} concludes.

\section{Linear fractional relative risk aversion and the utility function}
\label{sec:LFRRA}

In this section, we derive the family of utility functions that have linear fractional relative risk aversion (LFRRA), which is formally defined as follows.

\setcounter{theorem}{0}
\begin{definition}[Linear fractional relative risk aversion (LFRRA)]
\label{def}
Let $u$ and $q$ denote a utility function and a quantity. The linear fractional relative risk aversion is defined as
\begin{equation}
\label{eq:LFRRA_general}
- \frac{q u''(q)}{u'(q)} = \frac{\alpha q+ \beta }{\alpha \sigma q+1},
\end{equation}
where $q \ge 0$ and $\alpha \in (-\infty, \infty)$, $\beta \in [0,1]$, and $\sigma\in (-\infty, \infty)$ are parameters.
\end{definition}
\bigbreak

There are at least two reasons why considering this family is meaningful. First, as shown below, it encompasses the existing parametric approaches to the RRA, including the HARA ($\beta=0$) and CREMR ($\beta=1$) as the two corner cases. The former case further includes the CARA ($\sigma=0$), quadratic ($\sigma=-1$), and log ($\sigma=1$) subcases. Observe that the CRRA can be obtained from \eqref{eq:LFRRA_general} by letting $\alpha=0$ or $\beta=\frac{1}{\sigma}$. Indeed, if $\alpha=0$, then equation \eqref{eq:LFRRA_general} reduces to $- \frac{q u''(q)}{u'(q)} = \beta$. Thus, we have $u(q) = \frac{K q^{1-\beta}}{1-\beta} +C$ for $\beta \in [0,1)$ and $u(q) = K \ln q +C$ for $\beta =1$, where $K>0$ and $C$ are constants. The case with $\beta=\frac{1}{\sigma}$ will be incorporated into the analysis below. Note that the number 1 in the denominator of \eqref{eq:LFRRA_general} is a normalization.\footnote{To see this, let $- \frac{ q u''(q)}{u'(q)} = \frac{a q+b }{c q+d}  =  \frac{(a/d) q+b/d }{(c/d) q+1}$. Setting $a/d= \alpha$, $b/d=\beta $, and $c/d= \alpha \sigma $, we obtain \eqref{eq:LFRRA_general}.}

Second, the LFRRA in \eqref{eq:LFRRA_general} is the $[1/1]$--Pad\'{e} approximant (Pad\'{e}, 1892) to the formal power series $\sum_{j=0}^\infty a_j q^j$ and satisfies $\sum_{j=0}^\infty a_j q^j - \frac{\alpha q+\beta}{\alpha \sigma q+1} = O(q^3)$ (see Baker, 1975).\footnote{Indeed, multiplying the latter equation by $\alpha \sigma q+1$, we see that $a_0=\beta$, $a_1+ a_0\alpha\sigma=\alpha$, and $a_2+a_1 \alpha \sigma=0$, which can be solved uniquely for $\{\alpha, \beta, \sigma\}$, except for the following two cases. First, if $a_1=0$ and $a_2=0$, then $\beta=a_0$, $\sigma=\frac{1}{a_0}$, and $\alpha$ is indeterminate. Thus, the $[1/1]$-Pad\'{e} approximant exists, but is not unique. Second, if $a_1=0$ and $a_2 \neq 0$, then the solution for $\alpha$ and thus the $[1/1]$-Pad\'{e} approximant do not exist. Hence, the $[1/1]$-Pad\'{e} approximant is a second-order approximation of any formal power series except for the second case with $a_1=0$ and $a_2 \neq 0$.}
In the economically relevant case where $a_j$ are real numbers and the formal power series has a positive radius of convergence $R$ (which may be $+\infty$), we have a real analytic function $\varphi(q)= \sum_{j=0}^\infty a_j q^j$ with $a_j = \frac{\varphi^{(j)}(0)}{j!}$ for all $|q| <R$ (see, e.g., Theorem 8.4 in Rudin, 1976; Theorem 10.22 in Hunter, 2014). Thus, the $[1/1]$--Pad\'{e} approximant and the second-order Taylor expansion of essentially any real analytic function at $q=0$ share the information on $\{\varphi(0), \varphi'(0), \varphi''(0)\}$, or up to the fourth derivative of the utility function.\footnote{This remark can be applied to the case with $q_0>0$ by replacing $q$ with $q-q_0$ (see Seaborn, 1991).}
The $[1/1]$--Pad\'{e} approximant uses the rational function, whereas the second-order Taylor expansion relies on the polynomial. Hence, the former captures the curvature of $\varphi$ by the denominator $\alpha \sigma q+1$, unlike the latter that does so by the second-order term $\frac{\varphi''(0)}{2!} q^2$. We opt for the Pad\'{e} approximant since the Taylor expansion does not nest the HARA and CREMR cases.

To derive the utility function $u$ for $\alpha \neq 0$, we first rewrite \eqref{eq:LFRRA_general} as a second-order differential equation as follows:
\begin{equation}
\label{eq:LFRRAchange}
(\alpha \sigma q + 1) q u''(q) + (\alpha q + \beta) u'(q) = 0.
\end{equation}
In what follows, we first analyze the case with $\sigma \neq 0$ to establish Theorem~\ref{prop:T1} and Proposition~\ref{integral} for $\beta \in (0,1)$ and Theorem~\ref{prop:utility_NEW} and Proposition~\ref{integral2} for $\beta \in \{0,1\}$. Afterwards, we turn to the case with $\sigma=0$ for all $\beta \in [0,1]$ in Corollary~\ref{coroll_NEW}.

Assume that $\sigma \neq 0$. Letting $u(q) = v(z)$, where $z = -\alpha \sigma q$, we can express \eqref{eq:LFRRAchange} as the hypergeometric differential equation, whose solution can be readily obtained. Indeed, noting that $u'(q) = -\alpha\sigma v'(z)$ and $u''(q) = (-\alpha \sigma)^2 v''(z)$, equation \eqref{eq:LFRRAchange} can be rewritten as
\begin{equation}
\textstyle z (1-z) v''(z) + \bigl(\beta-\frac{z}{\sigma}\bigr) v'(z) = 0,
\label{eq:ourHGDE}
\end{equation}
which is the hypergeometric differential equation $z (1-z) v''(z) + [c-(a+b+1)z] v'(z) - ab v(z) = 0$ with $a$, $b$, and $c$ satisfying $c=\beta$, $a+b+1 = \frac{1}{\sigma}$, and $ab = 0$. Letting $c_1$ and $c_2$ denote arbitrary coefficients, the solution to this hypergeometric differential equation is given by $v(z) = \textstyle c_1\times {}_2 F_1 (a, b; c; z ) + c_2\times z^{1-c}{}_2 F_1(1+a-c, 1+b-c; 2-c; z),$ where ${}_2 F_{1}  (a, b; c; z ) = \sum_{n=0}^\infty \frac{(a)_n (b)_n }{(c)_n} \frac{z^n}{n!}$, with $\lvert z \rvert<1$, denotes the Gauss hypergeometric function and $(\cdot)_n$ is the (rising) Pochhammer symbol  (see Seaborn, 1991). Solving $c=\beta$, $a+b+1 = \frac{1}{\sigma}$, and $ab = 0$ for $a$, $b$, and $c$ yields two solutions $\{a, b,c\} = \{\frac{1-\sigma}{\sigma}, 0, \beta\}$ or $\{0, \frac{1-\sigma}{\sigma}, \beta\}$ that are symmetric in $a$ and $b$. Without loss of generality, we take the first one, so that the solution to \eqref{eq:ourHGDE} is given by
\begin{equation}
v(z) = \textstyle c_1\times {}_2 F_1\bigl(\frac{1-\sigma}{\sigma}, 0; \beta; z\bigr) + c_2\times z^{1-\beta}{}_2 F_1\bigl(1+\frac{1-\sigma}{\sigma}-\beta, 1-\beta; 2-\beta; z\bigr).
\label{eq:vz}
\end{equation}

Using this solution for $\alpha \neq 0$, as well as the solution for $\alpha = 0$ below Definition~\ref{def}, we can establish the relationship between the LFRRA and the utility function $u(q)$ as follows.

\medbreak
\setcounter{theorem}{0}
\begin{theorem}[LFRRA and utility function]
\label{prop:T1}
Assume that $\beta \in (0,1)$ and $\sigma \neq 0$. Then, the relative risk aversion is linear fractional if and only if the utility function is given by
\begin{equation}
\label{eq:prop4_subutility_general_case}
u(q) =  \textstyle \frac{K \sigma}{\sigma-1} (1+\alpha \sigma q)^{\beta-\frac{1}{\sigma}} q^{1-\beta} \left[ 1
+\frac{\beta -\frac{1}{\sigma}}{1-\beta} \,
{}_2 F_{1} \left(1- \frac{1}{\sigma}, 1; 2- \beta ; - \alpha \sigma q   \right)  \right]+ C,
\end{equation}
where ${}_2 F_{1} \left(1- \frac{1}{\sigma}, 1; 2- \beta ; - \alpha \sigma q   \right)  = \sum_{n=0}^\infty \frac{(1- \frac{1}{\sigma})_n (1)_n }{(2- \beta)_n} \frac{(- \alpha \sigma q)^n}{n!}$, with $\lvert-\alpha \sigma q \rvert<1$, denotes the Gauss hypergeometric function; $(\cdot)_n$ is the (rising) Pochhammer symbol; and $K>0$ and $C$ are constants.
\end{theorem}

\vspace{-.5cm}

\paragraph{Proof.} See \ref{app:theorem1}.

\bigbreak

For the utility function in \eqref{eq:prop4_subutility_general_case} to be well behaved, we assume that the marginal utility is positive and strictly decreasing. As shown in the proof of Theorem~\ref{prop:T1}, the marginal utility function is given by
\begin{equation}
u'(q) = K (1+ \alpha \sigma q)^{\beta-\frac{1}{\sigma}} q^{-\beta}.
\label{eq:marginal}
\end{equation}
For this to be positive (and real valued), we impose $K>0$ and $1+ \alpha \sigma q > 0$. If $\alpha \sigma \ge 0$, the inequality $1+ \alpha \sigma q > 0$ is satisfied for all $q \ge 0$. If $\alpha \sigma < 0$, the inequality implies $0\le q < -\frac{1}{\alpha \sigma}$ since $q \ge 0$ must hold by Definition~\ref{def}. Differentiating \eqref{eq:marginal} yields
\begin{equation}
u''(q) = - K (\alpha q+ \beta) (1+ \alpha \sigma q)^{-1+ \beta-\frac{1}{\sigma}} q^{-1-\beta}.
\label{eq:marginal_second}
\end{equation}
For the marginal utility to be strictly decreasing, we impose $\alpha q+ \beta > 0$, where $\beta \in (0,1)$ by the assumption in Theorem~\ref{prop:T1}. If $\alpha \ge 0$, the inequality $\alpha q+ \beta > 0$ is satisfied for all $q \ge 0$. If $\alpha < 0$, the inequality implies $0 \le q < -\frac{\beta}{\alpha}$ since $q \ge 0$ must hold.\footnote{\label{fnCRRA}In the CRRA case with $\beta =\frac{1}{\sigma}$, which requires $\sigma>1$ for $\beta \in (0,1)$, we have $u'(q)=K q^{-\frac{1}{\sigma}}>0$ and $u''(q) = - \frac{K}{\sigma} q^{-\frac{\sigma+1}{\sigma}}<0$ for all $q \in [0, \infty)$. Thus, in this case, the conditions $1+\alpha \sigma q>0$ and $\alpha q+\beta>0$ are immaterial for signing the derivatives \eqref{eq:marginal} and \eqref{eq:marginal_second}.}

Hence, we have three types of quantity restrictions for the utility function in \eqref{eq:prop4_subutility_general_case}, i.e., $u'(q)>0$, $u''(q) < 0$, and $\lvert-\alpha \sigma q \rvert<1$, where the last condition is required from the definition of ${}_2 F_{1} \left(1- \frac{1}{\sigma}, 1; 2- \beta ; - \alpha \sigma q   \right)$ in Theorem~\ref{prop:T1}. We first summarize the first two restrictions in Assumption~\ref{as} (see Table~\ref{tab:QR}). We then show in Proposition~\ref{integral} that we can use analytic continuation to extend the utility function \eqref{eq:prop4_subutility_general_case} to all quantities satisfying the restrictions in Table~\ref{tab:QR}, so that the third restriction $\lvert-\alpha \sigma q \rvert<1$ is immaterial.

\setcounter{theorem}{0}
\begin{assumption}
\label{as}
For the utility function $u$ in \eqref{eq:prop4_subutility_general_case} to be strictly increasing and strictly concave, $1+\alpha \sigma q > 0$ and $\alpha q+\beta > 0$ must hold, which implies the quantity restrictions in Table~\ref{tab:QR}.
\vskip .25cm
\begin{table}[h]
\caption{Ranges of quantities satisfying $q\ge 0$, $u'>0$, and $u'' < 0$.}
\label{tab:QR}
\centering
\begin{tabular}{c|cc}
\hline
&& \\[-4mm]
& \quad \quad $\alpha \ge 0$ & \quad $\alpha<0$  \\[.5mm]   \hline
&& \\[-4mm]
$\alpha \sigma \ge 0$ & \quad \quad $q \in [0,\infty)$ & \quad $q \in [0,  - \frac{\beta}{\alpha} )$ \\[1.25mm]
$\alpha \sigma<0$ &  \quad \quad $q \in [0,   -\frac{1}{\alpha \sigma} ) $  & \quad $q  \in [0,  -\frac{1}{\alpha \sigma}) \cap [0, -\frac{\beta}{\alpha} )$$^\dagger$ \\[.75mm] \hline
\multicolumn{3}{p{10.6cm}}{\scriptsize{{\it Notes}: $^\dagger$ The CRRA case with $\beta =\frac{1}{\sigma}$ requires $\sigma>1$ for $\beta \in (0,1)$, which implies that either the top-left cell or the bottom-right cell is applicable. In the latter case, $q  \in [0,  -\frac{1}{\alpha \sigma}) \cap [0, -\frac{\beta}{\alpha} )$ must be replaced with $q \in [0,\infty)$ (see footnote~\ref{fnCRRA}).}}
\end{tabular}
\vspace{-.25cm}
\end{table}
\end{assumption}

\setcounter{theorem}{0}
\begin{proposition}[Integral representation]
\label{integral}
Suppose that $\beta \in (0,1)$ and $\sigma \neq 0$. Then, the utility function \eqref{eq:prop4_subutility_general_case} can be rewritten for all $q$ satisfying the quantity restrictions in Table~\ref{tab:QR} as
\begin{equation}
\label{eq:utility_general_continuation}
\textstyle u(q)=  \frac{K \sigma}{\sigma-1} (1+\alpha \sigma q)^{\beta-\frac{1}{\sigma}} q^{1-\beta} \left[ 1
+\left(\beta -\frac{1}{\sigma} \right) \int_0^1  (1-t)^{- \beta}  (1 + t \alpha \sigma q)^{-(1- \frac{1}{\sigma})} {\rm d}t  \right]+ C.
\end{equation}
\end{proposition}

\vspace{-.5cm}

\paragraph{Proof.} See \ref{app:proposition1}.

\medbreak

Theorem~\ref{prop:T1} and Proposition~\ref{integral} do not include the two corner cases $\beta = 0$ and $\beta=1$. The next theorem and proposition extend Theorem~\ref{prop:T1} and Proposition~\ref{integral} to the corner cases and show that the utility function that has LFRRA nests the HARA and CREMR utility functions as special cases.\footnote{The HARA utility (e.g., Merton, 1971) is defined such that the inverse of the absolute risk aversion is linear in $q$, i.e., $-\frac{u'(q)}{u''(q)} = \frac{K (1+ \alpha \sigma  q)^{-\frac{1}{\sigma}}}{\alpha K (1+ \alpha \sigma  q)^{-\frac{1}{\sigma}-1}} =\frac{1 +\alpha \sigma  q}{\alpha}$ or that the prudence $-\frac{u'''(q)}{u''(q)}$ is proportional to the absolute risk aversion $-\frac{u''(q)}{u'(q)}$, i.e., $\frac{u'''(q) u'(q)}{[u''(q)]^2} = 1+ \sigma$.}

\setcounter{theorem}{1}
\begin{theorem}[Limit cases] 
\label{prop:utility_NEW}
Assume that $\alpha \neq 0$, and that $\sigma \neq 0$. Then, we obtain the following limit cases. First, as $\beta$ goes to 0 from above, the utility function \eqref{eq:prop4_subutility_general_case} reduces to the HARA form\emph{:}
\begin{equation}
\label{eq:prop4_subutility_HARA}
\lim_{\beta \to 0^+} u(q)= \textstyle  \frac{K}{\alpha (\sigma-1)} \bigl[ (1+\alpha \sigma q)^\frac{\sigma-1}{\sigma} -1  \bigr] + C.
\end{equation}
Second, as $\beta$ goes to 1 from below, the utility function \eqref{eq:prop4_subutility_general_case} reduces to the CREMR form\emph{:}
\begin{equation}
\label{eq:prop4_subutility_hypergeo}
\lim_{\beta \to 1^-} u(q) = \textstyle \frac{K \sigma}{\sigma-1} \frac{(1+\alpha \sigma q)^{1-\frac{1}{\sigma}} } {q} \bigl[ q
-\frac{\sigma-1}{\alpha \sigma}
{}_2 F_{1} \bigl(1, 1 ; 1+\frac{1}{\sigma}; - \frac{1}{\alpha \sigma q}   \bigr)  \bigr] + C,
\end{equation}
where $\sigma \neq 1$ and $ {}_2 F_{1} \bigl(1, 1; 1+ \frac{1}{\sigma}; -\frac{1}{ \alpha \sigma q} \bigr)=\sum_{n=0}^\infty \frac{(1)_n (1)_n}{(1+\frac{1}{\sigma})_n} \frac{(- \frac{1}{\alpha \sigma q})^n}{n!}$, with $\lvert - \frac{1}{\alpha \sigma q} \rvert<1$.
\end{theorem}

\paragraph{Proof.} See \ref{app:theorem2}.

\medbreak
For the HARA and CREMR utility functions in \eqref{eq:prop4_subutility_HARA} and \eqref{eq:prop4_subutility_hypergeo}  to be well behaved, we assume that the marginal utility is positive and strictly decreasing. As shown in the proof of Theorem~\ref{prop:utility_NEW}, the marginal utility functions for the HARA and CREMR cases
are given by
\begin{equation}
u'(q) = K (1+\alpha \sigma q)^{-\frac{1}{\sigma}} \quad \mbox{and} \quad
u'(q) = K (1+\alpha \sigma q)^{1-\frac{1}{\sigma}} q^{-1},
\label{eq:marginal_HARA_CREMR}
\end{equation}
where $\alpha \neq 0$ and $\sigma \neq 0$, and the latter requires $\sigma \neq 1$ by the assumption in Theorem~\ref{prop:utility_NEW}. For these to be positive (and real valued), we impose $K>0$ and $1+ \alpha \sigma q > 0$. If $\alpha \sigma > 0$, the inequality $1+ \alpha \sigma q > 0$ is satisfied for all $q \ge 0$. If $\alpha \sigma < 0$, the inequality implies $0\le q < -\frac{1}{\alpha \sigma}$ since $q \ge 0$ must hold by Definition~\ref{def}. Differentiating \eqref{eq:marginal_HARA_CREMR} yields
\begin{equation}
u''(q)= - K\alpha (1+\alpha \sigma q)^{-\frac{\sigma+1}{\sigma}} \quad \mbox{and} \quad
u''(q)= - K (1+\alpha q) (1+\alpha \sigma q)^{-\frac{1}{\sigma}} q^{-2}.
\label{eq:marginal_second_HARA_CREMR}
\end{equation}
For the marginal utility to be strictly decreasing, we impose $\alpha>0$ in the former and $1+ \alpha q>0$ in the latter.\footnote{These conditions correspond to $u''(q) = - K (\alpha q+ \beta) (1+ \alpha \sigma q)^{-1+ \beta-\frac{1}{\sigma}} q^{-1-\beta}<0$ that we assume for the case $\beta \in (0,1)$ in Theorem~\ref{prop:T1}. Indeed, letting $\beta \to 0^+$ or $\beta \to 1^-$ in this inequality, we obtain $\alpha>0$ or $1+ \alpha q>0$, respectively.}
If $\alpha > 0$, these conditions are satisfied for all $q \ge 0$. If $\alpha < 0$, the latter condition implies $0 \le q < -\frac{1}{\alpha}$ since $q \ge 0$ must hold. Thus, Table~\ref{tab:QR} is applicable to the limit cases with $\beta \to 0^+$ and $\beta \to 1^-$.

Hence, as in Theorem~\ref{prop:T1}, we have three types of quantity restrictions for the utility function \eqref{eq:prop4_subutility_hypergeo}, i.e., $u'(q)>0$, $u''(q) < 0$, and $\lvert-\frac{1}{\alpha \sigma q}\rvert<1$, where the last condition is required from the definition of $ {}_2 F_{1} \bigl(1, 1; 1+ \frac{1}{\sigma}; -\frac{1}{ \alpha \sigma q} \bigr)$ in Theorem~\ref{prop:utility_NEW}. We have shown that the first two restrictions are subsumed by Table~\ref{tab:QR}. The last restriction, $\lvert- \frac{1}{\alpha \sigma q} \rvert<1$, requires $\alpha \sigma>0$.\footnote{\label{fn14}We can show that $\lvert-\frac{1}{\alpha \sigma q}\rvert<1$ requires $\alpha \sigma>0$ as follows. If $\alpha \sigma < 0$ were to hold, then $\lvert - \frac{1}{\alpha \sigma q} \rvert <1$ would imply $-\frac{1}{\alpha \sigma q} < 1$, so that $1<-\alpha \sigma q$, thus violating our assumption $1+\alpha \sigma q > 0$. The case with $\alpha \sigma=0$ is excluded by the assumptions in Theorem~\ref{prop:utility_NEW}. We will elaborate on this case in Corollary~\ref{coroll_NEW}.}
These inequalities, together with $1+\alpha \sigma q>0$, imply $1+\frac{1}{\sigma}>0$. This allows us to exclude the possibility that $1+\frac{1}{\sigma}$ can be a non-positive integer, in which case $ {}_2 F_{1} \bigl(1, 1; 1+ \frac{1}{\sigma}; -\frac{1}{ \alpha \sigma q} \bigr)$ is not well defined.\footnote{\label{fn15}If $\sigma>0$, we readily obtain $1+\frac{1}{\sigma}>0$. If $\sigma < 0$, we have $\alpha<0$ since $\lvert-\frac{1}{\alpha \sigma q}\rvert<1$ requires $\alpha\sigma>0$ as shown in footnote~\ref{fn14}. In that case, the restriction $1+\alpha q>0$ implies $q < -\frac{1}{\alpha}$. Since $\lvert - \frac{1}{\alpha \sigma q} \rvert<1$ implies $\frac{1}{\alpha \sigma} < q$, it follows that $\frac{1}{\alpha \sigma} < q < -\frac{1}{\alpha}$, which is a non-empty interval only if
$\frac{1}{\alpha \sigma} <  -\frac{1}{\alpha}$. This, together with $\alpha<0$, implies $\frac{1}{\sigma} >  -1$. Hence, we obtain $1+ \frac{1}{\sigma} > 0$. Recall that $\sigma=0$ is excluded in Theorem~\ref{prop:utility_NEW}.}
We summarize these restrictions in Assumption~\ref{as2}.

\setcounter{theorem}{1}
\begin{assumption}
\label{as2}
For the HARA utility function $u$ in \eqref{eq:prop4_subutility_HARA} to be strictly increasing and strictly concave, $1+\alpha \sigma q > 0$ and $\alpha> 0$ must hold, which corresponds to the left column in Table~\ref{tab:QR}. For the CREMR utility function $u$ in \eqref{eq:prop4_subutility_hypergeo} to be strictly increasing and strictly concave, $1+\alpha \sigma q > 0$ and $1+ \alpha q> 0$ must hold. Furthermore, \eqref{eq:prop4_subutility_hypergeo} requires $\alpha \sigma>0$ and $1+\frac{1}{\sigma}>0$, thus corresponding to the first row in Table~\ref{tab:QR}.
\end{assumption}

Since $ {}_2 F_{1} \bigl(1, 1; 1+ \frac{1}{\sigma}; -\frac{1}{ \alpha \sigma q} \bigr)$ requires $\lvert-\frac{1}{\alpha \sigma q}\rvert<1$, which in turn requires $\alpha \sigma>0$, we have $\frac{1}{\alpha \sigma} < q$, which appears to be restrictive given the ranges of quantities in the first row in Table~\ref{tab:QR}. However, we can disregard the quantity restriction, $\frac{1}{\alpha \sigma} < q$, using the integral representation. Indeed, we can use analytic continuation to extend the utility function \eqref{eq:prop4_subutility_hypergeo} to all quantities satisfying the first row in Table~\ref{tab:QR} as follows.

\setcounter{theorem}{1}
\begin{proposition}[Integral representation]
\label{integral2}
Suppose that $\alpha \neq 0$, $\sigma \notin \{0,1\}$, and Assumption~\ref{as2} holds. If $\sigma>0$, then the utility function \eqref{eq:prop4_subutility_hypergeo} \mbox{can be rewritten for all $q \in [0, \infty)$ as}
\begin{equation}
\label{eq:utility_special_continuation}
\lim_{\beta \to 1^-} u(q) = \textstyle \frac{K \sigma}{\sigma-1} \frac{(1+\alpha \sigma q)^{1-\frac{1}{\sigma}} } {q} \bigl[ q -\frac{\sigma-1}{\alpha \sigma} \frac{1}{\sigma} \int_0^1 (1-t)^{\frac{1}{\sigma}-1} \bigl(1 + \frac{t}{\alpha \sigma q} \bigr)^{-1} {\rm d}t  \bigr] + C.
\end{equation}
If $\sigma<0$, then the utility function \eqref{eq:prop4_subutility_hypergeo} can be rewritten for all $q \in [0, -\frac{1}{\alpha})$ as
\begin{eqnarray}
\lim_{\beta \to 1^-} u(q) = \textstyle \frac{K \sigma}{\sigma-1} \frac{(1+\alpha \sigma q)^{-\frac{1}{\sigma}} } {q}\bigl[(1+\alpha\sigma q)q -(\sigma-1)q - \frac{\sigma-1}{\alpha\sigma}\frac{1}{\sigma}\int_0^1(1-t)^{\frac{1}{\sigma}}\bigl(1+\frac{t}{\alpha\sigma q}\bigr)^{-1}{\rm d}t\bigr] + C.
\label{eq:Lebedev}
\end{eqnarray}
\end{proposition}

\paragraph{Proof.} See \ref{app:proposition2}.

\bigbreak
We already pointed out that letting $\alpha=0$ or $\beta=\frac{1}{\sigma}$ in the LFRRA \eqref{eq:LFRRA_general} yields the CRRA. We can also obtain the CRRA case by letting $\alpha \to \infty$ in \eqref{eq:LFRRA_general}, which is useful when $\beta=0$ or $\beta=1$. Similarly, once we set $K=(1-\beta)(1+\alpha \sigma)^\frac{1}{\sigma}+\beta (1+\alpha \sigma)^{\frac{1}{\sigma}-1}$ in the HARA and CREMR utility functions, \eqref{eq:prop4_subutility_HARA} and \eqref{eq:prop4_subutility_hypergeo}, while assuming that $\beta=0$ and $\sigma>1$ in the former case and $\beta=1$, $\sigma>0$, and $\sigma \neq 1$ in the latter case, we obtain the CRRA as a limit case: $\lim_{\alpha \to \infty} u(q)= \frac{\sigma}{\sigma-1} q^{\frac{\sigma-1}{\sigma}}+C$. Hence, both the HARA and CREMR forms derived from our general utility function in \eqref{eq:prop4_subutility_general_case} nest the CRRA.

Starting from the LFRRA in \eqref{eq:LFRRA_general}, Theorem~\ref{prop:T1} considers the case with $\beta \in (0,1)$ while imposing $\sigma\neq0$. Assuming $\alpha \neq 0$, we further proceed from Theorem~\ref{prop:T1} to obtain the HARA case with $\beta=0$ and the CREMR case with $\beta=1$ in Theorem~\ref{prop:utility_NEW}, where the latter additionally imposes $\sigma\neq1$. The following corollary complements Theorems~\ref{prop:T1} and \ref{prop:utility_NEW} by covering those excluded cases and highlights some special cases derived from those two theorems.\footnote{The case with $\alpha=0$ and $\beta=0$ yields $u(q)= K q+C$, which violates $u''<0$ and is thus excluded.}

\setcounter{theorem}{0}
\begin{corollary}
Let $\gamma(\cdot, \cdot)$, $B(\cdot,\cdot,\cdot)$, and ${\rm Ei} (\cdot)$ denote the lower incomplete gamma, incomplete beta, and exponential integral functions, respectively. If $\alpha=0$, then the LFRRA in \eqref{eq:LFRRA_general} implies the CRRA utility functions, $u(q) = \frac{K q^{1-\beta}}{1-\beta}+C$ for $\beta \in  (0,1)$ and $u(q)= K \ln q  + C$ for $\beta=1$. If $\alpha \neq 0$, then \eqref{eq:LFRRA_general} admits the following cases depending on the values of $\sigma$, which complements Theorems~\ref{prop:T1} and \ref{prop:utility_NEW} and highlights some special forms derived from those two theorems.

\label{coroll_NEW}
\begin{itemize}[parsep=0pt]
\item[] \hskip -1cm \emph{A.} Assume that $\beta \in (0,1)$. The utility function of the general form \eqref{eq:prop4_subutility_general_case} is derived for $\sigma \neq 0$.

\hskip -.44cm It extends to $\sigma =0$ and admits special forms for $\sigma =1$ and $\sigma= \frac{1}{\beta}$ as follows:
\begin{itemize}
\item[] \hskip -1.25cm $\bullet$ If $\sigma = 0$, then $u(q) =   \frac{K}{\alpha^{1-\beta}} \gamma(1-\beta, \alpha q) + C,$
which is the lower incomplete gamma utility.
\item[] \hskip -1.25cm $\bullet$ If $\sigma = 1$, then $u(q)= \frac{K}{(-\alpha)^{1-\beta}} B (-\alpha q,1-\beta,\beta) +C,$ which is the incomplete beta utility.
\item[] \hskip -1.25cm $\bullet$ If $\sigma = \frac{1}{\beta}$, then $u(q)= \frac{K q^{1-\beta}}{1-\beta}  +C,$ which is the CRRA utility.
\end{itemize}

\item[]
\hskip -1cm \emph{B.} Assume that $\beta$ goes to $0$ from above. The utility function of the HARA form \eqref{eq:prop4_subutility_HARA}  is derived

\hskip -.44cm for $\alpha \neq0$ and $\sigma \neq 0$ and requires $\alpha>0$ by Assumption~\ref{as2}. It extends to $\sigma =0$ and admits

\hskip -.44cm  special forms for $\sigma =-1$ and $\sigma= 1$ as follows:

\begin{itemize}
\item[] \hskip -1.25cm $\bullet$ If $\sigma = 0$, then $u(q) = \frac{K}{\alpha} (1-{\rm e}^{-\alpha q})+C$, which is the CARA utility.
\item[] \hskip -1.25cm $\bullet$ If $\sigma = -1$, then $u(q) = - \frac{K}{2} (\alpha q^2- 2q)+C$, which is the quadratic utility.
\item[] \hskip -1.25cm $\bullet$ If $\sigma = 1$, then $u(q) = \frac{K}{\alpha} \ln (\alpha q+1)+C$, which is the (translated) log utility.
\end{itemize} 

\item[] \hskip -1cm \emph{C.} Assume that $\beta$ goes to $1$ from below. The utility function of the CREMR \mbox{form \eqref{eq:prop4_subutility_hypergeo} is derived}

\hskip -.44cm  for $\alpha \neq 0$ and $\sigma \notin \{0,1\}$ and requires $\alpha \sigma>0$ and $1+\frac{1}{\sigma}>0$ by  Assumption~\ref{as2}. It extends~to

\hskip -.44cm  $\sigma = \{0 ,1\}$ as follows:
\begin{itemize}
\item[] \hskip -1.25cm $\bullet$ If $\sigma = 0$, then $u(q) =  K\, {\rm Ei}(-\alpha q) + C$, which is the exponential integral utility.
\item[] \hskip -1.25cm $\bullet$ If $\sigma = 1$, then $u(q)= K \ln q  + C$, which is the log utility, \mbox{a special case of the CRRA utility.}
\end{itemize}
\end{itemize}
\vspace{-.2cm}
Table \ref{tab:special_cases} summarizes these cases.
\medbreak
\begin{table}[h]
\caption{Special cases of utility functions.}
\label{tab:special_cases}
\centering
\scalebox{0.85}{
\begin{tabular}{l|lll}
\hline
& \multicolumn{3}{c}{} \\[-3.5mm]
& \multicolumn{1}{c}{HARA} & \multicolumn{1}{c}{Gauss hypergeometric} & \multicolumn{1}{c}{CREMR} \\
& \multicolumn{1}{c}{$\beta = 0$} & \multicolumn{1}{c}{$\beta \in (0,1)$} & \multicolumn{1}{c}{$\beta = 1$}  \\[1mm]
\hline 
& \multicolumn{3}{c}{} \\[-3.5mm]
$\sigma=0$ & \ $u(q)= \frac{K}{\alpha} (1-{\rm e}^{-\alpha q})$  & \ \  $u(q)= \frac{K}{\alpha^{1-\beta}} \gamma(1-\beta, \alpha q)$  \quad &  \ \  $u(q)= K\, {\rm Ei}(-\alpha q) $ \quad \\[2.5mm]
$\sigma=1$ & \ $u(q)= \frac{K}{\alpha} \ln (\alpha q+1)$  & \ \  $u(q)= \frac{K}{(-\alpha)^{1-\beta}} B (-\alpha q,1-\beta,\beta) $ \quad & \ \  $u(q)= K \ln q  $ \quad \\[2.5mm]
$\sigma=-1$  & \  $u(q)= - \frac{K}{2} (\alpha q^2- 2q) $  &  &  \\[1.5mm]
\hline
\multicolumn{4}{p{16.5cm}}{\scriptsize{{\it Notes}: {\rm We suppress the additive constant $C$ in each expression.
}}}
\end{tabular}
}
\end{table}
\vspace{-.75cm}
\end{corollary}

\paragraph{Proof.} See \ref{app:corollary1}.

\bigbreak

We have characterized the utility function that gives rise to the LFRRA, and we have shown how to obtain special cases. Another question is how our CREMR utility relates to that proposed by Mr\'azov\'a et al. (2021). Recall that the second part of Theorem~\ref{prop:utility_NEW}, where $\beta$ goes to $1$ from below, does not apply to the case with $\alpha\sigma<0$ by Assumption~\ref{as2}, whereas Mr\'azov\'a et al. (2021) have no restriction on the sign of their $\gamma$ parameter. Yet, we can nest this case in our approach by relaxing our assumptions that $K>0$, and that $1+ \alpha \sigma q > 0$. Indeed, assuming that $K\in\mathbb{C}$, where $\mathbb{C}$ is the set of complex numbers, and that $1+ \alpha \sigma q<0$, we can retain the positive and real-valued marginal utility in \eqref{eq:marginal}. By making a change of variables we can subsume the CREMR utility function in Mr\'{a}zov\'{a} et al. (2021, their equation (12)) as follows:

\setcounter{theorem}{1}
\begin{corollary}[Another special case]
\label{corollaryMNP}
Let $\alpha = - \frac{1}{\gamma \sigma }$, $K=- \widetilde \beta \gamma(-\frac{1}{\gamma})^\frac{1}{\sigma}$, and $C = \kappa$ in \eqref{eq:prop4_subutility_hypergeo}, where $\widetilde \beta>0$, $\sigma>0$, $\gamma \in (-\infty, \infty)$, and $\kappa \in (-\infty, \infty)$ are parameters. Then, the utility function is given by
\begin{equation}
\textstyle u(q) =  \kappa + \widetilde\beta \frac{\sigma}{\sigma-1}  \frac{(q-\gamma)^{\frac{\sigma-1}{\sigma}} } {q} \bigl[ q +
\gamma(\sigma-1)
{}_2 F_{1} \bigl(1, 1 ; 1+\frac{1}{\sigma}; \frac{\gamma}{q}   \bigr)  \bigr],
\label{eq:LFRRA_utility_C2}
\end{equation}
where ${}_2 F_{1} \bigl(1, 1 ; 1+\frac{1}{\sigma}; \frac{\gamma}{q}   \bigr) = \sum_{n=0}^\infty \frac{(1)_n (1)_n }{(1+\frac{1}{\sigma})_n } \frac{(\frac{\gamma}{q})^n}{n!}$, with $\lvert \frac{\gamma}{q} \rvert<1$. Using the integral representation, \eqref{eq:LFRRA_utility_C2} can be rewritten as
\begin{equation}
\textstyle u(q) =  \kappa + \widetilde\beta \frac{\sigma}{\sigma-1}  \frac{(q-\gamma)^{\frac{\sigma-1}{\sigma}} } {q} \bigl[ q +
\frac{\gamma (\sigma-1)}{\sigma} \int_0^1 (1-t)^{\frac{1}{\sigma}-1} \bigl(1-\frac{t \gamma}{q} \bigr)^{-1}{\rm d}t  \bigr].
\label{eq:LFRRA_utility_C2_int}
\end{equation}
\end{corollary}

\paragraph{Proof.} See \ref{app:corollary2}.

\bigbreak
The restriction $\lvert \frac{\gamma}{q} \rvert<1$ for \eqref{eq:LFRRA_utility_C2} has the following two cases.\footnote{Strictly speaking, there is another case with $\gamma=0$, which we discuss in \ref{app:corollary2}.}
First, if $\gamma > 0$, then $q> \gamma$ must hold.\footnote{\label{fn20}Although $K$ can be complex and $\alpha \sigma<0$ when $\gamma>0$, this poses no problem because the complex term $(-\frac{1}{\gamma})^\frac{1}{\sigma}$ in $K$ is offset by the complex term $(1+\alpha \sigma q)^{1-\frac{1}{\sigma}}$ in \eqref{eq:prop4_subutility_hypergeo}, which equals $ \bigl(1-\frac{q}{\gamma}\bigr)^{1-\frac{1}{\sigma}} = \bigl[\bigl( -\frac{1}{\gamma} \bigr) ( q- \gamma) \bigr]^{1-\frac{1}{\sigma}},$ to yield a real-valued utility.}
This condition is always satisfied because $u'(q)= \widetilde \beta (q-\gamma)^\frac{\sigma-1}{\sigma} q^{-1}>0$ and $\widetilde \beta>0$ imply $q>\gamma$. Second, if $\gamma < 0$, then $q> - \gamma$ must hold. This case has a domain restriction: \eqref{eq:LFRRA_utility_C2} is not defined for $q \in  (0,  -\gamma]$. Equation \eqref{eq:LFRRA_utility_C2_int} allows us to disregard this restriction and to depict Figure 3(c) in Mr\'{a}zov\'{a} et al. (2021) for $q \in (0, -\gamma]$. Note also that Mr\'{a}zov\'{a} et al. (2021) impose $\sigma >1$ in~\eqref{eq:LFRRA_utility_C2}. The same restriction is required for monopolistic competition that we analyze in the next section, where profit maximizing prices depend on $\sigma$. However, it is sufficient to assume that $\sigma>0$ in different contexts (e.g., when equation~\eqref{eq:LFRRA_utility_C2} is embedded in a perfectly competitive model with intertemporally optimizing consumers).

\section{Monopolistic competition}
\label{sec:mc}

There is a mass $L$ of consumers in the economy. Let ${\cal U} \equiv \int_{\Omega} u (q(\omega)){\rm d}\omega$ denote an additively separable upper-tier utility function for the representative consumer, where $u$ is a utility function satisfying both $u' > 0$ and $u'' < 0$, $q(\omega) \ge 0$ is consumption of variety $\omega$, and $\Omega$ denotes the set of varieties available in the economy. The representative consumer maximizes $\cal U$ subject to the budget constraint $\int_{\Omega}p(\omega)q(\omega){\rm d}\omega = y$, where $p(\omega)$ is the price of variety $\omega$ and $y$ is the income of that consumer. This yields the first-order condition $u'(q(\omega)) = {\lambda} p(\omega)$, where ${\lambda}>0$ is the Lagrange multiplier.

There is a continuum of firms producing differentiated varieties sold in a monopolistically competitive market. Each variety $\omega$ is produced by a distinct firm, so that the variety index $\omega$ refers to a unique producer. Each firm $\omega$ incurs a common fixed cost $f \ge 0$ when it produces and has a firm-specific marginal cost $m (\omega)$ that does not depend on the output. We suppress $\omega$ when there is no confusion.

A firm with marginal cost $m$ maximizes operating profits $(p-m ) L q -f$, where $Lq$ is the aggregate demand for the firm's variety. Using the first-order condition for utility maximization, the firm's profit maximization problem is given by $\max_q  \bigl[\frac{u'(q)}{{\lambda}}-m \bigr] Lq -f$. Since there is a continuum of firms, no firm can affect ${\lambda}$. Thus, the first-order condition for profit maximization is given by $\frac{u''(q)}{{\lambda}}Lq+ \bigl[ \frac{u'(q)}{{\lambda}} -m \bigr] L=0$, and the profit maximizing price $p(m)$ satisfies:
\begin{equation}
\frac{p(m)-m}{p(m)} = 1-\mu(m) = - \frac{q(m) u''(q(m))}{u'(q(m))},
\label{eq:generalFOC}
\end{equation}
where $\mu (m)\equiv \frac{m}{p(m)}$ denotes the inverse of the markup and $1-\mu (m)$ is the Lerner index (Lerner, 1934). For the profit-maximizing price to be valid, we focus on the case with $p(m) =\frac{m}{\mu (m)} \in [m, \infty)$, which implies $\mu (m) \in (0,1]$. In what follows, we suppress $m$ to simplify the notation when there is no confusion.

The second-order condition (SOC) for profit maximization is given by $\frac{u'''(q)}{{\lambda}}Lq+\frac{u''(q)}{{\lambda}}L + \frac{u''(q)}{{\lambda}}L =\frac{u'''(q)}{{\lambda}}Lq+2\frac{u''(q)}{{\lambda}}L  =  - \frac{L u''(q)}{{\lambda}} \bigl[  - \frac{q u'''(q)}{u''(q)} - 2  \bigr] <0$. Since $- \frac{L u''(q)}{{\lambda}}>0$, the SOC becomes
\begin{equation}
2 -\left[- \frac{q u'''(q)}{u''(q)}  \right] >0,
\label{eq:SOC_general}
\end{equation}
which must hold for any $q$ satisfying the first-order condition for profit maximization.

\subsection{Price equilibrium}
\label{sec:eq}

To analyze the price equilibrium, we consider the following two cases. First, assume that either $\alpha = 0$ or $\beta = \frac{1}{\sigma}$ holds, each of which corresponds to the CRRA case. If $\alpha=0$, then $u'(q)= K q^{-\beta}$ by \eqref{eq:marginal}, which implies $\frac{p-m}{p}=1-\mu=- \frac{ q u''(q)}{u'(q)}=\beta$ by \eqref{eq:generalFOC}. Thus, we obtain the price equilibrium $p(m)=\frac{m}{\mu (m)} = \frac{m}{1-\beta}$, which requires the parameter restriction, $\beta \in (0,1)$, since $\mu>0$ and $u''<0$ must hold. Similarly, if $\beta=\frac{1}{\sigma}$, then $u'(q) = K q^{-\frac{1}{\sigma}}$ by \eqref{eq:marginal}. Thus, we obtain $\frac{p-m}{p}=1-\mu=- \frac{ q u''(q)}{u'(q)}= \frac{1}{\sigma}$ and $p(m)=\frac{m}{\mu (m)} = \frac{\sigma m}{\sigma-1}$, which requires the parameter restriction, $\sigma>1$. In both cases, $u'>0$, $u''<0$, $0<\mu \le 1$, and the second-order condition \eqref{eq:SOC_general} hold for all $q \ge 0$ under these parameter restrictions.

Next, assume that $\alpha \neq 0$ and $\beta \neq \frac{1}{\sigma}$. Since $1-\mu= - \frac{ q u''(q)}{u'(q)}$ holds by \eqref{eq:generalFOC}, we can use \eqref{eq:LFRRA_general} to obtain:
\vspace{-.25cm}
\begin{equation}
q= \frac{1}{\alpha} \frac{1-\beta -\mu}{1- \sigma (1-\mu) }.
\label{eq:FOCprof}
\end{equation}
As in the CRRA case, five inequality conditions must be satisfied. First, $\frac{1}{\alpha} \frac{1-\beta -\mu}{1- \sigma (1-\mu) } \ge 0$ must hold for $q$ in \eqref{eq:FOCprof} to be nonnegative. Second, \eqref{eq:marginal} and \eqref{eq:marginal_HARA_CREMR} imply that $u' > 0$ requires that $1+\alpha \sigma q > 0$, which implies $1+ \frac{\sigma(1-\beta -\mu)}{1- \sigma (1-\mu) }  = \frac{1- \sigma (1-\mu) + \sigma(1-\beta -\mu)}{1- \sigma (1-\mu) } = \frac{1 -  \beta \sigma}{1- \sigma (1-\mu) }  > 0$. Third, \eqref{eq:marginal_second} and \eqref{eq:marginal_second_HARA_CREMR} imply that $u'' < 0$ requires that $\alpha q+\beta>0$ for $\beta \in (0,1]$, and that $\alpha>0$ for $\beta=0$. Fourth, $0< \mu \le 1$ must hold for the markup to satisfy $1 \le \frac{1}{\mu}  < \infty$. Last, the second-order condition \eqref{eq:SOC_general} must be satisfied. Using \eqref{eq:marginal} and \eqref{eq:marginal_HARA_CREMR}, it can be expressed as follows:
\begin{equation}
\frac{\alpha q}{\alpha q+\beta} + \frac{1-\alpha q-\beta}{\alpha \sigma q+1} >0.
\label{eq:SOC_LFRRA}
\end{equation}
We first derive the ranges of markups $\frac{1}{\mu}$ and quantities $q$ that satisfy these five inequality conditions. We then establish the existence and uniqueness of the markup and quantity functions for each firm that relate $m$ to $\frac{1}{\mu (m)}$ and $q(m)$ in price equilibrium.

Combining these two cases, we obtain the following proposition.
\vspace{-.35cm}

\begin{proposition}[Ranges of markups and quantities]
\label{prop:markup_bounds}
If $\alpha \neq 0$ and $\beta \neq \frac{1}{\sigma}$, then the ranges of markups $\frac{1}{\mu}$ and quantities $q$ satisfying $q \ge 0$, $u' > 0$, $u'' < 0$, $0<\mu \le 1$, and the second-order condition \eqref{eq:SOC_LFRRA} are summarized in Table~\ref{tab:tab2}. Marginal cost pricing $\frac{1}{\mu}=1$ is admissible only when $\beta=0$. If $\alpha =0$ or $\beta = \frac{1}{\sigma}$, then the markup is given by $\frac{1}{\mu} = \frac{1}{1-\beta} \in (1, \infty)$ or $\frac{1}{\mu} = \frac{\sigma}{\sigma-1} \in (1, \infty)$ for all $q \ge 0$, respectively, and $u' > 0$, $u'' < 0$, $0<\mu \le 1$, and \eqref{eq:SOC_LFRRA} are satisfied for all $q \ge 0$.
\vspace{-.4cm}
\end{proposition}

\paragraph{Proof.} See \ref{app:proofs3}.

\begin{table}
\caption{The ranges of markups and quantities satisfying $q \ge 0$, $u' > 0$, $u'' < 0$, $0<\mu \le 1$, and the second-order condition \eqref{eq:SOC_LFRRA}.}
\label{tab:tab2}
\vspace{-.35cm}
\scalebox{0.8}{
\hskip -.65cm \begin{tabular}{l|c|c|c|cc}
\multicolumn{5}{c}{Case 1: $\alpha >0$.} \\
\hline
 & \multicolumn{2}{c|}{Markups} & \multicolumn{2}{c}{Quantities} \\ \cline{2-5}
 & $\sigma \le 1$  & $\sigma > 1 $ & $\sigma \le 1$  & $\sigma > 1 $    \\ \hline
\multirow{2}{*}{Case 1a. $1-\beta \sigma>0$ and $\beta \neq 1$} &
\multirow{2}{*}{$\frac{1}{1-\beta} \le \frac{1}{\mu} < \infty$} &
\multirow{2}{*}{$\frac{1}{1-\beta} \le \frac{1}{\mu} < \frac{\sigma}{\sigma-1}$} &
\multirow{2}{*}{$0 \le q < -\frac{1-\beta}{\alpha (\sigma-1)}$$^\dagger$} &
\multirow{2}{*}{$0 \le q < \infty$}  \\
&
&
&
&
\\ \hline
&
&
&
&
\\[-3mm]
\multirow{2}{*}{Case 1b.} \hskip .05cm $1-\beta \sigma<0$ and $\beta\in(0,1)$ &
\multirow{2}{*}{n.a.} &
$\frac{\sigma}{\sigma-1} < \frac{1}{\mu }  \le  \frac{1}{1-\beta}$ &
\multirow{2}{*}{n.a.} & $0 \le q < \infty$  \\[2mm]  
\hskip 1.73cm
$1-\beta \sigma<0$ and $\beta=1$
&
& $\frac{\sigma}{\sigma-1} < \frac{1}{\mu} < \infty$
&
& $0 < q < \infty$
\\[1.5mm] \hline 
\multicolumn{5}{c}{}   \\[-3mm]
\multicolumn{5}{c}{Case 2: $\alpha <0$.} \\ \hline
 & \multicolumn{2}{c|}{Markups} & \multicolumn{2}{c}{Quantities} \\ \cline{2-5}
 & $\sigma \le 1$  & $\sigma > 1 $ & $\sigma \le 1$  & $\sigma > 1 $    \\ \hline 
\multirow{2}{*}{Case 2a. $1-\beta \sigma>0$ and $\beta \in ( 0, 1)$} &
\multirow{2}{*}{$\frac{(1-\beta)\sigma+1}{(1-\beta)\sigma+\Delta} < \frac{1}{\mu} \le \frac{1}{1-\beta}$$^\ddagger$} &
\multirow{2}{*}{$\frac{(1-\beta)\sigma+1}{(1-\beta)\sigma+\Delta} < \frac{1}{\mu} \le \frac{1}{1-\beta}$} &
\multirow{2}{*}{$ 0 \le  q  <  -\frac{1 - \beta - \Delta}{\alpha (\sigma-1)}$$^\S$} & 
\multirow{2}{*}{$ 0 \le  q  <  -\frac{1 - \beta - \Delta}{\alpha (\sigma-1)} $} \\
&
&
&
&
\\ \hline 
\multirow{2}{*}{Case 2b. $1-\beta \sigma<0$ and $\beta \in ( 0, 1)$}
&
\multirow{2}{*}{n.a.} &
\multirow{2}{*}{$\frac{1}{1-\beta}  \le \frac{1}{\mu }  < \infty$} &
\multirow{2}{*}{n.a.} &
\multirow{2}{*}{$0  \le q  < -\frac{1-\beta}{\alpha (\sigma-1)}$}  \\ 
&
&
&
&
\\ \hline
\multicolumn{5}{p{21.4cm}}{\scriptsize {\it Notes}: We let $\Delta \equiv \sqrt{(1-\beta)(1-\beta \sigma)}$. $^{\dagger}$ means that when $\sigma=1$, $0 \le q < -\frac{1-\beta}{\alpha (\sigma-1)}$ must be replaced with $0 \le q < \infty$. $^{\ddagger}$ means that when $\sigma=1$, $ \frac{(1-\beta)\sigma+1}{(1-\beta)\sigma+\Delta}$ must be replaced with $ \frac{2-\beta}{2(1-\beta)}$. $^\S$ means that when $\sigma=1$, $ -\frac{1 - \beta - \Delta}{\alpha (\sigma-1)} $ must be replaced with $-\frac{\beta}{2\alpha}$. Marginal cost pricing $\frac{1}{\mu}=1$ is admissible only when $\beta=0$ in Case 1a because $\frac{\sigma}{\sigma-1} > 1$ holds in Case 1b, $\frac{(1-\beta)\sigma+1}{(1-\beta)\sigma+\Delta} >1$ holds in Case 2a, and $\beta=0$ is not admissible in Case 2b. The case with $\alpha=0$ corresponds to the CES and requires that $\beta \in (0,1)$. The case with $1-\beta \sigma=0$ also corresponds to the CES and requires that  $\sigma > 1 $. If $\alpha=0$ or $\beta =\frac{1}{\sigma}$, then the markup is given by $\frac{1}{\mu}  =\frac{1}{1-\beta} \in  (1, \infty)$ or $\frac{1}{\mu}  =\frac{\sigma}{\sigma-1} \in (1, \infty)$, respectively, and the range of quantities is given by $0\le q<\infty$. Since these two cases are well known, we omit them from the table.}
\end{tabular}
}
\vspace{-.25cm}
\end{table}

\smallbreak
Having derived the ranges of markups $\frac{1}{\mu}$ and quantities $q$ that satisfy the five inequality conditions, we now establish the existence and uniqueness of the markup and quantity functions for each firm that relate $m$ to $\frac{1}{\mu(m)}$ and $q(m)$ in price equilibrium.

Using the marginal utility function \eqref{eq:marginal}, which nests the HARA ($\beta=0$) and CREMR ($\beta=1$) cases in \eqref{eq:marginal_HARA_CREMR}, the first-order condition for utility maximization is given by $u'(q) =K(1+\alpha \sigma q)^{\beta-\frac{1}{\sigma}} q^{-\beta}= {\lambda} p$. We can rewrite it as $(1+ \alpha \sigma q)^{\beta-\frac{1}{\sigma}} q^{-\beta} = \frac{{\lambda} p}{K} = \frac{{\lambda} m}{K} \frac{p}{m}$. Letting $x \equiv  \frac{{\lambda} m}{K} > 0$ and $\mu = \frac{m}{p}$, we have
\vspace{-.35cm}
\begin{equation}
(1+ \alpha \sigma q)^{\beta-\frac{1}{\sigma}} q^{-\beta}
= \frac{x}{\mu}.
\label{eq:FOCutil}
\end{equation}
Since $\lim_{q\to 0^+} u'(q)=K(1+\alpha \sigma q)^{\beta-\frac{1}{\sigma}} q^{-\beta} =\infty$ for $\beta>0$ and the limit is finite for $\beta=0$, we know that $q>0$ must hold for $\beta>0$ whereas $q=0$ may hold for $\beta=0$. Combining \eqref{eq:FOCutil} with the first-order condition \eqref{eq:FOCprof} for profit maximization, as well as the ranges of markups and quantities in Table~\ref{tab:tab2}, we can show the following result:

\vspace{-.25cm}

\begin{proposition}[Existence and uniqueness of a markup function]
\label{prop:propEU}
If $\alpha\neq 0$ and $\beta \neq \frac{1}{\sigma}$, then there exists a unique markup function $\frac{1}{\mu}=\frac{1}{{\cal W} (x, \alpha, \beta, \sigma)} \ge 1$ satisfying the first-order condition in \mbox{Table~\ref{tab:FOC}} and the second-order condition \eqref{eq:SOC_LFRRA} conditional on the restrictions on $x$ in Table~\ref{tab:tab3}. If $\alpha = 0$ and $\beta \in (0,1)$, then there exists a unique markup function $\frac{1}{\mu} = \frac{1}{{\cal W} (x, 0 , \beta,\sigma) }  = \frac{1}{1-\beta} >1$ satisfying the first-order condition \eqref{eq:generalFOC} and the second-order condition \eqref{eq:SOC_LFRRA}. If $\beta = \frac{1}{\sigma}$ and $\sigma>1$, then there exists a unique markup function $\frac{1}{\mu} = \frac{1}{{\cal W} (x,\alpha, \frac{1}{\sigma},\sigma) }  = \frac{\sigma}{\sigma-1} >1$ satisfying the first-order condition
\eqref{eq:generalFOC}
and the second-order condition \eqref{eq:SOC_LFRRA}.\footnote{\label{fn24}Alternatively, assume that $\sigma > 1$. Then, recalling $x= \frac{{\lambda} m}{K}$, using $K=(1-\beta)(1+\alpha \sigma)^\frac{1}{\sigma}+\beta (1+\alpha \sigma)^{\frac{1}{\sigma}-1}$, and letting $\alpha \to \infty$ in the bottom-right expression of Table~\ref{tab:FOC}, we can show that there exists a unique markup function $\frac{1}{\mu}= \lim_{\alpha \to \infty} \frac{1}{{\cal W} (x(\alpha), \alpha, \beta, \sigma)} =  \frac{\sigma}{\sigma-1} > 1$. Furthermore,
when $\alpha \to \infty$, the second-order condition \eqref{eq:SOC_LFRRA} reduces to $\frac{\sigma-1}{\sigma}>0$, which holds for all $\sigma>1$.}
We refer to ${\cal W} (x, \alpha, \beta, \sigma)$ as a generalized Lambert $W$ function.
\end{proposition}

\vspace{-.35cm}

\begin{table}[h]
\caption{First-order condition whose solution yields the markup function $\frac{1}{\mu}=\frac{1}{{\cal W} (x, \alpha, \beta,\sigma)}$.}
\label{tab:FOC}
\centering
\begin{tabular}{l|lll}
\hline
 & \multicolumn{2}{c}{ } \\[-4mm]
& \multicolumn{1}{c}{$\beta=0$} & &  \multicolumn{1}{c}{$\beta \in (0,1]$} \\[1mm] \hline
\\ [-2.5mm]
$\sigma  = 0$ &  $\textstyle 
1-\mu =   \ln \left( \frac{\mu}{x} \right)$ & & $\textstyle 1-\mu = \ln \left( \frac{\mu}{x} \right) + \beta \left[ 1 -  \ln \left( \frac{1-\beta-\mu}{\alpha} \right) \right]^\dagger$
\\ [4mm]
$\sigma \neq 0$ &  $\textstyle 1-\mu = \frac{1- \left( \frac{\mu}{x} \right)^{-\sigma}}{\sigma}$ & & $\textstyle 1-\mu=  \frac{1- \left( \frac{\mu}{x} \right)^{-\sigma} (1 -\beta \sigma) \left[  \frac{1-\beta -\mu}{\alpha ( 1 -\beta \sigma) } \right]^{\beta \sigma}}{\sigma}$  \\   [2.5mm]  \hline
\multicolumn{4}{p{12cm}}{\scriptsize {\it Note}: $^\dagger$ means that $\beta=1$ must be excluded since $\beta=1$ is possible only when $\sigma>1$ (see Case~1b in Table~\ref{tab:tab2}). If $\alpha=0$ or $\beta=\frac{1}{\sigma}$, then we cannot use \eqref{eq:FOCprof} as it relies on $\alpha\neq0$ or $\beta \neq \frac{1}{\sigma}$. However, we can combine \eqref{eq:LFRRA_general} and \eqref{eq:generalFOC} to obtain the first-order condition $1-\mu = \beta$ or $1-\mu = \frac{1}{\sigma}$.}
\end{tabular}
\vspace{-1cm}
\end{table}

\begin{table}[t]
\caption{The admissible range of $x$ ensuring the existence of a unique markup function.}
\label{tab:tab3}
\begin{center}
\vskip -.5cm
{
\begin{tabular}{c|c|c}
\multicolumn{3}{c}{Case 1: $\alpha >0$.} \\
\hline 
& $\sigma \le 1$  & $\sigma >1$    \\ \hline
\multirow{2}{*}{Case 1a.} $1-\beta \sigma>0$ and $\beta\in(0,1)$ & $0  < x  < \infty $  & $0 < x< \infty$\\ 
\hskip 1cm $1-\beta \sigma>0$ and $\beta = 0$ &  $0 < x \le 1$  & $0 < x \le 1$ \\ \hline
\multirow{2}{*}{Case 1b.} $1-\beta \sigma<0$ and $\beta\in(0,1)$ & \multirow{2}{*}{n.a.}& $0 < x < \infty$ \\ 
\hskip 1cm $1-\beta \sigma<0$ and $\beta = 1$  & 
 & $0 < x < \alpha(\sigma-1)$ \\  \hline
\multicolumn{3}{c}{} \\[-2mm]
\multicolumn{3}{c}{Case 2: $\alpha <0$.} \\
\hline
 & $\sigma \le 1$  & $\sigma >1$    \\ \hline
  &&\\[-4mm]
Case 2a. $1-\beta \sigma>0$ and $\beta\in(0,1)$ & $
\underline x <  x <\infty$ & $
\underline x <  x <\infty$ \\[1mm]
 \hline
   &&\\[-4mm]
Case 2b. $1-\beta \sigma<0$ and $\beta\in(0,1)$ & n.a.& $0  < x < \infty$ \\[1mm]  
\hline
\multicolumn{3}{p{12.7cm}}{\scriptsize {\it Notes}: We let $\underline x \equiv \overline \mu \left[ \frac{1 -\beta \sigma}{1- \sigma ( 1-\overline \mu ) }  \right]^{\beta- \frac{1}{\sigma}} \left[ \frac{1}{\alpha} \frac{1-\beta -\overline \mu }{1- \sigma (1-\overline \mu) } \right]^{-\beta}$, where $\overline \mu \equiv \frac{(1-\beta)\sigma + \Delta}{(1-\beta)\sigma+1}$ and $\Delta \equiv \sqrt{(1-\beta)(1-\beta \sigma)}$. The case with $\alpha=0$ corresponds to the CES and requires that $\beta \in (0,1)$. The case with $1-\beta \sigma=0$ also corresponds to the CES and requires that $\sigma > 1$. If $\alpha=0$ or $\beta =\frac{1}{\sigma}$, then the markup is given by $\frac{1}{\mu}  =\frac{1}{1-\beta} \in  (1, \infty)$ or $\frac{1}{\mu}  =\frac{\sigma}{\sigma-1} \in (1, \infty)$, respectively, and the admissible range of $x$ is given by $0 <  x<\infty$. Since these two cases are well known, we omit them from the table.}
\end{tabular}
}
\end{center}
\vspace{-.45cm}
\end{table}

\paragraph{Proof.} See \ref{app:proofs4}.

\medbreak

In what follows, we discuss our generalized Lambert $W$ function, provide some historical remarks, and derive the demand function for the LFRRA family of utility functions.

\vspace{-.3cm}

\paragraph{Generalized Lambert $W$ function.} The Lambert $W$ function analyzed in Corless et~al. (1996) corresponds to the solution $\mu= {\cal W} (x, \alpha, 0, 0)$ to the first-order condition, $1-\mu = \ln \bigl( \frac{\mu}{x} \bigr)$ in the top-left cell in Table~\ref{tab:FOC}. In the context of monopolistic competition, $\frac{1}{\mu}=\frac{1}{{\cal W} (x, \alpha, 0, 0)}$ corresponds to the equilibrium markup in the CARA model (Behrens et al., 2014, 2017, 2020; Egger and Huang, 2025). We generalize the Lambert $W$ function and the markup function in three directions: the direction of $\sigma$, $\frac{1}{\mu} = \frac{1}{{\cal W} (x, \alpha, 0, \sigma)}$, which is the HARA case; the direction of $\beta$, $\frac{1}{\mu} =\frac{1}{ {\cal W} (x, \alpha, \beta, 0)}$, which is the incomplete gamma case; and the direction of both $\beta$ and $\sigma$, $\frac{1}{\mu} =\frac{1}{ {\cal W} (x, \alpha, \beta, \sigma)}$, which are the CREMR and Gauss hypergeometric cases.\footnote{The special case of our generalized Lambert function, ${\cal W} (x, \alpha, \beta, 0)$, defined as the solution to the equation in the top-right cell of Table~\ref{tab:FOC}, i.e.,  $\frac{\mu}{(\frac{1-\beta-\mu}{\alpha})^{\beta}} {\rm e}^{-(1-\beta- \mu)} =x$, is similar to, but differs from $\frac{P(\mu)}{Q(\mu)} {\rm e}^{R(\mu)} =x$ in Mez\H{o} (2022), where $P(\mu)$, $Q(\mu)$, and $R(\mu)$ are polynomials. In our case, $(\frac{1-\beta-\mu}{\alpha})^{\beta}$ cannot be a polynomial because in the top-right cell of Table~\ref{tab:FOC}, the case with $\beta=1$ must be excluded by that table's note.}
We discuss each case below.

\vspace{-.2cm}

\begin{itemize}
\item {\bf HARA case.}
As shown in Theorem~\ref{prop:utility_NEW}, when $\beta \to 0^+$, we get  the HARA utility function of the   form $u(q) = \frac{K}{\alpha (\sigma-1)} \bigl[(1+\alpha \sigma q)^\frac{\sigma-1}{\sigma} -1 \bigr]+C$, where $K>0$ and $C$ are constants. Solving the bottom-left expression of Table~\ref{tab:FOC}, we obtain the associated markup function $\frac{1}{\mu} = \frac{1}{{\cal W}(x,\alpha, 0, \sigma)}$, which nests the quadratic ($\sigma=-1$), log ($\sigma=1$), and CARA  ($\sigma \to 0$) utility functions. Imposing $\sigma=-1$, $\sigma=1$, or $\sigma \to 0$ yields the markup function $\frac{1}{\mu} = \frac{1}{{\cal W} (x, \alpha, 0, -1)} = \frac{1+x}{2x} \ge 1$ for all $0< x \le 1$,  $\frac{1}{\mu} = \frac{1}{{\cal W} (x, \alpha, 0, 1)} = \frac{1}{\sqrt{x}}  \ge 1$  for all $0< x \le 1$, or $\frac{1}{\mu} = \frac{1}{{\cal W} (x, \alpha, 0, 0)} = \frac{1}{W({\rm e} x)} \ge 1$ for all $0<x \le 1$, respectively, where $W$ is the original Lambert $W$ function. Thus, this case generalizes the Lambert $W$ function in the direction of $\sigma$.

\vspace{-.2cm}

\item {\bf Incomplete gamma case.}
As shown in Corollary~\ref{coroll_NEW}, when $\beta \in (0,1)$ and $\sigma \to 0$, we get the lower incomplete gamma utility of the form $u(q) = \frac{K}{\alpha^{1-\beta}} \gamma(1-\beta, \alpha q) + C$, where $K>0$ and $C$ are constants. Solving the top-right expression of Table~\ref{tab:FOC}, we obtain the associated markup function $\frac{1}{\mu} = \frac{1}{{\cal W}(x,\alpha, \beta, 0)}$, which nests the CARA utility function ($\beta \to 0$). Thus, this case generalizes the Lambert $W$ function in the direction of $\beta$.

\vspace{-.2cm}

\item {\bf CREMR case.} As shown in Theorem~\ref{prop:utility_NEW}, when $\beta \to 1^-$, we get the CREMR utility of the form $u(q) = \textstyle \frac{K \sigma}{\sigma-1} \frac{(1+\alpha \sigma q)^{1-\frac{1}{\sigma}} } {q} \bigl[ q -\frac{\sigma-1}{\alpha \sigma} {}_2 F_{1} \bigl(1, 1 ; 1+\frac{1}{\sigma}; - \frac{1}{\alpha \sigma q}   \bigr)  \bigr] + C$, where $K>0$ and $C$ are constants. Recall from Table~\ref{tab:tab2} that $\beta =1$ is possible only in Case 1b, which requires $\alpha>0$ and $\sigma>1$.\footnote{Thus, the two special cases of the CREMR given in Table~\ref{tab:special_cases}, i.e., $u(q)= K\, {\rm Ei}(-\alpha q) $ and $u(q)= K \ln q  $, must be excluded from the analysis of monopolistic competition.}
Solving the bottom-right expression of Table~\ref{tab:FOC} by imposing $\alpha>0$, $\beta=1$, and $\sigma>1$, we obtain the associated markup function
\begin{equation*}
\textstyle
\frac{1}{\mu} = \frac{1}{ {\cal W} (x, \alpha, 1, \sigma)}
= \left\{1- \frac{1-  (1 - \sigma) \left[  \frac{x}{\alpha ( \sigma-1) } \right]^{\sigma}}{\sigma} \right\}^{-1}
= \frac{\sigma}{\sigma -1} \left\{ 1-  \left[  \frac{x}{\alpha ( \sigma-1) } \right]^{\sigma} \right\}^{-1} > \frac{\sigma}{\sigma -1} >1,
\end{equation*}
where $x < \alpha(\sigma-1)$ holds from Table~\ref{tab:tab3}.

\vspace{-.2cm}

\item{\bf Gauss hypergeometric case.}
The HARA and CREMR cases used in the literature impose either $\beta = 0$ or $\beta = 1$. The incomplete gamma case allows for intermediate values $\beta\in(0,1)$ but requires that $\sigma\to 0$. We have already shown in Theorem~\ref{prop:T1} how to deal with more general cases where $\beta\in(0,1)$ and $\sigma \neq 0$, thereby extending the results from the existing literature. As shown in Theorem~\ref{prop:T1}, when $\beta \in (0,1)$ and $\sigma \neq 0$, we get the utility of the general form $u(q) =  \textstyle \frac{K \sigma}{\sigma-1} (1+\alpha \sigma q)^{\beta-\frac{1}{\sigma}} q^{1-\beta} \bigl[ 1 +\frac{\beta -\frac{1}{\sigma}}{1-\beta} \, {}_2 F_{1} \left(1- \frac{1}{\sigma}, 1; 2- \beta ; - \alpha \sigma q   \right)  \bigr]+ C,$ where $K>0$ and $C$ are constants. Solving the bottom-right expression of Table~\ref{tab:FOC}, we obtain the associated markup function $\frac{1}{\mu} =\frac{1}{ {\cal W} (x, \alpha, \beta, \sigma)}$ in terms of the generalized Lambert $W$ function.

\end{itemize}

\vspace{-.5cm}

\paragraph{Some historical remarks.}
Lambert (1758) analyzes the equation $Z^\xi -  Z =\check x$, which is tranformed by Euler (1783) into $Z^\xi -Z^{\check \xi} = (\xi -\check \xi) x Z^{\xi+\check \xi}$.  Corless et al.~(1996) divide the latter by $\xi -\check \xi$ and let $\check \xi \to \xi$ to get $\ln Z =x Z^\xi$, and then consider the case with $\xi=-1$ and $Z={\rm e}^W$ to obtain the Lambert $W$ function defined as $x= W {\rm e}^W$. Letting $\tau(Z)$ and $\psi (Z)$ denote some functions of $Z$, we extend Euler's specification to $(Z/x)^\xi \tau(Z)  -(Z/x)^{\check \xi} = (\xi -\check \xi) x^{-\check \xi} \psi(Z) Z^{\xi+\check \xi}$, which implies $(Z/x)^{\xi - \check \xi}  \tau(Z) -1 = (\xi -\check \xi)\psi(Z) Z^{\xi}$. Setting $\xi=1$, $\check \xi =1+ \sigma$, $\psi(Z) = 1/Z-1$, $\tau(Z)=(1 -\beta \sigma) \bigl[  \frac{1-\beta -Z}{\alpha ( 1 -\beta \sigma) } \bigr]^{\beta \sigma}$, and $Z={\cal W}$, we obtain $({\cal W}/x)^{-\sigma}  (1 -\beta \sigma) \bigl[  \frac{1-\beta - {\cal W}}{\alpha ( 1 -\beta \sigma) } \bigr]^{\beta \sigma} -1 =- \sigma (1- {\cal W})$, which is equivalent to the bottom-right expression of Table~\ref{tab:FOC} with $\mu= {\cal W}$.

\vspace{-.25cm}

\paragraph{The LFRRA demand function.}
As mentioned above, the first-order condition for utility maximization is given by $(1+ \alpha \sigma q)^{\beta-\frac{1}{\sigma}} q^{-\beta} = \frac{{\lambda} p}{K}$, which can be rewritten as $1+ \alpha \sigma q = (\frac{{\lambda} p}{K} q^{\beta})^\frac{1}{\beta-\frac{1}{\sigma}} = (\frac{{\lambda}}{K} q^{\beta})^{-\frac{\sigma}{1- \beta \sigma}} p^{-\frac{\sigma}{1- \beta \sigma}}$. Hence, the LFRRA demand function is implicitly given by
\begin{equation}
\textstyle q = \frac{1}{ \alpha \sigma}  \left(\frac{{\lambda}}{K} q^{\beta} \right)^{-\frac{\sigma}{1- \beta \sigma}} p^{-\frac{\sigma}{1- \beta \sigma}} - \frac{1}{ \alpha \sigma},
\label{eq:imp_demand}
\end{equation}
which nests two important cases.

First, letting $\beta \to 0^+$, we obtain the HARA demand function explicitly as
\begin{equation}
\textstyle
q = \frac{1}{ \alpha \sigma}  \left(\frac{{\lambda}}{K}\right)^{-\sigma} p^{-\sigma} - \frac{1}{ \alpha \sigma},
\label{eq:imp_demand_HARA}
\end{equation}
which is isomorphic to the ``translated'' CES function in Mr\'{a}zov\'{a} and Neary (2017) and the ``Pollak family" in  Arkolakis et al. (2019).\footnote{The demand function in Mr\'{a}zov\'{a} and Neary (2017, MN) is given by $q = A^{\rm MN} p^{-\sigma} +B^{\rm MN}$, where $A^\text{MN} = \delta$ and  $B^\text{MN} = \gamma$ (in their notation), whereas that in Arkolakis et al. (2019, ACDR), is given by $q = A^{\rm ACDR} p^\frac{1}{\gamma} +B^{\rm ACDR}$, where  $A^\text{ACDR} = P^{-\frac{1}{\gamma}}$ and $B^\text{ACDR} = -\alpha$ (in their notation).}
It also nests the CARA demand function $q=-\frac{1}{\alpha} \ln \bigl(\frac{{\lambda} p}{K} \bigr)$ in Behrens and Murata (2007) as seen from letting $\sigma \to 0$.

Second, letting $\beta \to 1^-$, we obtain the CREMR demand function
\begin{equation}
\textstyle q = \frac{1}{ \alpha \sigma}  \left(\frac{{\lambda}}{K} q \right)^{-\frac{\sigma}{1- \sigma}} p^{-\frac{\sigma}{1- \sigma}} -  \frac{1}{ \alpha \sigma}.
\label{eq:imp_demand_CREMR}
\end{equation}
Solving this for $p$, we obtain the inverse demand function $p = \frac{K/{\lambda}}{q} (\alpha \sigma q +1 )^{\frac{\sigma-1}{\sigma}},$ which is isomorphic to the CREMR inverse demand function in Mr\'{a}zov\'{a} et al. (2021).\footnote{Our inverse demand function can be rewritten as $p = \frac{K/{\lambda}}{q} (\alpha \sigma)^{\frac{\sigma-1}{\sigma}} (q + \frac{1}{\alpha \sigma})^{\frac{\sigma-1}{\sigma}}$ if $\alpha \sigma$ and $q + \frac{1}{\alpha \sigma}$ are not both negative. This is isomorphic to Mr\'{a}zov\'{a} et al.'s (2021) inverse demand function $p= \frac{\beta}{q} (q -\gamma)^{\frac{\sigma-1}{\sigma}}$, where $\beta= (K/{\lambda}) (\alpha \sigma)^{\frac{\sigma-1}{\sigma}}$ and $\gamma = - \frac{1}{\alpha \sigma}$ (in their notation). There are two subcases depending on the sign of $\alpha \sigma$ and $q + \frac{1}{\alpha \sigma}$. First, assume that $\alpha \sigma>0$, which implies $q + \frac{1}{\alpha \sigma}>0$. Then, $(\alpha \sigma)^{\frac{\sigma-1}{\sigma}}$ is real valued, which is consistent with our assumption that $K$ is positive real. Second, assume that $\alpha \sigma<0$, which requires the quantity restriction $q + \frac{1}{\alpha \sigma}>0$ by Corollary~\ref{corollaryMNP}. Then, $(\alpha \sigma)^{\frac{\sigma-1}{\sigma}}$ is complex, which can be offset by relaxing our assumption of a positive real $K$ and including a complex term in $K$ (see footnote~\ref{fn20}).}

\subsection{Comparative statics results}
\label{subsec:compstat}

We now establish comparative statics results on prices, quantities, and markups with respect to marginal cost $m$. Our results show that the three parameters $\{\alpha, \beta, \sigma\}$ in the LFRRA utility function are crucial for whether the utility function has increasing, decreasing, or constant RRA, which in turn determines whether the markup is decreasing, increasing, or constant with respect to marginal cost.

\vspace{-.25cm}
\begin{proposition}[Comparative statics results for LFRRA]
\label{prop:compstat}
If $\alpha (1- \beta \sigma) >0$ holds (resp., if $\alpha (1- \beta \sigma)<0$ holds), then the markups are decreasing (resp., increasing) in $m$. If either $\alpha = 0$ or $\beta =\frac{1}{\sigma}$ holds, then the markups are constant in $m$. The prices are increasing in $m$ and the quantities are decreasing in $m$ regardless of parameter values. The results are summarized in Table~\ref{tab:tab4}.
\end{proposition}

\vspace{-.5cm}

\begin{table}[h]
\caption{Comparative statics.}
\label{tab:tab4}
\centering
\begin{tabular}{c|c|c|c}
\multicolumn{4}{c}{Case 1: $\alpha >0$.} \\
\hline
&&& \\[-4mm]
 & markups & prices  & quantities    \\[1mm] \hline
 &&& \\[-4mm]
Case 1a. $1-\beta \sigma>0$ & $\frac{\partial}{\partial m} \bigl( \frac{1}{\mu} \bigr)<0  $ & $\frac{\partial}{\partial m} \bigl( \frac{m}{\mu} \bigr)>0$ & $ \frac{\partial}{\partial m} \left[\frac{1}{\alpha} \frac{1-\beta -\mu}{1-\sigma(1-\mu)} \right] <0$\\[2.5mm] 
Case 1b. $1-\beta \sigma<0$ &  $\frac{\partial}{\partial m} \bigl( \frac{1}{\mu} \bigr)>0  $  & $\frac{\partial}{\partial m} \bigl( \frac{m}{\mu} \bigr) >0$ & $ \frac{\partial}{\partial m} \left[\frac{1}{\alpha} \frac{1-\beta -\mu}{1-\sigma(1-\mu)} \right] <0$ \\[2mm]
\hline
\multicolumn{4}{c}{} \\[-2mm] 
\multicolumn{4}{c}{Case 2: $\alpha <0$.} \\
\hline
&&& \\[-4mm]
 & markups & prices  & quantities    \\[1mm] \hline
  &&& \\[-4mm]
Case 2a. $1-\beta \sigma>0$ & $\frac{\partial}{\partial m} \bigl( \frac{1}{\mu} \bigr)>0  $ &$\frac{\partial}{\partial m} \bigl( \frac{m}{\mu} \bigr)>0$ & $ \frac{\partial}{\partial m} \left[\frac{1}{\alpha} \frac{1-\beta -\mu}{1-\sigma(1-\mu)} \right] <0$ \\ [2.5mm]
Case 2b. $1-\beta \sigma<0$ &  $\frac{\partial}{\partial m} \bigl( \frac{1}{\mu} \bigr)<0  $  & $\frac{\partial}{\partial m} \bigl( \frac{m}{\mu} \bigr) > 0$ & $ \frac{\partial}{\partial m} \left[\frac{1}{\alpha} \frac{1-\beta -\mu}{1-\sigma(1-\mu)} \right] <0$ \\ [2mm]
\hline
\multicolumn{4}{p{12.5cm}}{\scriptsize {\it Notes}: The case with $\alpha=0$ corresponds to the CES and requires that $\beta \in (0,1)$. The case with $1-\beta \sigma=0$ also corresponds to the CES and requires that $\sigma > 1$. If $\alpha=0$ or $\beta =\frac{1}{\sigma}$, then the price is given by $p(m)  =\frac{m}{1-\beta}$ or $p (m) =\frac{\sigma m}{\sigma-1}$, respectively, and thus $\frac{\partial}{\partial m} \bigl( \frac{1}{\mu} \bigr)=0$, $\frac{\partial}{\partial m} \bigl( \frac{m}{\mu} \bigr)>0$, and $\frac{\partial}{\partial m} \bigl( q(m) \bigr)= \frac{\partial}{\partial m} \bigl( (u')^{-1} ({\lambda} p(m)) \bigr)<0$. Since these two cases are well known, we omit them from the table.}
\end{tabular}
\vspace{-.25cm}
\end{table}

\paragraph{Proof.} See \ref{app:proofs5}.

\medbreak

The derivative of the RRA is given by $\frac{\alpha (1-\beta \sigma)}{(1+\alpha \sigma q)^2}$, and as seen from Table~\ref{tab:tab4}, the comparative statics results regarding markups depend solely on the sign of $\alpha (1-\beta \sigma)$.\footnote{Furthermore, Table~\ref{tab:tab4} shows that prices increase in $m$ in all cases, which is in line with Zhelobodko et al.~(2012). Hence, even when more productive firms charge higher markups, prices cannot increase with productivity.}
Thus, our LFRRA utility function constitutes a parametric version of Zhelobodko et al.~(2012) and captures all possible cases---increasing, decreasing, and constant RRA.

It is worth emphasizing that our parametric approach allows us to take the model to data and estimate the three parameters $\{\alpha, \beta, \sigma\}$ of the LFRRA utility function. These estimates can then be used to test whether the RRA is increasing, decreasing, and constant, and thus whether markups are decreasing, increasing, and constant with respect to marginal costs. We present these results in the next section.

\section{Fitting the monopolistic competition model to data}
\label{sec:fit}

We fit our monopolistic competition model to the data from De Loecker et al. (2016), who provide quantities and estimated markups at the product level for a sample of firms from eleven Indian manufacturing sectors from 1989 to 2001. Their markup data are suitable for our purpose as they are obtained without imposing any restrictions on consumer demand and thus on preferences. To stay close to our theoretical setting, we focus on single-product firms only. Hence, a variety $\omega$ in our model corresponds to the single product of a firm in the dataset. Dropping all observations where markups are either missing or less than one---and trimming the top and bottom three percent of the sectoral markup and quantity distributions (as in De Loecker et al., 2016)---yields 9,232 firm-year observations.

We illustrate a parsimonious way to quantify the model using the LFRRA specification~\eqref{eq:LFRRA_general} and
the first-order condition for profit maximization \eqref{eq:generalFOC}, which implies $1-\mu(\omega) = \frac{\alpha q (\omega)+\beta}{\alpha \sigma q(\omega)+1}$. Using observed firm-level markups $\frac{1}{\mu(\omega)}$ and quantities $q(\omega)$, we estimate the values of $\alpha$, $\beta$, and $\sigma$ by minimizing the following residual sum of squares (RSS):
\begin{equation}
\label{eq:estimation}
\min_{\alpha, \beta, \sigma}\, {\rm RSS} = \sum_{\omega=1}^\Omega \left[1- \mu(\omega)  - \frac{\alpha q (\omega)+\beta}{\alpha \sigma q(\omega)+1} \right]^2
\end{equation}
subject to the theoretical constraints $\alpha q (\omega)+\beta > 0$, $\alpha \sigma q(\omega)+1>0$, and $0 \le \frac{\alpha q  (\omega)+\beta}{\alpha \sigma q(\omega)+1} < 1$ (see \ref{app:empirics} for more details). We solve the minimization problem \eqref{eq:estimation} at the sectoral and aggregate levels to obtain the fitted values $\{\widehat \alpha, \widehat \beta, \widehat \sigma\}$ that are consistent with our LFRRA preferences. We also verify that the second-order conditions \eqref{eq:SOC_LFRRA} evaluated at $\{\widehat \alpha, \widehat \beta, \widehat \sigma\}$ are satisfied for all quantities $q(\omega) \ge 0$.

To conduct statistical inference, we construct confidence intervals for $\{ \alpha,  \beta,  \sigma\}$ by bootstrapping as follows. We first randomly draw, with replacement, a sample of markup and quantity observations from the data such that the generated and observed sample sizes are identical. We then compute $\{\widetilde \alpha, \widetilde \beta, \widetilde \sigma\}$ by solving the minimization problem \eqref{eq:estimation} using the procedure explained in \ref{app:empirics}. We repeat this process 200 times at the sectoral and aggregate levels to obtain the distributions of parameters $\{\widetilde \alpha_b, \widetilde \beta_b, \widetilde \sigma_b\}_{b=1}^{200}$, which allow us to construct the $95\%$ confidence interval using the 2.5th and 97.5th percentiles of the distribution of each parameter, respectively.

Table~\ref{tab:tab_est_3pct_comb_rounded} summarizes our results. As shown, the estimated values of $\beta$ are positive and their bootstrapped confidence intervals fall into $(0,1)$ for almost all sectors including the aggregate economy. The only exception is sector 15 (``Food products and beverages''), where we cannot reject the null hypothesis of the HARA preferences, i.e., $\beta=0$. The estimated values of $\alpha$ are positive for all sectors and their bootstrapped confidence intervals do not include zero except for sector 17 (``Textiles, apparel'') and sector 31 (``Electrical machinery and communications''). In those two sectors, we cannot reject the null hypothesis that $\alpha=0$, in which case the RRA is constant, $-\frac{q u''(q)}{u'(q)} = \beta$, and thus the derivative of the RRA with respect to quantity is zero.

We can further test whether the RRA for each sector is constant using the confidence interval for $\beta -\frac{1}{\sigma}$ and that for $\alpha (1-\beta \sigma)$, where the latter comes from the sign of the derivative of the RRA, $\frac{\partial \, {\rm RRA}}{\partial q} = \frac{\alpha (1-\beta \sigma)}{(1+\alpha \sigma q)^2}$. The former confidence interval implies that we cannot reject the null hypothesis that sector 25 (``Rubber and plastic''), as well as sectors 17 and 31, displays the CRRA, i.e., $\beta -\frac{1}{\sigma}=0$. Since $\frac{\partial \, {\rm RRA}}{\partial q}=0$ for sectors with either $\alpha =0$ or $\beta=\frac{1}{\sigma}$,
we compute the latter confidence intervals for all sectors other than 17, 25, and 31 and confirm that there are no other sectors displaying the CRRA.

Since the sign of the derivative of the RRA is given by the sign of $\alpha (1-\beta \sigma)$ for sectors satisfying $\alpha \neq 0$ and $\beta \neq \frac{1}{\sigma}$, the fitted values $\{\widehat  \alpha, \widehat \beta, \widehat \sigma\}$ can tell us whether the remaining sectors and the aggregate economy display increasing or decreasing RRA, which in turn determines whether markups are decreasing or increasing with respect to marginal costs. Column 8 of Table~\ref{tab:tab_est_3pct_comb_rounded}, together with Table~\ref{tab:tab4}, shows that there is substantial heterogeneity across sectors in how markups change with marginal costs. The numbers of sectors displaying increasing RRA, decreasing RRA, and constant RRA are four, four, and three, respectively. The result for all sectors pooled is consistent with IRRA.

Our results suggest that starting from the unified framework encompassing not only the HARA ($\beta=0$) and CREMR ($\beta=1$) preferences  but also other specifications with $\beta \in (0,1)$ is important, since this parameter, as well as $\alpha$ and $\sigma$, is a priori unknown. Setting $\beta=0$ or $\beta=1$, as usually done in the literature, need not be consistent with the underlying data.\footnote{\ref{app:empirics} provides the results for the alternative CREMR specification in Corollary~\ref{corollaryMNP}.}

\begin{sidewaystable}
\caption{Sectoral and aggregate estimates of the LFRRA parameters.}
\label{tab:tab_est_3pct_comb_rounded}
\scalebox{0.8}{
\begin{tabular}{llcccccccc}
\hline\hline
& Name & RSS & $\widehat \alpha$ & $\widehat \beta$ & $\widehat \sigma$ & $\widehat \beta-1/\widehat \sigma$ & $\widehat \alpha(1-\widehat \beta \widehat \sigma)$ & Type & Obs \\ \hline
$15$ & Food products and beverages & $53.705$ & $492.910$ & $0.000$ & $2.700$ & $-0.370$ & $492.910$ & IRRA & $1044$ \\ \vspace{1mm}
 &  &  & $[286.931$, $859.482]$ & $[0.000, 0.100]$ & $[2.585, 2.804]$ & $[-0.387, -0.263]$ & $[269.433, 859.482]$ & {\small (HARA)} &  \\
$17$ & Textiles, apparel & $22.692$ & $0.127$ & $0.257$ & $5.760$ & $0.084$ &n.a.$^\dagger$ & CRRA & $1534$ \\ \vspace{1mm}
 &  &  & $[-0.053, 42.486]$ & $[0.243, 0.262]$ & $[-3.012, 16.783]$ & $[-0.046, 0.259]$ & &  &  \\
$21$ & Paper and paper products & $6.479$ & $20.436$ & $0.288$ & $2.907$ & $-0.056$ & $3.326$ & IRRA & $546$ \\ \vspace{1mm}
 &  &  & $[6.947, 91.974]$ & $[0.256, 0.306]$ & $[2.652, 3.138]$ & $[-0.097, -0.015]$ & $[0.670, 21.255]$ &  &  \\
$24$ & Chemicals & $35.152$ & $118.472$ & $0.240$ & $2.774$ & $-0.120$ & $39.523$ & IRRA & $1684$ \\ \vspace{1mm}
 &  &  & $[59.345, 245.973]$ & $[0.222, 0.253]$ & $[2.609, 2.929]$ & $[-0.143, -0.098]$ & $[19.874, 85.745]$ &  &  \\
$25$ & Rubber and plastic & $11.809$ & $4.341$ & $0.273$ & $3.172$ & $-0.042$ & n.a.$^\dagger$  & CRRA & $676$ \\ \vspace{1mm}
&  &  & $[0.033, 15.674]$ & $[0.247, 0.313]$ & $[3.060, 3.699]$ & $[-0.076, 0.041]$ &  &  &  \\
$26$ & Nonmetallic mineral products & $23.243$ & $3.831$ & $0.501$ & $1.429$ & $-0.199$ & $1.088$ & IRRA & $744$ \\ \vspace{1mm}
 &  &  & $[1.916, 7.662]$ & $[0.485, 0.516]$ & $[1.276, 1.537]$ & $[-0.286, -0.150]$ & $[0.613, 2.021]$ & &  \\
$27$ & Basic metals & $20.660$ & $17.830$ & $0.281$ & $5.304$ & $0.093$ & $-8.756$ & DRRA & $988$ \\ \vspace{1mm}
 &  &  & $[6.372, 137.895]$ & $[0.255, 0.354]$ & $[4.770, 5.934]$ & $[0.062, 0.160]$ & $[-90.196, -2.753]$ &  &  \\
$28$ & Fabricated metal products  & $6.812$ & $66.155$ & $0.364$ & $3.673$ & $0.092$ & $-22.317$ & DRRA & $445$ \\ \vspace{1mm}
 &  &  & $[21.364, 179.554]$ & $[0.333, 0.404]$ & $[3.497, 3.924]$ & $[0.063, 0.136]$ & $[-77.008, -7.153]$ &  &  \\
$29$ & Machinery and equipment  & $15.208$ & $36.339$ & $0.312$ & $3.747$ & $0.045$ & $-6.150$ & DRRA & $745$ \\ \vspace{1mm}
 &  &  & $[14.509, 232.300]$ & $[0.299, 0.325]$ & $[3.497, 4.154]$ & $[0.021, 0.075]$ & $[-33.285, -2.016]$ &  &  \\
$31$ & Electrical machinery  & $8.182$ & $0.426$ & $0.280$ & $2.219$ & $-0.170$ & n.a.$^\dagger$ & CRRA & $495$ \\ \vspace{1mm}
 &  \quad  and communications  &  & $[-0.011, 7.683]$ & $[0.268, 0.293]$ & $[-5.947, 3.918]$ & $[-2.052, 1.880]$ &
 &  &  \\
$34$ & Motor vehicles, trailers & $8.281$ & $36.672$ & $0.380$ & $3.654$ & $0.106$ & $-14.224$ & DRRA & $331$ \\ \vspace{1mm}
 &  &  & $[0.593, 79.573]$ & $[0.351, 0.403]$ & $[3.313, 5.082]$ & $[0.063, 0.177]$ & $[-32.810, -0.674]$ &  &  \\
$0$ & All sectors pooled  & $272.448$ & $37.336$ & $0.295$ & $2.818$ & $-0.060$ & $6.263$ & IRRA & $9232$ \\ \vspace{1mm}
 &  &  & $[20.041, 65.529]$ & $[0.291, 0.301]$ & $[2.732, 2.914]$ & $[-0.071, -0.047]$ & $[3.243, 11.187]$ & &  \\
\hline \hline
\multicolumn{10}{p{28.5cm}}{\small {\it Notes}: We use markup and quantity data from De Loecker et al. (2016). We restrict ourselves to single-product firms and products for which the markup exceeds one. Following De Loecker et al. (2016), we trim the top- and bottom-3\% of quantities and markups by sector. We run 200 bootstrap replications with replacement, and the 95\% confidence intervals are provided below the coefficients. RSS denotes the residual sum of squares, i.e., the value of our objective function in \eqref{eq:estimation}. Obs stands for the number of observations. IRRA, DRRA, and CRRA stand for increasing, decreasing, and constant relative risk aversion, respectively. Details on the numerical procedure are provided in~\ref{app:empirics}. $^\dagger$ means that the derivative of RRA is zero when $\widehat \alpha$ or $\widehat \beta - {1}/{\widehat \sigma}$ is not different from zero and is therefore not applicable.
} \\
\end{tabular}
}
\end{sidewaystable}

\section{Extensions}
\label{sec:extensions}

In the previous section, we have analyzed the monopolistic competition model by embedding the LFRRA family of utility functions into the directly additively separable upper-tier utility function. In this section, we first apply the analysis to the case of implicitly additive preferences. We then extend the LFRRA family to include non-linear fractional forms.

\subsection{Implicitly additive preferences}
\label{sec:extensions_implicit_additive}

Consider the implicitly additively separable utility function $U$, defined as $\int_{\Omega} \Upsilon \bigl( \frac{q(\omega)}{U} \bigr){\rm d}\omega=1$, where $\Omega$ denotes the set of varieties, $q(\omega) \ge 0$ is consumption of variety $\omega$, and $\Upsilon$ is an increasing and concave sub-function (Kimball, 1995; Mr\'{a}zov\'{a} et al., 2021). Minimizing expenditure $\int_{\Omega} p(\omega) q(\omega) {\rm d} \omega$ subject to the utility constraint $\int_{\Omega} \Upsilon \bigl( \frac{q(\omega)}{U} \bigr){\rm d}\omega=1$ gives the first-order condition $p(\omega) = \frac{{\lambda}^{\rm imp} }{U} \Upsilon' \bigl( \frac{q(\omega)}{U} \bigr) $, where ${\lambda}^{\rm imp}>0$ is the Lagrange multiplier.

Using this condition, the profit maximization problem in the monopolistic competition model in Section~\ref{sec:mc} can be replaced with $\max_q  \bigl[ \frac{{\lambda}^{\rm imp}}{U} \Upsilon' \bigl( \frac{q}{U} \bigr)-m \bigr] LU \frac{q}{U} -f$. Let ${\theta} = \frac{q}{U}$ denote scaled quantities. We can then rewrite the problem as $\max_{{\theta}}  \bigl[ \frac{{\lambda}^{\rm imp}}{U} \Upsilon' ({\theta})-m \bigr] LU {\theta} -f$. Since there is a continuum of firms, no firm can affect ${\lambda}^{\rm imp}$ and $U$. Thus, the first-order condition for profit maximization is given by $\frac{{\lambda}^{\rm imp} }{U} \Upsilon'' ({\theta}) LU{\theta}+ \bigl[ \frac{{\lambda}^{\rm imp}}{U} \Upsilon' ({\theta}) -m \bigr] LU=0$. The profit maximizing price $p(m)$ hence satisfies:
\begin{equation}
\frac{p(m)-m}{p(m)} = 1-\mu (m) = - \frac{{\theta}(m) \Upsilon''({\theta}(m))}{\Upsilon'({\theta}(m))},
\label{eq:generalFOC_implicit}
\end{equation}
where $\mu(m)$ denotes the inverse of the markup and $1-\mu (m)$ is the Lerner index.\footnote{The second-order condition (SOC) for profit maximization is given by $\textstyle \frac{{\lambda}^{\rm imp}}{U} \Upsilon''' ({\theta}) LU{\theta} + \frac{{\lambda}^{\rm imp}}{U} \Upsilon'' ({\theta}) LU+  \frac{{\lambda}^{\rm imp}}{U} \Upsilon''({\theta})  LU = {\lambda}^{\rm imp} \Upsilon'''({\theta}) L {\theta}+ 2 {\lambda}^{\rm imp} \Upsilon'' ({\theta}) L = \textstyle -L  {\lambda}^{\rm imp} \Upsilon'' ({\theta})
\bigl[ -  \frac{  {\theta} \Upsilon''' ({\theta})}{  \Upsilon'' ({\theta}) }  -2   \bigr]  <0.$ Since $-L  {\lambda}^{\rm imp}\Upsilon''({\theta}) >0$, the SOC reduces to $2 - \bigl[ - \frac{  {\theta} \Upsilon''' ({\theta})}{  \Upsilon'' ({\theta}) } \bigr]>0$.}
In what follows, we suppress $m$ to simplify the notation.

We assume that $- \frac{ {\theta} \Upsilon''({\theta})}{\Upsilon'({\theta})}$ is of the linear fractional form:
\begin{equation}
- \frac{ {\theta} \Upsilon''({\theta})}{\Upsilon'({\theta})}= \frac{\alpha {\theta}+ \beta }{\alpha\sigma {\theta}+1},
\label{eq:LFRRA_general_imp}
\end{equation} 
which is isomorphic to equation \eqref{eq:LFRRA_general} in Definition~\ref{def}. Thus, we can rewrite \eqref{eq:LFRRA_general_imp} as a second-order differential equation, $ (\alpha\sigma {\theta}+1) {\theta} \Upsilon''({\theta})+  (\alpha {\theta}+ \beta )\Upsilon'({\theta})= 0$, as we do in equation~\eqref{eq:LFRRAchange}. Solving this yields the general form of $\Upsilon({\theta})$, as well as its special cases, as in Section~\ref{sec:LFRRA}.

Furthermore, \eqref{eq:generalFOC_implicit} and \eqref{eq:LFRRA_general_imp} imply
\begin{equation}
1-\mu= \frac{\alpha {\theta}+ \beta }{\alpha\sigma {\theta}+1}.
\label{eq:linear_fractional_implicit}
\end{equation}
Recalling $\theta= \frac{q}{U}$ and letting $\widetilde \alpha = \frac{\alpha}{U}$, it can be rewritten as $1-\mu =  \frac{\alpha \frac{q}{U}+ \beta }{\alpha\sigma \frac{q}{U}+1} =  \frac{ \widetilde \alpha q + \beta }{\widetilde \alpha \sigma q+1}$, from which we obtain $q= \frac{1}{\widetilde \alpha} \frac{1-\beta -\mu}{1- \sigma (1-\mu)}.$ Since this quantity is the same as \eqref{eq:FOCprof} except that $\alpha$ in \eqref{eq:FOCprof} is replaced with $\widetilde \alpha$, we can analyze the price equilibrium and conduct comparative statics analysis as in Section~\ref{sec:mc}. We can also solve the same minimization problem \eqref{eq:estimation} as in Section~\ref{sec:fit} by replacing $\alpha$ in \eqref{eq:estimation} with $\widetilde \alpha$ and relabel the resulting estimates $\{\widehat \alpha, \widehat \beta, \widehat \sigma\}$ as $\{\widehat {\widetilde \alpha}, \widehat \beta, \widehat \sigma\}$.

\subsection{Non-linear fractional forms}
\label{sec:nonlinear_RRA}

We have so far shown that our linear fractional form can be applied to both directly and implicitly additive preferences. In each case, we can back out $u(q)$ or $\Upsilon (\theta)$ by solving the second-order differential equation in Section~\ref{sec:LFRRA} or in Section~\ref{sec:extensions_implicit_additive}, which yields the solution in terms of the Gauss hypergeometric functions (or the confluent hypergeometric functions, which are a degenerate form of the Gauss hypergeometric functions).

We now generalize our linear fractional form to include four types of non-linear fractional forms used in the literature. Each type can also be expressed as a second-order differential equation as in Section~\ref{sec:LFRRA} or as in Section~\ref{sec:extensions_implicit_additive}, which allows us to solve it in a similar way.

Let $\phi$ denote some general function of quantity $q$. Consider the general RRA of the form: $-\frac{q u''(q)}{u'(q)} = \phi(q)$, where $\phi(q) \ge 0$. This implies
\begin{equation}
\label{eq:diff_before_transform}
q u''(q) + \phi(q) u'(q) = 0.
\end{equation}
Let $u(q) = v(z)$, where $z=\zeta (q)$ is strictly monotone in $q$. We then have $u'(q) = v'(z) \zeta' (q)$ and $u''(q) = v''(z) [\zeta' (q)]^2 + v'(z) \zeta''(q)$, which allow us to rewrite \eqref{eq:diff_before_transform} as: $v''(z) [\zeta' (q)]^2 + v'(z) \zeta''(q)+ \frac{\phi(q)}{q} v'(z) \zeta' (q) = 0$, so that $\textstyle [\zeta' (q)]^2 v''(z) + \bigl[ \zeta''(q)+ \frac{\phi(q)}{q}  \zeta' (q)  \bigr]v'(z) = 0.$ Multiplying this by $z\bigl(1-\frac{z}{\iota}\bigr) [\zeta' (q)]^{-2}$ yields
\begin{equation}
\textstyle z\left(1-\frac{z}{\iota}\right) v''(z) +  z\left(1-\frac{z}{\iota}\right)  \left\{ \frac{\zeta''(q)}{ [\zeta' (q)]^2}+ \frac{\phi(q)}{q}  \frac{1}{\zeta' (q) } \right\} v'(z) = 0.
\label{eq:SODE}
\end{equation}
Let
\begin{equation}
\textstyle z\left(1-\frac{z}{\iota}\right)  \left\{ \frac{\zeta''(q)}{ [\zeta' (q)]^2}+ \frac{\phi(q)}{q}  \frac{1}{\zeta' (q) } \right\} = c-(a+b+\iota) \frac{z}{\iota}, 
\quad \mbox{where} \quad ab=0.
\label{eq:impf}
\end{equation}
Then, \eqref{eq:SODE} can be rewritten as
\begin{equation}
\textstyle z\left(1-\frac{z}{\iota}\right) v''(z) + \left[  c-(a+b+\iota) \frac{z}{\iota} \right] v'(z) = 0,
\quad \mbox{where} \quad ab=0,
\label{eq:SODE_re}
\end{equation}
which is a hypergeometric differential equation if  $\iota=1$ and a confluent hypergeometric differential equation if $\iota \to \infty$.

Solving \eqref{eq:impf} for $\phi(q)$ and using $z=\zeta (q)$ yield
\begin{equation}
\textstyle \phi(q)
=  \frac{c-(a+b+\iota) \frac{\zeta (q)}{\iota}}{1-\frac{\zeta (q)}{\iota}} \frac{q \zeta' (q) }{\zeta(q)}- \frac{q \zeta''(q)}{ \zeta' (q)},
\label{eq:fofq}
\end{equation}
which reduces to the LFRRA, $\phi(q) = \frac{\alpha q+\beta}{\alpha \sigma q+1}$, if $\iota=1$, $a=0$, $b=\frac{1}{\sigma}-1$, $c =\beta$, and $z= \zeta(q) =  -\alpha \sigma q$. Furthermore, equation \eqref{eq:fofq} nests four non-linear fractional cases used in the literature. All four cases share the common form $\phi(q) = \frac{\alpha q^\epsilon+\beta}{\gamma q^\epsilon+\delta}$, which is linear fractional in terms of $q^\epsilon$ and reduces to the LFRRA \mbox{when $\epsilon=1$, and have up to essentially three parameters.}\footnote{In addition to the four degenerated cases analyzed below, we can solve $-\frac{q u''(q)}{u'(q)}=\frac{\alpha q^\epsilon+\beta}{\gamma q^\epsilon+\delta}$ for $u(q)$ as follows. First, since this case corresponds to $\iota=1$, $a=\frac{\alpha}{\gamma \epsilon} -\frac{1}{\epsilon}$, $b=0$, $c=1+\frac{\beta}{\delta \epsilon} -\frac{1}{\epsilon}$, and $z=\zeta(q)= -\frac{\gamma}{\delta} q^\epsilon$ in \eqref{eq:fofq}, one can rewrite \eqref{eq:SODE_re} as a hypergeometric differential equation:
\begin{equation*}
\textstyle z(1-z) v''(z) + [1 + \frac{\beta}{\delta \epsilon} -\frac{1}{\epsilon}
- ( 1+\frac{\alpha}{\gamma \epsilon} -\frac{1}{\epsilon} )z ] v'(z)=0,
\end{equation*}
which generalizes \eqref{eq:ourHGDE} as it reduces to \eqref{eq:ourHGDE} when $\gamma =\alpha \sigma$, $\delta=1$, and $\epsilon=1$. Second, solving this as in Section~\ref{sec:LFRRA} yields
\begin{equation*}
\textstyle v(z)=c_1+c_2 z^\frac{\delta-\beta}{\delta \epsilon} {}_2F_1 (\frac{\alpha \delta -\beta \gamma}{\gamma \delta \epsilon} , \frac{\delta -\beta}{\delta \epsilon}; 1+\frac{\delta -\beta}{\delta \epsilon}; z ),
\end{equation*}
which generalizes \eqref{eq:vz}. Last, as in \ref{app:theorem1}, one can obtain the corresponding utility function as follows:
\begin{equation*}
\textstyle u(q) = \frac{K \gamma}{\gamma- \alpha} (1+\frac{\gamma}{\delta} q^\epsilon)^{-\frac{\alpha}{\gamma \epsilon} + \frac{\beta}{\delta \epsilon}} q^\frac{\delta -\beta}{\delta} [ 1+ \frac{\beta \gamma -\alpha \delta}{(\delta -\beta) \gamma} \, {}_2 F_1 ( \frac{\gamma-\alpha}{\gamma \epsilon} ,1 ; 1 + \frac{\delta-\beta}{\delta \epsilon} ; -\frac{\gamma}{\delta} q^{\epsilon}) ]+C,
\end{equation*}
where we assume that $1 + \frac{\delta-\beta}{\delta \epsilon}$ is not a negative integer for ${}_2 F_1 ( \frac{\gamma-\alpha}{\gamma \epsilon} ,1 ; 1 + \frac{\delta-\beta}{\delta \epsilon} ; -\frac{\gamma}{\delta} q^{\epsilon})$ to be well defined (see Lebedev, 1965) and where $K>0$, $1+  \frac{\gamma q^\epsilon}{\delta}>0$, and $-\frac{q u''(q)}{u'(q)}=\frac{\alpha q^\epsilon+\beta}{\gamma q^\epsilon+\delta}>0$ must hold for $u'(q)>0$ and $u''(q)<0$. This utility function reduces to \eqref{eq:prop4_subutility_general_case} in Theorem~\ref{prop:T1} when $\gamma =\alpha \sigma$, $\delta=1$, and $\epsilon=1$.}

\vspace{-.3cm}
\paragraph{Bipower case.}
We first consider the bipower inverse demand function~$p  = A q^{-\eta}+B q^{-\vartheta}>0$ in Mr\'{a}zov\'{a} and Neary (2017), where $q>0$ and $A \eta q^{-\eta}  + B \vartheta q^{-\vartheta}>0$. Noting the first-order condition for utility maximization, $u'(q)={\lambda} p$, the RRA and its derivative are given by
\begin{equation*}
-\frac{q u''(q)}{u'(q)} = \frac{\frac{A}{B} \eta q^{\vartheta-\eta} + \vartheta }{\frac{A}{B} q^{\vartheta-\eta} + 1} \quad \mbox{and} \quad
\frac{\partial}{\partial q} \left\{ -\frac{q u''(q)}{u'(q)} \right\}
= - \frac{\frac{A}{B}(\eta-\vartheta)^2 q^{-1-\eta+\vartheta}}{(\frac{A}{B} q^{\vartheta-\eta}+1)^2},
\end{equation*}
which correspond to the case with $\iota=1$,  $a=0$, $b=\frac{1-\eta}{\eta-\vartheta}$, $c=2+ \frac{1-\eta}{\eta-\vartheta}$, and $z= -\frac{A}{B} q^{\vartheta-\eta}$ in \eqref{eq:fofq}. In this case, we can rewrite \eqref{eq:SODE_re} as a hypergeometric differential equation:
\begin{equation}
\textstyle z (1-z) v''(z) + \bigl[ 2+ \frac{1-\eta}{\eta-\vartheta} - \bigl(1+ \frac{1-\eta}{\eta-\vartheta} \bigr) z \bigr] v'(z) = 0.
\label{eq:SODE_bipower}
\end{equation}
Since the solution to \eqref{eq:SODE_re} for $\iota=1$ is given by $v(z) = \textstyle c_1\times z^{-a} {}_2 F_1 (a, a-c+1; a-b+1; \frac{1}{z} ) + c_2\times z^{-b} {}_2 F_1(b, b-c+1; b-a+1; \frac{1}{z})$, where $c_1$ and $c_2$ denote arbitrary coefficients (see, e.g., Sneddon, 1961), we have
\begin{equation}
v(z) = \textstyle c_1
+ c_2\times z^{-\frac{1-\eta}{\eta-\vartheta}} {}_2 F_1(\frac{1-\eta}{\eta-\vartheta}, -1; \frac{1-\vartheta}{\eta-\vartheta}; \frac{1}{z}),
\label{eq:vz_nonlinear_bipower}
\end{equation}
where we use ${}_2 F_1 (0, - 1- \frac{1-\eta}{\eta-\vartheta}; 1 - \frac{1-\eta}{\eta-\vartheta} ; \frac{1}{z} )=1$ and assume that $\frac{1-\vartheta}{\eta-\vartheta}$ is not a negative integer for ${}_2 F_1(\frac{1-\eta}{\eta-\vartheta}, -1; \frac{1-\vartheta}{\eta-\vartheta}; \frac{1}{z})$ to be well defined (see Lebedev, 1965). Noting that ${}_2 F_1(a, -1; c; \frac{1}{z}) = 1-\frac{a}{c} \frac{1}{z}$ and recalling $u(q)=v(z)$ and $z= -\frac{A}{B} q^{\vartheta-\eta}$, we obtain the bipower utility function of the form:
\begin{equation}
u(q) = \textstyle C+  K ( \frac{A q^{1-\eta}}{1-\eta} +\frac{B q^{1-\vartheta}}{1-\vartheta}), 
\label{eq:vz_nonlinear2_bipower}
\end{equation}
where $C=c_1$ and $K=  c_2 \frac{1-\eta}{A} \bigl(- \frac{A}{B} \bigr)^{-\frac{1-\eta}{\eta-\vartheta}}>0$.

\vspace{-.2cm}
\paragraph{The constant proportional of pass-through case.}
Second, we consider the constant proportional of pass-through (CPPT) inverse demand function $p  = \beta (q^\frac{k-1}{k}+\gamma)^\frac{k}{k-1} q^{-1}>0$ in Mr\'{a}zov\'{a} and Neary (2017), where $q>0$, $\beta>0$, $\gamma>0$, $k>0$, and $k\neq 1$ (in their notation). Noting the first-order condition for utility maximization, $u'(q)={\lambda} p$, the RRA and its derivative are given by:
\vspace{-.1cm}
\begin{equation*}
-\frac{q u''(q)}{u'(q)} = \frac{\gamma}{q^\frac{k-1}{k} +\gamma}
\quad \mbox{and} \quad
\frac{\partial}{\partial q} \left\{ -\frac{q u''(q)}{u'(q)}  \right\}
= -\frac{k-1}{k}    \frac{\gamma q^{-\frac{1}{k}}}{(q^\frac{k-1}{k} +\gamma)^2},
\vspace{-.1cm}
\end{equation*}
which correspond to the case with $\iota=1$,  $a=0$, $b=\frac{1}{1-k}-1$, $c=1$, and $z= -\frac{1}{\gamma} q^\frac{k-1}{k}<0$ in \eqref{eq:fofq}. In this case, we can rewrite \eqref{eq:SODE_re} as a hypergeometric differential equation:
\vspace{-.1cm}
\begin{equation}
\textstyle z (1- z) v''(z) + \left(  1- \frac{z}{1-k} \right) v'(z) = 0.
\label{eq:SODE_re2}
\vspace{-.1cm}
\end{equation}
Since the solution to \eqref{eq:SODE_re} for $\iota=1$ is given by $v(z) = \textstyle c_1\times z^{-a} {}_2 F_1 (a, a-c+1; a-b+1; \frac{1}{z} ) + c_2\times z^{-b} {}_2 F_1(b, b-c+1; b-a+1; \frac{1}{z})$, where $c_1$ and $c_2$ denote arbitrary coefficients (see, e.g., Sneddon, 1961), we have
\vspace{-.1cm}
\begin{equation}
v(z) = \textstyle c_1
+ c_2\times z^{-\frac{k}{1-k}} {}_2 F_1(\frac{k}{1-k}, \frac{k}{1-k}; \frac{1}{1-k}; \frac{1}{z}),
\label{eq:vz_nonlinear}
\vspace{-.1cm}
\end{equation}
where we use $ {}_2 F_1 \bigl(0, 0; 1- \frac{k}{1-k}; \frac{1}{z} \bigr)=1$ and assume that $\frac{1}{1-k}$ is not a negative integer for ${}_2 F_1(\frac{k}{1-k}, \frac{k}{1-k}; \frac{1}{1-k}; \frac{1}{z})$ to be well defined (see Lebedev, 1965). Recalling $u(q)=v(z)$ and $z= -\frac{1}{\gamma} q^\frac{k-1}{k}$, we obtain the CPPT utility function of the form:
\vspace{-.1cm}
\begin{equation}
u(q) = \textstyle C
+ K q \, {}_2 F_1\bigl(\frac{k}{1-k}, \frac{k}{1-k}; \frac{1}{1-k}; -\gamma q^\frac{1-k}{k} \bigr),
\label{eq:vz_nonlinear2}
\vspace{-.1cm}
\end{equation}
where $C=c_1$ and $K=  c_2  \bigl(-\frac{1}{\gamma} \bigr)^{-\frac{k}{1-k}} > 0$.

\vspace{-.2cm}
\paragraph{The expo-power utility case.}

We next consider the expo-power utility in Dhingra and Morrow (2019).~The RRA and its derivative for this case are given by (see, e.g., Saha, 1993; Holt and Laury, 2002):
\begin{equation*}
-\frac{q u''(q)}{u'(q)} = \rho+ (1-\rho) \alpha q^{1-\rho}
\quad \mbox{and} \quad
\frac{\partial}{\partial q} \left\{ -\frac{q u''(q)}{u'(q)} \right\} = \alpha (1-\rho)^2 q^{-\rho}.
\end{equation*}
Letting $\iota \to \infty$ in \eqref{eq:fofq} and setting $c=0$ and $z= -\alpha q^{1-\rho}$, we have
$\phi(q) =\rho+ (1-\rho) \alpha q^{1-\rho}$. In this case, \eqref{eq:SODE_re} may be viewed as a (degenerate) confluent hypergeometric differential equation:
\vspace{-.6cm}
\begin{equation}
\textstyle
z  v''(z)  - z v'(z) = 0,
\label{eq:ourHGDE_expo_power}
\vspace{-.1cm}
\end{equation}
which can be readily solved for $v(z)$ as
\vspace{-.1cm}
\begin{equation}
v(z) =  \textstyle c_1 + c_2  {\rm e}^z.
\label{eq:vz_expo_power}
\vspace{-.1cm}
\end{equation}
Setting $u(q) = v(z)$, $c_1=\frac{1}{\alpha}$, and $c_2 =-\frac{1}{\alpha}$, and using $z= -\alpha q^{1-\rho}$, we obtain the expo-power utility of the form:
\vspace{-.1cm}
\begin{equation*}
\textstyle u(q) = \frac{1}{\alpha} (1-  {\rm e}^{-\alpha q^{1-\rho}} ),
\vspace{-.1cm}
\end{equation*}
which requires $1-\rho>0$ for $u'(q)>0$. One can verify that $1-\rho>0$ and $-\frac{q u''(q)}{u'(q)} = \rho+ (1-\rho) \alpha q^{1-\rho}>0$ ensure $u''(q) = -(1-\rho) q^{-1-\rho} {\rm e}^{-\alpha q^{1-\rho}} [\rho+ (1-\rho) \alpha q^{1-\rho}]  <0$.

\vspace{-.3cm}
\paragraph{The implicitly additive case.}
Recall the first-order condition for \mbox{expenditure minimization} $p = \frac{{\lambda}^{\rm imp} }{U} \Upsilon' \bigl( \frac{q}{U} \bigr)$ in Section~\ref{sec:extensions_implicit_additive}. Consider the inverse demand function $\Upsilon'(\theta)  = \frac{\overline \sigma-1}{\overline \sigma} {\rm exp} \bigl(\frac{1- \theta^{\varepsilon /\overline \sigma}}{\varepsilon} \bigr)$, where $\theta = \frac{q}{U}$ and $\overline \sigma>1$ (see, e.g., Klenow and Willis, 2016; Edmond et al., 2023). \mbox{We then have}
\vspace{-.3cm}
\begin{equation*}
-\frac{\theta \Upsilon''(\theta)}{\Upsilon'(\theta)} = \frac{1}{\overline \sigma \theta^{-\frac{\varepsilon}{\overline \sigma}}}
\quad \mbox{and} \quad
\frac{\partial}{\partial \theta} \left\{ -\frac{\theta \Upsilon''(\theta)}{\Upsilon'(\theta)}  \right\}
= \frac{\varepsilon}{\overline \sigma^2 \theta^{-\frac{\varepsilon}{\overline \sigma} +1} },
\vspace{-.1cm}
\end{equation*}
which correspond to the case with $\iota \to \infty$,  $a=0$, $c=1-\frac{\overline \sigma}{\varepsilon}$, $z= -\frac{1}{\varepsilon} q^\frac{\varepsilon}{\overline \sigma}$, and $q=\theta$ in \eqref{eq:fofq}. In this case, we can rewrite \eqref{eq:SODE_re} as a confluent hypergeometric differential equation:
\vspace{-.1cm}
\begin{equation}
\textstyle
z  v''(z) + \bigl(1-\frac{\overline \sigma}{\varepsilon} -z \bigr) v'(z) = 0.
\label{eq:ourHGDE_nonlinear_ed}
\vspace{-.1cm}
\end{equation}
Since the solution to \eqref{eq:SODE_re} for $\iota \to \infty$ is given by $v(z) = \textstyle c_1\times {}_1 F_1 (a; c; z ) + c_2\times z^{1-c} {}_1 F_1(1+a-c; 2-c; z)$, where ${}_1 F_1 (a; c; z)= \sum_{n=0}^\infty \frac{(a)_n}{(c)_n} \frac{z^n}{n!}$ is the confluent hypergeometric function (see Seaborn, 1991), we have
\vspace{-.1cm}
\begin{equation}
v(z) = \textstyle c_1
+ c_2 \, z^{\frac{\overline \sigma}{\varepsilon}} {}_1 F_1(\frac{\overline \sigma}{\varepsilon}; 1+\frac{\overline \sigma}{\varepsilon}; z),
\label{eq:vz_nonlinear_ed}
\vspace{-.1cm}
\end{equation}
where we use  ${}_1 F_1 (0; 1-\frac{\overline \sigma}{\varepsilon}; z )=1$. Setting $v(z)= \Upsilon(\theta)$ and using $z= -\frac{1}{\varepsilon} \theta^\frac{\varepsilon}{\overline \sigma}$, we obtain the sub-function of the form:
\vspace{-.1cm}
\begin{equation}
\Upsilon(\theta) = \textstyle c_1
+ c_2 (-\frac{1}{\varepsilon} )^{\frac{\overline \sigma}{\varepsilon}}  \theta \, {}_1 F_1(\frac{\overline \sigma}{\varepsilon}; 1+\frac{\overline \sigma}{\varepsilon}; -\frac{1}{\varepsilon} \theta^\frac{\varepsilon}{\overline \sigma}).
\label{eq:vz_nonlinear2_ed}
\vspace{-.1cm}
\end{equation}
Noting ${}_1 F_1(\frac{\overline \sigma}{\varepsilon}; 1+\frac{\overline \sigma}{\varepsilon}; -\frac{1}{\varepsilon} \theta^\frac{\varepsilon}{\overline \sigma})
= \frac{\overline \sigma}{\varepsilon} (\frac{1}{\varepsilon} \theta^{\frac{\varepsilon}{\overline \sigma}})^{-\frac{\overline \sigma}{\varepsilon}}\gamma(\frac{\overline \sigma}{\varepsilon},  \frac{1}{\varepsilon} \theta^\frac{\varepsilon}{\overline \sigma})$, equation \eqref{eq:vz_nonlinear2_ed} can be rewritten as
$\Upsilon(\theta) = \textstyle c_1
+ c_2 (-\frac{1}{\varepsilon} )^{\frac{\overline \sigma}{\varepsilon}}   \frac{\overline \sigma}{\varepsilon} (\frac{1}{\varepsilon} )^{-\frac{\overline \sigma}{\varepsilon}}\gamma(\frac{\overline \sigma}{\varepsilon},  \frac{1}{\varepsilon} \theta^\frac{\varepsilon}{\overline \sigma})$.\footnote{See \url{https://dlmf.nist.gov/13.6.5} (NIST Digital Library of Mathematical Functions).}
Using $\gamma(\frac{\overline \sigma}{\varepsilon},  \frac{1}{\varepsilon} \theta^\frac{\varepsilon}{\overline \sigma})= \Gamma(\frac{\overline \sigma}{\varepsilon}) - \Gamma(\frac{\overline \sigma}{\varepsilon},  \frac{1}{\varepsilon} \theta^\frac{\varepsilon}{\overline \sigma})$,
we have $\Upsilon(\theta) = \textstyle c_1
+ c_2 (-\frac{1}{\varepsilon} )^{\frac{\overline \sigma}{\varepsilon}}   \frac{\overline \sigma}{\varepsilon} (\frac{1}{\varepsilon} )^{-\frac{\overline \sigma}{\varepsilon}}
\left[
\Gamma(\frac{\overline \sigma}{\varepsilon}) - \Gamma(\frac{\overline \sigma}{\varepsilon},  \frac{1}{\varepsilon} \theta^\frac{\varepsilon}{\overline \sigma})
\right]$.\footnote{See \url{https://dlmf.nist.gov/8.2.3} (NIST Digital Library of Mathematical Functions).}
Setting $c_1$ and $c_2$ to satisfy $c_2 (-\frac{1}{\varepsilon} )^{\frac{\overline \sigma}{\varepsilon}}   \frac{\overline \sigma}{\varepsilon} (\frac{1}{\varepsilon} )^{-\frac{\overline \sigma}{\varepsilon}}
=(\overline \sigma-1) {\rm e}^\frac{1}{\varepsilon} \varepsilon^{\frac{\overline \sigma}{\varepsilon}-1}$
and $c_1+ c_2 (-\frac{1}{\varepsilon} )^{\frac{\overline \sigma}{\varepsilon}}   \frac{\overline \sigma}{\varepsilon} (\frac{1}{\varepsilon} )^{-\frac{\overline \sigma}{\varepsilon}}
\Gamma(\frac{\overline \sigma}{\varepsilon})=1+ (\overline \sigma-1) {\rm e}^\frac{1}{\varepsilon} \varepsilon^{\frac{\overline \sigma}{\varepsilon}-1}
\Gamma(\frac{\overline \sigma}{\varepsilon},  \frac{1}{\varepsilon})$, we obtain $\Upsilon (\theta)$, which is isomorphic to $\Upsilon (q)$ in footnote 9 in Edmond et al.~(2023).

\medbreak
As in Section~\ref{sec:fit}, we fit these four non-linear fractional forms to data (see \ref{app:empirics}). We find that the value of the RSS for each sector and the aggregate economy using the LFRRA (see  Table~\ref{tab:tab_est_3pct_comb_rounded}) is smaller than the corresponding values for these non-linear fractional forms (see  Tables~\ref{tab:tab_app_est_bipower}--\ref{tab:tab_app_est_edmond} in \ref{app:empirics}). Put differently, for the data from De Loecker et al.~(2016), the LFRRA provides a better fit than these four cases.

\section{Conclusion}
\label{sec:conclusion}

In this paper, we first characterize the LFRRA family of utility functions in terms of the Gauss hypergeometric functions. This family, which is parametrized by $\{\alpha, \beta, \sigma\}$, nests the CRRA (CES), CARA, quadratic, log, HARA, CREMR, incomplete gamma, incomplete beta, and exponential integral utility functions as special cases. It also provides a flexible RRA that is a second-order approximation of essentially any real analytic function of quantity. We then embed this family into a monopolistic competition model and derive the price equilibrium. We show that the inverse markup function is expressed as the generalized Lambert $W$ function, ${\cal W}(x, \alpha, \beta, \sigma)$, which reduces to the original Lambert $W$ function when $\beta$ and $\sigma$ are set to zero. We further take the relationship between the inverse markup and the LFRRA to firm-level data and illustrate a parsimonious way to estimate $\{\widehat \alpha, \widehat \beta, \widehat \sigma\}$. The estimates reveal which sectors display IRRA, DRRA, or CRRA, which in turn determines whether markups are decreasing, increasing, or constant in marginal costs. Our results suggest that all cases are present in the data, thus highlighting the importance of taking a flexible parametric approach. We finally extend the LFRRA to non-linear fractional forms, which allows us to incorporate an even wider range of specifications into a unified framework. Notably, all examples analyzed in this paper can be viewed through the lens of hypergeometric differential equations.

We believe that the LFRRA would be useful to fill the gap between the empirical literature on markups, which shows that markups are heterogeneous across firms in each sector, and quantitative trade and urban models, where markups are typically assumed to be constant. Applying the LFRRA family to such general equilibrium analysis will require to replace the inverse markup function or the Lambert $W$ function in Behrens et al.~(2014, 2017, 2020) with our generalized Lambert $W$ function. Observe that the benefit of using the LFRRA is not limited to the price equilibrium in trade and urban models. It would also spur the development of qualitative and quantitative analysis in various other contexts such as economic growth and the economics of uncertainty, where the RRA matters.

\setcounter{section}{0}
\numberwithin{equation}{section}
\renewcommand{\thesection}{Appendix \Alph{section}}
\renewcommand\thefigure{\Alph{section}.\arabic{figure}}
\renewcommand\thetable{\Alph{section}.\arabic{table}}
\renewcommand\theequation{\Alph{section}.\arabic{equation}}

\section{Proofs}

\vspace{-.25cm}

\subsection{Proof of Theorem~\ref{prop:T1}}
\label{app:theorem1}
\vspace{-.1cm}
Assume that $\alpha=0$. Then, we can establish the if part as follows. Setting $\alpha=0$ in \eqref{eq:prop4_subutility_general_case} and noting that ${}_2 F_{1} \left(1- \frac{1}{\sigma}, 1; 2- \beta ; 0   \right)=1$, we have $u(q)= \frac{K q^{1-\beta}}{1-\beta}+C$. Differentiating this yields $u'(q)= K q^{-\beta}$ and $u''(q)= -K \beta q^{-1-\beta}$, so that $- \frac{q u''(q)}{u'(q)} =\beta$, which is equivalent to the LFRRA in \eqref{eq:LFRRA_general} evaluated at $\alpha=0$. We can also establish the only if part using the definition of the LFRRA. Indeed, setting $\alpha=0$ in \eqref{eq:LFRRA_general} yields $- \frac{q u''(q)}{u'(q)} =\beta$. We can solve it for $u$ as $u(q)= \frac{K q^{1-\beta}}{1-\beta}+C$, which is equivalent to \eqref{eq:prop4_subutility_general_case} evaluated at $\alpha=0$.

Assume now that $\alpha \neq 0$.
Then, we can establish the if part by using \eqref{eq:prop4_subutility_general_case}. Using the differentiation formula for the Gauss hypergeometric function
$\frac{\partial}{\partial (-\alpha \sigma q )} \, {}_2 F_{1} \bigl(1- \frac{1}{\sigma}, 1; 2- \beta ; - \alpha \sigma q   \bigr)
=\frac{1-\frac{1}{\sigma}}{2-\beta} \, {}_2 F_{1} \bigl(2- \frac{1}{\sigma}, 2; 3- \beta ; - \alpha \sigma q   \bigr) 
$, we have
$\frac{\partial}{\partial q} \, {}_2 F_{1} \bigl(1- \frac{1}{\sigma}, 1; 2- \beta ; - \alpha \sigma q   \bigr)
= (- \alpha \sigma) \frac{1-\frac{1}{\sigma}}{2-\beta} \, {}_2 F_{1} \bigl(2- \frac{1}{\sigma}, 2; 3- \beta ; - \alpha \sigma q   \bigr)$.\footnote{See \url{https://dlmf.nist.gov/15.5.1} (NIST Digital Library of Mathematical Functions).}
We can thus differentiate \eqref{eq:prop4_subutility_general_case} with respect to $q$ to obtain
\begin{eqnarray}
u'(q) &=& \textstyle \frac{K \sigma}{\sigma-1} [1-\beta + \alpha (\sigma-1) q] (1+\alpha \sigma q)^{-1+ \beta-\frac{1}{\sigma}} q^{-\beta} \left[ 1
+\frac{\beta -\frac{1}{\sigma}}{1-\beta} \,
{}_2 F_{1} \left(1- \frac{1}{\sigma}, 1; 2- \beta ; - \alpha \sigma q   \right)  \right]  \notag\\
&&  \textstyle + \frac{K \sigma}{\sigma-1} (1+\alpha \sigma q)^{\beta-\frac{1}{\sigma}} q^{1-\beta}  \frac{\beta -\frac{1}{\sigma}}{1-\beta}
 (- \alpha \sigma) \frac{1-\frac{1}{\sigma}}{2-\beta} \, {}_2 F_{1} \bigl(2- \frac{1}{\sigma}, 2; 3- \beta ; - \alpha \sigma q   \bigr) 
\notag \\
&=&  \textstyle \frac{K \sigma}{\sigma-1} (1+\alpha \sigma q)^{-1+ \beta-\frac{1}{\sigma}} q^{-\beta} \left\{ 
[1-\beta + \alpha (\sigma-1) q] \left[ 1
+\frac{\beta -\frac{1}{\sigma}}{1-\beta}  \,
{}_2 F_{1} \left(1- \frac{1}{\sigma}, 1; 2- \beta ; - \alpha \sigma q   \right)  \right] \right. \notag\\
&& \left. \textstyle + (1+\alpha \sigma q) q  \frac{\beta -\frac{1}{\sigma}}{1-\beta}
 (- \alpha \sigma) \frac{1-\frac{1}{\sigma}}{2-\beta} \, {}_2 F_{1} \bigl(2- \frac{1}{\sigma}, 2; 3- \beta ; - \alpha \sigma q   \bigr) \right\} \notag \\
 &=&  \textstyle \frac{K \sigma}{\sigma-1} (1+\alpha \sigma q)^{-1+ \beta-\frac{1}{\sigma}} q^{-\beta} \left\{ 
[1-\beta + \alpha (\sigma-1) q] \left[ 1
+\frac{\beta -\frac{1}{\sigma}}{1-\beta}  \,
{}_2 F_{1} \left(1- \frac{1}{\sigma}, 1; 2- \beta ; - \alpha \sigma q   \right)  \right] \right. \notag\\
&& \left. \textstyle + \frac{\beta -\frac{1}{\sigma}}{1-\beta}
\frac{1}{2-\beta}  (- \alpha \sigma q) (1+\alpha \sigma q) (1-\frac{1}{\sigma})    \, {}_2 F_{1} \bigl(2- \frac{1}{\sigma}, 2; 3- \beta ; - \alpha \sigma q   \bigr) \right\}.
 \label{app:hg}
\end{eqnarray}
The contiguous relationship of the Gauss hypergeometric functions implies\footnote{See \url{https://dlmf.nist.gov/15.5.19} (NIST Digital Library of Mathematical Functions).}
\begin{eqnarray*}
&& \hskip -.5cm \textstyle  -  \alpha \sigma q  (1+ \alpha \sigma q )   \bigl(1- \frac{1}{\sigma} \bigr) \times 1 \times  \, 
{}_2 F_{1} \bigl(2- \frac{1}{\sigma}, 2; 3- \beta ; - \alpha \sigma q   \bigr) \notag \\
&& \textstyle  +\bigl[(1-\beta)- \bigl(-\frac{1}{\sigma}  + 0+1 \bigr) (- \alpha \sigma q) \bigr] (2- \beta ) {}_2 F_{1} \bigl(1- \frac{1}{\sigma}, 1; 2- \beta ; - \alpha \sigma q   \bigr) \\
&& \textstyle - (1-\beta) (2- \beta ) {}_2 F_{1} \bigl(- \frac{1}{\sigma}, 0; 1- \beta ; - \alpha \sigma q   \bigr)  =0.
\end{eqnarray*}
Noting that $ {}_2 F_{1} \bigl(- \frac{1}{\sigma}, 0; 1- \beta ; - \alpha \sigma q   \bigr)=1$, the foregoing equation can be rewritten as
\begin{eqnarray*}
&&  \textstyle  -  \alpha \sigma q  (1+ \alpha \sigma q )   \bigl(1- \frac{1}{\sigma} \bigr)   \, 
{}_2 F_{1} \bigl(2- \frac{1}{\sigma}, 2; 3- \beta ; - \alpha \sigma q   \bigr) \notag \\
&=& \textstyle  - \bigl[1-\beta + (\sigma-1) \alpha q  \bigr] (2- \beta ) {}_2 F_{1} \bigl(1- \frac{1}{\sigma}, 1; 2- \beta ; - \alpha \sigma q   \bigr)  \textstyle + (1-\beta) (2- \beta ). 
\end{eqnarray*}
Plugging this into \eqref{app:hg}, we have the marginal utility function as follows:
\begin{eqnarray*}
u'(q)
&=&  \textstyle \frac{K \sigma}{\sigma-1} (1+\alpha \sigma q)^{-1+ \beta-\frac{1}{\sigma}} q^{-\beta} \left\{ 
[1-\beta + \alpha (\sigma-1) q] \left[ 1
+\frac{\beta -\frac{1}{\sigma}}{1-\beta}  \,
{}_2 F_{1} \left(1- \frac{1}{\sigma}, 1; 2- \beta ; - \alpha \sigma q   \right)  \right] \right. \notag\\
&& \left. \textstyle + \frac{\beta -\frac{1}{\sigma}}{1-\beta}
\frac{1}{2-\beta}  (- \alpha \sigma q) (1+\alpha \sigma q) (1-\frac{1}{\sigma})    \, {}_2 F_{1} \bigl(2- \frac{1}{\sigma}, 2; 3- \beta ; - \alpha \sigma q   \bigr) \right\} \notag \\
&=&  \textstyle \frac{K \sigma}{\sigma-1} (1+\alpha \sigma q)^{-1+ \beta-\frac{1}{\sigma}} q^{-\beta} \left\{ 
[1-\beta + \alpha (\sigma-1) q] \left[ 1
+\frac{\beta -\frac{1}{\sigma}}{1-\beta}  \,
{}_2 F_{1} \left(1- \frac{1}{\sigma}, 1; 2- \beta ; - \alpha \sigma q   \right)  \right] \right. \notag\\
&& \left. \textstyle - \frac{\beta -\frac{1}{\sigma}}{1-\beta}
\frac{1}{2-\beta}
 \bigl[1-\beta + (\sigma-1) \alpha q  \bigr] (2- \beta ) {}_2 F_{1} \bigl(1- \frac{1}{\sigma}, 1; 2- \beta ; - \alpha \sigma q   \bigr)  \textstyle + \beta -\frac{1}{\sigma}
\right\} \notag \\
   &=&    \textstyle \frac{K \sigma}{\sigma-1} (1+\alpha \sigma q)^{-1+ \beta-\frac{1}{\sigma}} q^{-\beta} \bigl[ 
1-\beta + \alpha (\sigma-1) q
  +  \beta -\frac{1}{\sigma}  \bigr]\notag \\
   &=&    \textstyle \frac{K \sigma}{\sigma-1} (1+\alpha \sigma q)^{-1+ \beta-\frac{1}{\sigma}} q^{-\beta} \bigl[ 
1  -\frac{1}{\sigma} + \alpha (\sigma-1) q
  \bigr]\notag  \\
     &=&    \textstyle K (1+\alpha \sigma q)^{-1+ \beta-\frac{1}{\sigma}} q^{-\beta} (1+   \alpha \sigma q )
     =K (1+\alpha \sigma q)^{\beta-\frac{1}{\sigma}} q^{-\beta}, \notag 
\end{eqnarray*}
whose derivative yields $u''(q) = - K (\alpha q+ \beta) (1+ \alpha \sigma q)^{-1+ \beta-\frac{1}{\sigma}} q^{-1-\beta}.$ Using the expressions for $u'(q)$ and $u''(q)$, we obtain the LFRRA in \eqref{eq:LFRRA_general}. This completes the proof of the if part.

We can also establish the only if part using \eqref{eq:vz}. Since ${}_2 F_1\bigl(\frac{1-\sigma}{\sigma}, 0; \beta; z\bigr)=1$, $z = -\alpha \sigma q$, and $v(z)=u(q)$, equation \eqref{eq:vz} can be rewritten as
\begin{equation}
u(q)
= \textstyle c_1 + c_2 (-\alpha\sigma q)^{1-\beta}{}_2 F_1\bigl(\frac{1}{\sigma}-\beta,1-\beta; 2-\beta; -\alpha\sigma q\bigr). \label{eq:u1}
\end{equation}
We can then use Euler's hypergeometric transformation
to obtain\footnote{See \url{https://mathworld.wolfram.com/EulersHypergeometricTransformations.html}.}
\begin{eqnarray*}
&& \textstyle {}_2 F_1\bigl(\frac{1}{\sigma}-\beta,1-\beta; 2-\beta; -\alpha\sigma q\bigr) \\
&=& \textstyle(1+\alpha\sigma q)^{2-\beta-(\frac{1}{\sigma}-\beta)-(1-\beta)} {}_2 F_1\bigl(2-\beta - (\frac{1}{\sigma}-\beta), 2-\beta-(1-\beta); 2-\beta; -\alpha\sigma q\bigr) \\
&=& \textstyle (1+\alpha\sigma q)^{1-\frac{1}{\sigma}+\beta} {}_2 F_1\bigl(2-\frac{1}{\sigma}, 1; 2-\beta; -\alpha\sigma q\bigr).
\end{eqnarray*}
Applying this transformation to \eqref{eq:u1} yields
\begin{eqnarray}
u(q) = \textstyle c_1 + c_2  (-\alpha\sigma q)^{1-\beta} (1+\alpha\sigma q)^{1-\frac{1}{\sigma}+\beta} {}_2 F_1\bigl(2-\frac{1}{\sigma}, 1; 2-\beta; -\alpha\sigma q\bigr).
\label{eq:u2}
\end{eqnarray}
We next use the recurrence relationship\footnote{See \url{https://dlmf.nist.gov/15.5.13} (NIST Digital Library of Mathematical Functions).}
\begin{eqnarray*}
&& \hskip -.8cm \textstyle \bigl[2-\beta-(1-\frac{1}{\sigma})-1\bigr] {}_2 F_1\bigl(1-\frac{1}{\sigma}, 1; 2-\beta; -\alpha\sigma q\bigr) \\
&& \hskip -.8cm + \textstyle \bigl(1-\frac{1}{\sigma}\bigr)(1+\alpha\sigma q) \textstyle {}_2 F_1\bigl(2-\frac{1}{\sigma}, 1; 2-\beta; -\alpha\sigma q\bigr) -
(2-\beta-1)  {}_2 F_1\bigl(1-\frac{1}{\sigma}, 0; 2-\beta; -\alpha\sigma q\bigr) = 0.
\end{eqnarray*}
Solving it for $\textstyle (1+\alpha\sigma q) {}_2 F_1\bigl(2-\frac{1}{\sigma}, 1; 2-\beta; -\alpha\sigma q\bigr)$,
plugging the resulting expression into \eqref{eq:u2}, and noting that $ {}_2 F_1\bigl(1-\frac{1}{\sigma}, 0; 2-\beta; -\alpha\sigma q\bigr) = 1$ yield
\begin{eqnarray*}
u(q) &=& \textstyle c_1 + c_2 (-\alpha\sigma q)^{1-\beta} (1+\alpha\sigma q)^{1-\frac{1}{\sigma}+\beta} {}_2 F_1\bigl(2-\frac{1}{\sigma}, 1; 2-\beta; -\alpha\sigma q\bigr) \\
&=& \textstyle c_1 + c_2 (-\alpha\sigma q)^{1-\beta} (1+\alpha\sigma q)^{-\frac{1}{\sigma}+\beta} \frac{\sigma}{\sigma-1}  \left[1-\beta + \bigl(\beta-\frac{1}{\sigma}\bigr){}_2 F_1\bigl(1-\frac{1}{\sigma}, 1; 2-\beta; -\alpha\sigma q\bigr) \right] \\
&=& \textstyle c_1 + \frac{c_2 \sigma }{\sigma-1}  (-\alpha\sigma)^{1-\beta} (1+\alpha\sigma q)^{-\frac{1}{\sigma}+\beta}q^{1-\beta}  \left[1-\beta + \bigl(\beta-\frac{1}{\sigma}\bigr){}_2 F_1\bigl(1-\frac{1}{\sigma}, 1; 2-\beta; -\alpha\sigma q\bigr) \right] \\
&=& \textstyle c_1 + \frac{c_2  \sigma  }{\sigma-1}
(1-\beta)(-\alpha\sigma)^{1-\beta}
(1+\alpha\sigma q)^{-\frac{1}{\sigma}+\beta}q^{1-\beta}  \left[1 + \frac{\beta-\frac{1}{\sigma}}{1-\beta}{}_2 F_1\bigl(1-\frac{1}{\sigma}, 1; 2-\beta; -\alpha\sigma q\bigr) \right],
\end{eqnarray*}
where the third equality follows from 
$(s_1 s_2)^j=s_1^{j} s_2^{j}$ for any real number $j$ if real numbers $s_1$ and $s_2$ are not both negative
(Pennisi, 1963,
Theorem 4.4.5, together with Theorem 4.4.2).
Setting $C=c_1$ and $K = c_2 (1-\beta) (-\alpha\sigma)^{1-\beta}$ and rearranging terms then yield
the utility function \eqref{eq:prop4_subutility_general_case}.\footnote{Even when $\alpha\sigma>0$, we can choose a complex constant $c_2$ such that $K = c_2 (1-\beta) (-\alpha\sigma)^{1-\beta}$ is real valued.}
This completes the only if part.
\hfill $\square$

\subsection{Proof of Proposition~\ref{integral}}
\label{app:proposition1}
\vspace{-.1cm}
The Gauss hypergeometric function
${}_2 F_{1} \bigl(1- \frac{1}{\sigma}, 1; 2- \beta ; - \alpha \sigma q   \bigr)$ in \eqref{eq:prop4_subutility_general_case} is defined for $\lvert - \alpha \sigma q \rvert <1$. Thus, there are two cases depending on the sign of $\alpha \sigma$.\footnote{Strictly speaking, there is another case with $\alpha \sigma=0$, which implies $\alpha=0$ by the assumption that $\sigma \neq 0$ in Theorem~\ref{prop:T1} and Proposition~\ref{integral}.
Since ${}_2 F_{1} \bigl(1- \frac{1}{\sigma}, 1; 2- \beta ; - \alpha \sigma q   \bigr)=1$ for $\alpha=0$,
we do not have to replace ${}_2 F_{1} \bigl(1- \frac{1}{\sigma}, 1; 2- \beta ; - \alpha \sigma q   \bigr)$ in \eqref{eq:prop4_subutility_general_case}  with the integral in \eqref{eq:utility_general_continuation}. However, we can let $\alpha \to 0$ in \eqref{eq:utility_general_continuation} to obtain $u(q) = \frac{K q^{1-\beta}}{1-\beta}+C$ for all $q \in [0,\infty)$, which is consistent with Table~\ref{integral} and equivalent to \eqref{eq:prop4_subutility_general_case} evaluated at $\alpha=0$. We will return to this case in Corollary~\ref{coroll_NEW}.}
First, when $\alpha \sigma <0$, $\lvert - \alpha \sigma q \rvert <1$ implies $0 \le q< -\frac{1}{\alpha \sigma}$, which is not restrictive given Table~\ref{tab:QR}. Second, when $\alpha \sigma > 0$, $\lvert - \alpha \sigma q \rvert <1$ implies $0\le q<\frac{1}{\alpha \sigma}$, which appears to be restrictive given Table~\ref{tab:QR}.  However, since $2-\beta>1>0$ holds in Theorem~\ref{prop:T1}, we can extend ${}_2 F_{1} \bigl(1- \frac{1}{\sigma}, 1; 2- \beta ; - \alpha \sigma q   \bigr)$ in \eqref{eq:prop4_subutility_general_case} to $q \ge \frac{1}{\alpha \sigma}$
by analytic continuation (see, e.g.,~(9.1.6) in Lebedev, 1965, where $\lvert {\rm arg} (1-z)\rvert =\lvert {\rm arg} (1+\alpha \sigma q)\rvert =0$ since $1+\alpha \sigma q > 0$). Thus, in both cases, we can use the integral representation ${}_2 F_{1} \bigl(1- \frac{1}{\sigma}, 1; 2- \beta ; - \alpha \sigma q   \bigr) = \frac{\Gamma(2- \beta)}{\Gamma(1)\Gamma(1-\beta)}  \int_0^1 (1-t)^{- \beta}  (1 + t \alpha \sigma q)^{-(1- \frac{1}{\sigma})} {\rm d}t$, where $\Gamma(\cdot)$ is the gamma function. Using the recurrence relation $\Gamma(2- \beta)=(1-\beta) \Gamma(1-\beta)$ and $\Gamma (1)=1$, the foregoing expression reduces to $ {}_2 F_{1} \bigl(1- \frac{1}{\sigma}, 1; 2- \beta ; - \alpha \sigma q   \bigr) = (1-\beta) \int_0^1 (1-t)^{- \beta}  (1 + t \alpha \sigma q)^{-(1- \frac{1}{\sigma})} {\rm d}t$. Hence, for all $q$ satisfying the quantity restrictions in Table~\ref{tab:QR}, we can rewrite \eqref{eq:prop4_subutility_general_case}  as \eqref{eq:utility_general_continuation}.
\hfill $\square$

\subsection{Proof of Theorem~\ref{prop:utility_NEW}}
\label{app:theorem2}
\vspace{-.1cm}
To establish \eqref{eq:prop4_subutility_HARA} and \eqref{eq:prop4_subutility_hypergeo}, we first rewrite  \eqref{eq:prop4_subutility_general_case}.
Recall that \eqref{eq:prop4_subutility_general_case} is derived from
a linear combination of the two generally linearly independent solutions
$ {}_2 F_1\bigl(\frac{1-\sigma}{\sigma}, 0; \beta; z \bigr)$ and $z^{1-\beta}{}_2 F_1\bigl(1+\frac{1-\sigma}{\sigma}-\beta, 1-\beta; 2-\beta; z \bigr)$. However, when $\beta$ goes to~1 from below, these solutions are no longer linearly independent. Thus, to obtain both \eqref{eq:prop4_subutility_HARA} and \eqref{eq:prop4_subutility_hypergeo} from a single expression, we use the transformation of the Gauss hypergeometric function for $z= - \alpha \sigma q \notin (1,\infty)$ as follows:\footnote{See  \url{https://functions.wolfram.com/07.23.17.0079.01} for that transformation. Observe that the condition $z = - \alpha \sigma q \notin (1,\infty)$ always holds
for all quantities satisfying the restrictions in Table~\ref{tab:QR}.}
\begin{eqnarray*}
&& \textstyle {}_2 F_{1} \left(1- \frac{1}{\sigma}, 1; 2- \beta ; - \alpha \sigma q   \right) \\
&=&  \textstyle \frac{\Gamma (  \frac{1}{\sigma} ) \Gamma( 2- \beta )}{\Gamma (1) \Gamma ( 1- \beta+ \frac{1}{\sigma}   ) }
\bigl(  \frac{1}{\alpha \sigma q} \bigr)^{1- \frac{1}{\sigma}} \bigl( 1+  \frac{1}{\alpha \sigma q}  \bigr)^{- \beta + \frac{1}{\sigma}}
 {}_2 F_{1} \bigl(0, 1- \beta; \frac{1}{\sigma} - \beta +1   ; 1 +\frac{1}{ \alpha \sigma q}   \bigr) \\
&&  \textstyle + \frac{\Gamma (  \frac{1}{\sigma} ) \Gamma( 2- \beta )}{\Gamma ( \frac{1}{\sigma}  +1 ) \Gamma ( 1- \beta   ) }
\bigl( - \frac{1}{\alpha \sigma q} \bigr) 
 {}_2 F_{1} \bigl(1,  \beta;   \frac{1}{\sigma}  +1   ; -\frac{1}{ \alpha \sigma q}   \bigr) \\ 
 &=&  \textstyle \frac{\Gamma (  \frac{1}{\sigma} ) \Gamma( 2- \beta )}{ \Gamma ( 1- \beta+ \frac{1}{\sigma}   ) }
\bigl(  \frac{1}{\alpha \sigma q} \bigr)^{1- \frac{1}{\sigma}} \bigl( 1+  \frac{1}{\alpha \sigma q}  \bigr)^{- \beta + \frac{1}{\sigma}}  + \frac{\Gamma (  \frac{1}{\sigma} ) \Gamma( 2- \beta )}{\Gamma ( \frac{1}{\sigma}  +1 ) \Gamma ( 1- \beta   ) }
\bigl( - \frac{1}{\alpha \sigma q} \bigr) 
 {}_2 F_{1} \bigl(1,  \beta;   1+ \frac{1}{\sigma}     ; -\frac{1}{ \alpha \sigma q}   \bigr),
\end{eqnarray*}
where $\Gamma(\cdot)$ is the gamma function and we use $\Gamma(1)=1$ and $ {}_2 F_{1} \bigl(0, 1- \beta; \frac{1}{\sigma} - \beta +1   ; 1 +\frac{1}{ \alpha \sigma q}   \bigr)=1$.
Plugging the foregoing expression into \eqref{eq:prop4_subutility_general_case}, we have
\begin{eqnarray*}
u(q) &=&  \textstyle \frac{K \sigma}{\sigma-1} (1+\alpha \sigma q)^{\beta-\frac{1}{\sigma}} q^{1-\beta} \left[ 1
+\frac{\beta -\frac{1}{\sigma}}{1-\beta} \,
{}_2 F_{1} \left(1- \frac{1}{\sigma}, 1; 2- \beta ; - \alpha \sigma q   \right)  \right] + C \\
&=&  \textstyle \frac{K \sigma}{\sigma-1} (1+\alpha \sigma q)^{\beta-\frac{1}{\sigma}} q^{1-\beta} \left[ 1
+\frac{\beta -\frac{1}{\sigma}}{1-\beta}
\frac{\Gamma (  \frac{1}{\sigma} ) \Gamma( 2- \beta )}{ \Gamma ( 1- \beta+ \frac{1}{\sigma}   ) }
\bigl(  \frac{1}{\alpha \sigma q} \bigr)^{1- \frac{1}{\sigma}} \bigl( 1+  \frac{1}{\alpha \sigma q}  \bigr)^{- \beta + \frac{1}{\sigma}}  \right. \\
&& \textstyle \left. +
\frac{\beta -\frac{1}{\sigma}}{1-\beta}
\frac{\Gamma (  \frac{1}{\sigma} ) \Gamma( 2- \beta )}{\Gamma ( \frac{1}{\sigma}  +1 ) \Gamma ( 1- \beta   ) }
\bigl( - \frac{1}{\alpha \sigma q} \bigr) 
 {}_2 F_{1} \bigl(1,  \beta;  1+  \frac{1}{\sigma}    ; -\frac{1}{ \alpha \sigma q}   \bigr)  \right] + C.
\end{eqnarray*}
\vskip .1cm
\noindent
Since $\frac{\beta -\frac{1}{\sigma}}{1-\beta}
\frac{\Gamma (  \frac{1}{\sigma} ) \Gamma( 2- \beta )}{ \Gamma ( 1- \beta+ \frac{1}{\sigma}   ) }= - \frac{\Gamma( 1- \beta ) \Gamma (  \frac{1}{\sigma} ) }{ \Gamma ( - \beta+ \frac{1}{\sigma}   ) }$ and  $\frac{\beta -\frac{1}{\sigma}}{1-\beta}
\frac{\Gamma (  \frac{1}{\sigma} ) \Gamma( 2- \beta )}{\Gamma ( \frac{1}{\sigma}  +1 ) \Gamma ( 1- \beta   ) }= -1+\beta \sigma$,
we can rewrite it as
\begin{eqnarray}
\textstyle u(q) 
&=&  \textstyle \frac{K \sigma (1+\alpha \sigma q)^{\beta-\frac{1}{\sigma}} q^{1-\beta}}{\sigma-1}  \left[ 1
- \frac{\Gamma( 1- \beta ) \Gamma (  \frac{1}{\sigma} ) }{ \Gamma ( - \beta+ \frac{1}{\sigma}   ) }
\bigl(  \frac{1}{\alpha \sigma q} \bigr)^{1- \frac{1}{\sigma}} \bigl( 1+  \frac{1}{\alpha \sigma q}  \bigr)^{- \beta + \frac{1}{\sigma}}  \right.  \notag \\
&& \textstyle \left. + (-1+\beta \sigma)
\bigl( - \frac{1}{\alpha \sigma q} \bigr) 
 {}_2 F_{1} \bigl(1,  \beta;   1 + \frac{1}{\sigma}    ; -\frac{1}{ \alpha \sigma q}   \bigr)  \right] + C \notag \\
  &=&  \textstyle \frac{K \sigma (1+\alpha \sigma q)^{\beta-\frac{1}{\sigma}} q^{1-\beta}}{\sigma-1}  
- \frac{K \sigma (1+\alpha \sigma q)^{\beta-\frac{1}{\sigma}} q^{1-\beta}}{\sigma-1} 
\frac{\Gamma( 1- \beta ) \Gamma (  \frac{1}{\sigma} ) }{ \Gamma ( - \beta+ \frac{1}{\sigma}   ) }
\bigl(  \frac{1}{\alpha \sigma q} \bigr)^{1- \frac{1}{\sigma}} \bigl(  \frac{1+\alpha \sigma q}{\alpha \sigma q}  \bigr)^{- \beta + \frac{1}{\sigma}}   \label{eq:step}  \\
&& \textstyle 
+ \frac{K \sigma (1+\alpha \sigma q)^{\beta-\frac{1}{\sigma}} q^{1-\beta}}{\sigma-1}  \frac{1- \beta \sigma}{\alpha \sigma q}  
 {}_2 F_{1} \bigl(1,  \beta;  1+  \frac{1}{\sigma}    ; -\frac{1}{ \alpha \sigma q}   \bigr)   + C. \notag
\end{eqnarray}
Noting that $s^{j_1} s^{j_2} = s^{j_1+j_2}$ for any real numbers $s$, $j_1$, and $j_2$,
and that $s_1^{j} s_2^{j}=(s_1 s_2)^j$ for any real number $j$ if real numbers $s_1$ and $s_2$ are not both negative
(Pennisi, 1963, Theorem 4.4.4; and Theorem 4.4.5, together with Theorem 4.4.2),
we can simplify the second term \mbox{as follows:}
\begin{eqnarray*}
&& \textstyle  - \frac{K \sigma (1+\alpha \sigma q)^{\beta-\frac{1}{\sigma}} q^{1-\beta}}{\sigma-1} 
\frac{\Gamma( 1- \beta ) \Gamma (  \frac{1}{\sigma} ) }{ \Gamma ( - \beta+ \frac{1}{\sigma}   ) }
\bigl(  \frac{1}{\alpha \sigma q} \bigr)^{1- \frac{1}{\sigma}} \bigl( \frac{1+\alpha \sigma q}{\alpha \sigma q}  \bigr)^{- \beta + \frac{1}{\sigma}}  \\
&=& \textstyle  - \frac{K \sigma (1+\alpha \sigma q)^{\beta-\frac{1}{\sigma}} q^{1-\beta}}{\sigma-1} 
\frac{\Gamma( 1- \beta ) \Gamma (  \frac{1}{\sigma} ) }{ \Gamma ( - \beta+ \frac{1}{\sigma}   ) }
\bigl(  \frac{1}{\alpha \sigma q} \bigr)^{1- \frac{1}{\sigma}} ({1+\alpha \sigma q})^{- \beta + \frac{1}{\sigma}}
\bigl( \frac{1}{\alpha \sigma q}  \bigr)^{- \beta + \frac{1}{\sigma}}  \\
&=& \textstyle  - \frac{K \sigma q^{1-\beta}}{\sigma-1} 
\frac{\Gamma( 1- \beta ) \Gamma (  \frac{1}{\sigma} ) }{ \Gamma ( - \beta+ \frac{1}{\sigma}   ) }
\bigl(  \frac{1}{\alpha \sigma q} \bigr)^{1- \frac{1}{\sigma}} 
\bigl( \frac{1}{\alpha \sigma q}  \bigr)^{- \beta + \frac{1}{\sigma}}  \\
&=& \textstyle  - \frac{K \sigma q^{1-\beta}}{\sigma-1} 
\frac{\Gamma( 1- \beta ) \Gamma (  \frac{1}{\sigma} ) }{ \Gamma ( - \beta+ \frac{1}{\sigma}   ) }
\bigl(  \frac{1}{\alpha \sigma q} \bigr)^{1- \beta}   
= \textstyle  - \frac{K \sigma }{\sigma-1} 
\frac{\Gamma( 1- \beta ) \Gamma (  \frac{1}{\sigma} ) }{ \Gamma ( - \beta+ \frac{1}{\sigma}   ) }
\bigl(  \frac{1}{\alpha \sigma} \bigr)^{1- \beta}.
\end{eqnarray*}
Plugging this expression into the second term of \eqref{eq:step}
we obtain 
\begin{eqnarray}
u(q)
 &=&   \textstyle \frac{K \sigma (1+\alpha \sigma q)^{\beta-\frac{1}{\sigma}} q^{1-\beta}}{\sigma-1}  
- \frac{K \sigma }{\sigma-1} 
\frac{\Gamma( 1- \beta ) \Gamma (  \frac{1}{\sigma} ) }{ \Gamma ( - \beta+ \frac{1}{\sigma}   ) }
\bigl(  \frac{1}{\alpha \sigma} \bigr)^{1- \beta}  \notag \\
&& \textstyle 
+ \frac{K \sigma (1+\alpha \sigma q)^{\beta-\frac{1}{\sigma}} q^{1-\beta}}{\sigma-1}  \frac{1- \beta \sigma}{\alpha \sigma q}  
 {}_2 F_{1} \bigl(1,  \beta;   1+\frac{1}{\sigma}    ; -\frac{1}{ \alpha \sigma q}   \bigr)   + C  \notag \\
 &=&   \textstyle \frac{K \sigma (1+\alpha \sigma q)^{\beta-\frac{1}{\sigma}} q^{1-\beta}}{\sigma-1}
 \bigl[1+  \frac{1- \beta \sigma}{\alpha \sigma q}  
 {}_2 F_{1} \bigl(1,  \beta;   1+\frac{1}{\sigma}    ; -\frac{1}{ \alpha \sigma q}   \bigr) \bigr]  
- \frac{K \sigma }{\sigma-1} 
\frac{\Gamma( 1- \beta ) \Gamma (  \frac{1}{\sigma} ) }{ \Gamma ( - \beta+ \frac{1}{\sigma}   ) }
\bigl(  \frac{1}{\alpha \sigma} \bigr)^{1- \beta}     + C, \qquad \ \ 
\label{eq:prop1_general}
\end{eqnarray}
where $ {}_2 F_{1} \bigl(1, \beta; 1+ \frac{1}{\sigma}; -\frac{1}{ \alpha \sigma q} \bigr)=\sum_{n=0}^\infty \frac{(1)_n (\beta)_n}{(1+\frac{1}{\sigma})_n} \frac{(- \frac{1}{\alpha \sigma q})^n}{n!}$, with $\lvert - \frac{1}{\alpha \sigma q} \rvert<1$, 
denotes the Gauss hypergeometric function.

Using \eqref{eq:prop1_general}, we now establish \eqref{eq:prop4_subutility_HARA} and \eqref{eq:prop4_subutility_hypergeo}. First, when $\beta$ goes to $0$ from above, we can disregard the restriction $\lvert-\frac{1}{\alpha \sigma q}\rvert<1$ because
we have $\lim_{\beta \to 0^+} {}_2 F_{1} \bigl(1,  \beta;   1+ \frac{1}{\sigma} ; -\frac{1}{ \alpha \sigma q}   \bigr)=1$, regardless of the value of $-\frac{1}{\alpha \sigma q}$.  Hence, letting $\beta$ in \eqref{eq:prop1_general} go to 0 from above and by noting that 
$\lim_{\beta \to 0^+}  {}_2 F_{1} \bigl(1,  \beta;  1+ \frac{1}{\sigma}; -\frac{1}{ \alpha \sigma q}   \bigr) =1$ and $\Gamma(1)=1$, we obtain  \eqref{eq:prop4_subutility_HARA}  as follows:
\begin{eqnarray*}
\lim_{\beta \to 0^+} u(q) &=&
\textstyle \frac{K \sigma (1+\alpha \sigma q)^{-\frac{1}{\sigma}} q}{\sigma-1}
 \bigl(1+  \frac{1}{\alpha \sigma q}    \bigr)  
- \frac{K \sigma }{\sigma-1} 
\bigl(  \frac{1}{\alpha \sigma} \bigr)     + C \\
 &=&
\textstyle \frac{K  (1+\alpha \sigma q)^{-\frac{1}{\sigma}} }{ \alpha(\sigma-1)}
(\alpha \sigma q+1)
- \frac{K }{\alpha (\sigma-1)}     + C
= \textstyle \frac{K   }{ \alpha(\sigma-1)} \bigl[ (1+\alpha \sigma q)^{1-\frac{1}{\sigma}} -1 \bigr]     + C.
\end{eqnarray*}
Differentiating this with respect to $q$, we obtain $u'(q) = K (1+\alpha \sigma q)^{-\frac{1}{\sigma}}$ and $u''(q)= - K\alpha (1+\alpha \sigma q)^{-\frac{\sigma+1}{\sigma}}$, which yield the HARA: $- \frac{u'(q)}{u''(q)}
=- \frac{K (1+\alpha \sigma q)^{-\frac{1}{\sigma}}}{- K\alpha (1+\alpha \sigma q)^{-\frac{\sigma+1}{\sigma}}}
= \frac{1+\alpha \sigma q}{\alpha}$.

Second, to establish \eqref{eq:prop4_subutility_hypergeo}, let $C=\frac{K \sigma }{\sigma-1} 
\frac{\Gamma( 1- \beta ) \Gamma (  \frac{1}{\sigma} ) }{ \Gamma ( - \beta+ \frac{1}{\sigma}   ) }
\bigl(  \frac{1}{\alpha \sigma} \bigr)^{1- \beta}+\widetilde C$ in \eqref{eq:prop1_general}. We then have
\begin{equation*}
u(q) =
  \textstyle \frac{K \sigma (1+\alpha \sigma q)^{\beta-\frac{1}{\sigma}} q^{1-\beta}}{\sigma-1}
 \bigl[1+  \frac{1- \beta \sigma}{\alpha \sigma q}  
 {}_2 F_{1} \bigl(1,  \beta;   1+\frac{1}{\sigma}    ; -\frac{1}{ \alpha \sigma q}   \bigr) \bigr]  
+ \widetilde C.
\end{equation*}
Letting $\beta$ go to 1 from below and relabeling $\widetilde C$ as $C$, we obtain \eqref{eq:prop4_subutility_hypergeo} as follows:
\begin{equation*}
\lim_{\beta \to 1^-} u(q) = \textstyle \frac{K \sigma}{\sigma-1} \frac{(1+\alpha \sigma q)^{1-\frac{1}{\sigma}} } {q} \bigl[ q
-\frac{\sigma-1}{\alpha \sigma}
{}_2 F_{1} \bigl(1, 1 ; 1+\frac{1}{\sigma}; - \frac{1}{\alpha \sigma q}   \bigr)  \bigr] + C,
\end{equation*}
which requires the restriction $\lvert-\frac{1}{\alpha \sigma q}\rvert<1$. Differentiating this with respect to $q$, we obtain $u'(q) = K (1+\alpha \sigma q)^{1-\frac{1}{\sigma}} q^{-1}$ and $u''(q)= - K  (1+\alpha q) (1+\alpha \sigma q)^{-\frac{1}{\sigma}} q^{-2}$, which yield the constant revenue elasticity of marginal revenue (CREMR): $-\frac{{\rm d} \ln [ u'(q) + q u''(q) ]}{{\rm d} \ln [ u'(q) q] }= \frac{1}{\sigma-1}$, where the denominator
is the derivative of the log revenue and the numerator
is the derivative of the log of marginal revenue under the assumption that firms cannot affect consumers' marginal utility of income (see Mr\'{a}zov\'{a} et al., 2021).
\hfill $\square$

\subsection{Proof of Proposition~\ref{integral2}}
\label{app:proposition2}
\vspace{-.1cm}
First, assume that $\sigma>0$, which corresponds to the top-left cell in Table~\ref{tab:QR} since the utility function  \eqref{eq:prop4_subutility_hypergeo} requires $\alpha \sigma>0$ by Assumption~\ref{as2}. Then,
$1+\frac{1}{\sigma} > 1$ holds, so that
we can disregard the restriction $\frac{1}{\alpha \sigma} < q$ and extend the utility function to $q \le \frac{1}{\alpha \sigma}$
by analytic continuation
(9.1.6) in Lebedev (1965), where $\vert {\rm arg} (1-z) \rvert = \vert {\rm arg} (1 + \frac{1}{\alpha \sigma q}) \rvert =0$.
Thus, we can use the integral representation 
$\textstyle {}_2 F_1 \bigl(1, 1; 1+\frac{1}{\sigma}; -\frac{1}{\alpha \sigma q} \bigr) = \frac{\Gamma (1+\frac{1}{\sigma})}{\Gamma(1) \Gamma(\frac{1}{\sigma})} \int_0^1
(1-t)^{\frac{1}{\sigma}-1} \bigl(1 + \frac{t}{\alpha \sigma q} \bigr)^{-1} {\rm d}t$ for all $q$ satisfying the quantity restrictions in the
top-left cell in Table~\ref{tab:QR}.\footnote{If $\alpha \sigma<0$, we cannot use the integral representation even when $1 +\alpha \sigma q>0$ because the integrand is not continuous at $t =- \alpha \sigma q\in (0,1)$ and not bounded (see Lebedev, 1965, Section 9.1). We will overcome this limitation in Corollary~\ref{corollaryMNP}.}
Hence,
noting that
$\frac{\Gamma (1+\frac{1}{\sigma})}{\Gamma(\frac{1}{\sigma})}=\frac{1}{\sigma}$ and $\Gamma(1)=1$, we can rewrite \eqref{eq:prop4_subutility_hypergeo} for all $q \in [0,\infty)$
as
\eqref{eq:utility_special_continuation}.

Second, assume that $\sigma<0$, which corresponds to the top-right cell in Table~\ref{tab:QR} since the utility function  \eqref{eq:prop4_subutility_hypergeo} requires $\alpha \sigma>0$ by Assumption~\ref{as2}.
Then, using
the following recurrence relationship:\footnote{See \url{https://dlmf.nist.gov/15.5.16} (NIST Digital Library of Mathematical Functions).}
\begin{eqnarray*}
&&\textstyle (1+\frac{1}{\sigma} )
(1+\frac{1}{\alpha\sigma q} ) \,
{}_2 F_1(1,1;1+\frac{1}{\sigma}; - \frac{1}{\alpha\sigma q})
- (1+\frac{1}{\sigma} ) \,
{}_2 F_1(0,1;1+\frac{1}{\sigma}; - \frac{1}{\alpha\sigma q})  \\
& &  \textstyle
- \frac{1}{\alpha\sigma^2 q} \, 
{}_2 F_1(1,1;2+\frac{1}{\sigma}; - \frac{1}{\alpha\sigma q}) = 0
\end{eqnarray*}
and noting that ${}_2 F_1(0,1;1+\frac{1}{\sigma}; - \frac{1}{\alpha\sigma q})=1$, we have
\begin{eqnarray*}
\textstyle {}_2 F_1(1,1;1+\frac{1}{\sigma}; - \frac{1}{\alpha\sigma q}) &=&
\textstyle \frac{1}{(1+\frac{1}{\sigma}) (1+\frac{1}{\alpha\sigma q} )}
[(1+\frac{1}{\sigma} )+\frac{1}{\alpha\sigma^2 q} \, 
{}_2 F_1(1,1;2+\frac{1}{\sigma}; - \frac{1}{\alpha\sigma q})] \\
&=&
\textstyle
\frac{\alpha\sigma q}{1 + \alpha\sigma q}
[1+  \frac{\sigma}{\sigma+1} \frac{1}{\alpha\sigma^2 q} \, 
{}_2 F_1(1,1;2+\frac{1}{\sigma}; - \frac{1}{\alpha\sigma q})].
\end{eqnarray*}
Thus, we can rewrite \eqref{eq:prop4_subutility_hypergeo} as follows:
\begin{eqnarray*}
\lim_{\beta \to 1^-} u(q) &=& \textstyle \frac{K \sigma}{\sigma-1} \frac{(1+\alpha \sigma q)^{1-\frac{1}{\sigma}} } {q} \bigl[ q
-\frac{\sigma-1}{\alpha \sigma}
{}_2 F_{1} \bigl(1, 1 ; 1+\frac{1}{\sigma}; - \frac{1}{\alpha \sigma q}   \bigr)  \bigr] + C \\
&=& \textstyle \frac{K \sigma}{\sigma-1} \frac{(1+\alpha \sigma q)^{1-\frac{1}{\sigma}} } {q} \big\{ q
-\frac{\sigma-1}{\alpha \sigma}
\frac{\alpha\sigma q}{1 + \alpha\sigma q}
[1+  \frac{\sigma}{\sigma+1} \frac{1}{\alpha\sigma^2 q} \, 
{}_2 F_1(1,1;2+\frac{1}{\sigma}; - \frac{1}{\alpha\sigma q})]   \bigr\} + C \\
&=& \textstyle \frac{K \sigma}{\sigma-1} \frac{(1+\alpha \sigma q)^{-\frac{1}{\sigma}} } {q} \big\{ (1+\alpha \sigma q) q
- (\sigma-1) q 
[1+  \frac{\sigma}{\sigma+1} \frac{1}{\alpha\sigma^2 q} \, 
{}_2 F_1(1,1;2+\frac{1}{\sigma}; - \frac{1}{\alpha\sigma q})]   \bigr\} + C \\
&=& \textstyle \frac{K \sigma}{\sigma-1} \frac{(1+\alpha \sigma q)^{-\frac{1}{\sigma}} } {q} \big[ (1+\alpha \sigma q) q
- (\sigma-1) q   -   \frac{1}{\sigma+1} \frac{\sigma-1}{\alpha\sigma} \, 
{}_2 F_1(1,1;2+\frac{1}{\sigma}; - \frac{1}{\alpha\sigma q})   \bigr] + C,
\end{eqnarray*}
where ${}_2 F_1(1,1;2+\frac{1}{\sigma}; - \frac{1}{\alpha\sigma q})$ is defined for $\lvert-\frac{1}{\alpha \sigma q}\rvert<1$.
We can extend it to $q \le \frac{1}{\alpha \sigma}$
by analytic continuation
(9.1.6) in Lebedev (1965) since $2+\frac{1}{\sigma} > 1$ holds by Assumption~\ref{as2}.
Thus, we can use the integral representation 
$\textstyle {}_2 F_1 \bigl(1, 1; 2+\frac{1}{\sigma}; -\frac{1}{\alpha \sigma q} \bigr) = \frac{\Gamma (2+\frac{1}{\sigma})}{\Gamma(1) \Gamma(1+\frac{1}{\sigma})} \int_0^1 (1-t)^{\frac{1}{\sigma}} \bigl(1 + \frac{t}{\alpha \sigma q} \bigr)^{-1} {\rm d}t = \frac{\sigma+1}{\sigma} \int_0^1 (1-t)^{\frac{1}{\sigma}} \bigl(1 + \frac{t}{\alpha \sigma q} \bigr)^{-1} {\rm d}t$ for all $q$ satisfying the quantity restrictions in the top-right cell in Table~\ref{tab:QR}. We thus obtain
\begin{eqnarray*}
\lim_{\beta \to 1^-} u(q) &=&
 \textstyle \frac{K \sigma}{\sigma-1} \frac{(1+\alpha \sigma q)^{-\frac{1}{\sigma}} } {q} \big[ (1+\alpha \sigma q) q
- (\sigma-1) q   -   \frac{1}{\sigma+1} \frac{\sigma-1}{\alpha\sigma} \, 
{}_2 F_1(1,1;2+\frac{1}{\sigma}; - \frac{1}{\alpha\sigma q})   \bigr] + C \\
&=&
 \textstyle \frac{K \sigma}{\sigma-1} \frac{(1+\alpha \sigma q)^{-\frac{1}{\sigma}} } {q} \big[ (1+\alpha \sigma q) q
- (\sigma-1) q   -   \frac{1}{\sigma+1} \frac{\sigma-1}{\alpha\sigma}
\frac{\sigma+1}{\sigma} \int_0^1 (1-t)^{\frac{1}{\sigma}} \bigl(1 + \frac{t}{\alpha \sigma q} \bigr)^{-1} {\rm d}t   \bigr] + C \\
&=&
 \textstyle \frac{K \sigma}{\sigma-1} \frac{(1+\alpha \sigma q)^{-\frac{1}{\sigma}} } {q} \big[ (1+\alpha \sigma q) q
- (\sigma-1) q   -    \frac{\sigma-1}{\alpha \sigma} \frac{1}{\sigma} \int_0^1 (1-t)^{\frac{1}{\sigma}} \bigl(1 + \frac{t}{\alpha \sigma q} \bigr)^{-1} {\rm d}t   \bigr] + C.
\end{eqnarray*}
Hence, we can rewrite \eqref{eq:prop4_subutility_hypergeo} for all $q$ satisfying the top-right cell in Table~\ref{tab:QR}  as \eqref{eq:Lebedev}.
\hfill$\square$

\subsection{Proof of Corollary~\ref{coroll_NEW}}
\label{app:corollary1}
\vspace{-.1cm}
First, assume that $\alpha =0$. Letting $\alpha =0$ in \eqref{eq:prop4_subutility_general_case} and noting ${}_2 F_{1} \left(1- \frac{1}{\sigma}, 1; 2- \beta ; 0 \right)=1$, we obtain $u(q) = \frac{K q^{1-\beta}}{1-\beta}+C$ for $\beta \in  (0,1)$. Similarly, letting $\alpha =0$ and $\beta = 1$ in \eqref{eq:LFRRA_general}, we obtain $- \frac{q u''(q)}{u'(q)}=1$, which implies $u(q)= K \ln q  + C$ for $\beta=1$.

\bigbreak
\noindent
Assume now that $\alpha \neq 0$. The proofs for Cases A to C are given as follows.
\vspace{-.3cm}

\paragraph{Case A. $\beta\in(0,1)$.} There are three subcases.

\vspace{-.3cm}

\paragraph{$\sigma = 0$ (Incomplete gamma).} In this case, equation \eqref{eq:LFRRA_general}
reduces to $-\frac{qu''(q)}{u'(q)} =\alpha q+\beta$, which can be rewritten as $q u''(q) + (\alpha q+\beta) u'(q)=0$. Let $z= -\alpha q$ and $v(z)= u(q)$, so that $u'(q)= -\alpha v'(z)$ and $u''(q)= (-\alpha)^2 v''(z)$. Thus, we have
$- \frac{z}{\alpha}  (-\alpha)^2 v''(z) + (-z+\beta) (-\alpha) v'(z) =0$. This implies
\begin{equation*}
z  v''(z) + (\beta-z)  v'(z) =0,
\end{equation*}
which is the confluent hypergeometric differential equation $z v''(z) + (c -z)  v'(z)  - a v(z)=0$ with coefficients $a$ and $c$ satisfying $a=0$ and $c=\beta$.
Letting $c_1$ and $c_2$ denote arbitrary coefficients,
the solution to this differential equation is
given by $v(z) = c_1 \times {}_1 F_1 (a;c;z)+c_2 \times z^{1-c} {}_1 F_1 (1+a-c; 2-c; z)$, where ${}_1 F_1 (a; c; z)  = \sum_{n=0}^\infty \frac{(a)_n}{(c)_n} \frac{z^n}{n!}$,
is the confluent hypergeometric function (see Seaborn, 1991).
Noting that ${}_1 F_1 (0; \beta; z)=1$, we have
\begin{equation*}
u (q) = c_1 + c_2  (-\alpha q)^{1-\beta} {}_1 F_1 (1-\beta; 2-\beta; -\alpha q).
\end{equation*}
Denote the lower incomplete gamma function by $\gamma(1-\beta, \alpha q )$.
We can then apply the relationship $\gamma(1-\beta, \alpha q )=(1-\beta)^{-1} (\alpha q)^{1-\beta}  {}_1 F_1 (1-\beta; 2-\beta; -\alpha q)$ to obtain\footnote{See \url{https://dlmf.nist.gov/8.5.1} (NIST Digital Library of Mathematical Functions).}
\begin{eqnarray*}
u (q) &=&  c_1 + c_2  (-\alpha q)^{1-\beta} (1-\beta) (\alpha q)^{-(1-\beta)} \gamma(1-\beta, \alpha q ) \\
&=&  c_1 + c_2  (-\alpha)^{1-\beta} (1-\beta) (\alpha)^{-(1-\beta)} \gamma(1-\beta, \alpha q ) 
= \textstyle  \frac{K}{\alpha^{1-\beta}} \gamma(1-\beta, \alpha q ) +C,
\end{eqnarray*}
where
$C= c_1$ and $K= c_2  (-\alpha)^{1-\beta}  (1-\beta)$.\footnote{Even when $\alpha>0$, we can choose a complex constant $c_2$ such that $K= c_2  (-\alpha)^{1-\beta}  (1-\beta)$ is real valued.}
Hence, we obtain the utility function for $\beta\in(0,1)$ and $\sigma = 0$.

It can be verified that $u$ is a real-valued function for any $\alpha \neq 0$ because $\frac{K}{\alpha^{1-\beta}} \gamma(1-\beta, \alpha q ) = \frac{K}{\alpha^{1-\beta}} \int_0^{\alpha q} t^{-\beta} {\rm e}^{-t}  {\rm d}t =  \frac{K}{\alpha\alpha^{-\beta}} \int_0^{\alpha q} t^{-\beta} {\rm e}^{-t}  {\rm d}t  =\frac{K}{\alpha} \int_0^{\alpha q} \frac{t^{-\beta}}{\alpha^{-\beta}} {\rm e}^{-t}  {\rm d}t = \frac{K}{\alpha}  \int_0^{\alpha q} (\frac{t}{\alpha})^{-\beta} {\rm e}^{-t} {\rm d}t$, where the first, second,
third, and last equalities hold by the integral representation of the lower incomplete gamma function $\gamma(1-\beta, \alpha q )=\int_0^{\alpha q} t^{-\beta} {\rm e}^{-t} {\rm d}t$, by Pennisi (1963, Theorem 4.4.4), by Berg (2012, equation (3) in Section 2.1), and by Pennisi (1963, Theorem 4.4.5, together with Theorem 4.4.2), respectively.
Furthermore, the integral converges since $\frac{K}{\alpha} \int_0^{\alpha q} (\frac{t}{\alpha})^{-\beta}  {\rm e}^{-t} {\rm d}t
\le \frac{K}{\alpha} {\rm e}^0 \int_0^{\alpha q}(\frac{t}{\alpha})^{-\beta}  {\rm d}t = \frac{K q^{1-\beta}}{1-\beta}$ when $\alpha>0$, and $\frac{K}{\alpha} \int_0^{\alpha q} (\frac{t}{\alpha})^{-\beta}  {\rm e}^{-t} {\rm d}t
= -\frac{K}{\alpha} \int_{\alpha q}^0 (\frac{t}{\alpha})^{-\beta}  {\rm e}^{-t} {\rm d}t 
\le \linebreak
-\frac{K}{\alpha} {\rm e}^{-\alpha q} \int_{\alpha q}^0 (\frac{t}{\alpha})^{-\beta}  {\rm d}t =  {\rm e}^{-\alpha q} \frac{K q^{1-\beta} }{1-\beta}$ when $\alpha<0$.

\vspace{-.3cm}

\paragraph{$\sigma = 1$  (Incomplete beta).}
Letting $\sigma = 1$ in \eqref{eq:u2}, we have
\begin{equation*}
u(q) = \textstyle c_1 + c_2  (-\alpha q)^{1-\beta} (1+\alpha q)^{\beta} {}_2 F_1\bigl(1, 1; 2-\beta; -\alpha q\bigr).
\end{equation*}
By the hypergeometric representation of the incomplete beta function, we have $B(-\alpha q, 1-\beta, \beta) =  \frac{(-\alpha q)^{1-\beta} (1+\alpha q)^\beta }{1-\beta} {}_2 F_1\bigl(1, 1; 2-\beta; -\alpha q\bigr)$, so that\footnote{See \url{https://dlmf.nist.gov/8.17.8} (NIST Digital Library of Mathematical Functions).}
\begin{eqnarray*}
u(q) &=& \textstyle c_1 + c_2  (-\alpha q)^{1-\beta} (1+\alpha q)^{\beta}  \frac{1-\beta}{(-\alpha q)^{1-\beta} (1+\alpha q)^\beta} B(-\alpha q, 1-\beta, \beta) \\
&=& \textstyle c_1 + c_2  (-\alpha q)^{1-\beta}   \frac{1-\beta}{(-\alpha q)^{1-\beta} } B(-\alpha q, 1-\beta, \beta) \\
&=& \textstyle c_1 + c_2  (-\alpha)^{1-\beta}   \frac{1-\beta}{(-\alpha)^{1-\beta} } B(-\alpha q, 1-\beta, \beta) \\
&=& \textstyle  \frac{K}{(-\alpha)^{1-\beta} } B(-\alpha q, 1-\beta, \beta) +C ,
\end{eqnarray*}
where we set $C=c_1$ and  $K = c_2  (-\alpha)^{1-\beta} (1-\beta)$. Hence, we obtain the utility function for $\beta\in(0,1)$ and $\sigma = 1$.

It can be verified that $u$ is a real-valued function for any $\alpha \neq 0$ because $\frac{K}{(-\alpha)^{1-\beta} } B(-\alpha q, 1-\beta, \beta) = \frac{K}{(-\alpha)^{1-\beta} } \int_0^{-\alpha q} t^{-\beta} (1-t)^{\beta-1} {\rm d}t = \frac{K}{-\alpha(-\alpha)^{-\beta} } \int_0^{-\alpha q} t^{-\beta} (1-t)^{\beta-1} {\rm d}t = - \frac{K}{\alpha} \int_0^{-\alpha q} \frac{t^{-\beta}}{(-\alpha)^{-\beta}} (1-t)^{\beta-1} {\rm d}t =- \frac{K}{\alpha}  \int_0^{-\alpha q} (-\frac{t}{\alpha})^{-\beta} (1-t)^{\beta-1} {\rm d}t$, where the first, second,
third, and last equalities hold by the integral representation of the incomplete beta function $B(-\alpha q, 1-\beta, \beta)= \int_0^{-\alpha q} t^{-\beta} (1-t)^{\beta-1} {\rm d}t$, by Pennisi (1963, Theorem 4.4.4), by Berg (2012, equation (3) in Section 2.1), and by Pennisi (1963, Theorem 4.4.5, together with Theorem 4.4.2), respectively.
Furthermore, the integral converges since  $- \frac{K}{\alpha}  \int_0^{-\alpha q} (-\frac{t}{\alpha})^{-\beta} (1-t)^{\beta-1} {\rm d}t = \frac{K}{\alpha}  \int_{-\alpha q}^0  (-\frac{t}{\alpha})^{-\beta} (1-t)^{\beta-1} {\rm d}t \le   \frac{K}{\alpha} (1-0 )^{\beta-1}  \int_{-\alpha q}^0 (-\frac{t}{\alpha})^{-\beta}  {\rm d}t
= \frac{K q^{1-\beta}}{1-\beta}$
when $\alpha>0$, and
$- \frac{K}{\alpha}  \int_0^{-\alpha q} (-\frac{t}{\alpha})^{-\beta} (1-t)^{\beta-1} {\rm d}t
\le  - \frac{K}{\alpha} (1+\alpha q)^{\beta-1}  \int_0^{-\alpha q} (-\frac{t}{\alpha})^{-\beta}  {\rm d}t
=- \frac{K}{\alpha} (1+\alpha q)^{\beta-1}  \bigl( - \frac{\alpha q^{1-\beta}}{1-\beta} \bigr)
=(1+\alpha q)^{\beta-1} \frac{K q^{1-\beta}}{1-\beta} 
$ when $\alpha<0$, which is finite for any $q$ satisfying the quantity restriction in the bottom-right cell of Table~\ref{tab:QR} with $\alpha<0$, $\beta \in (0,1)$, and $\sigma=1$, i.e., for any $q \in [0, -\frac{\beta}{\alpha})$.

\vspace{-.3cm}

\paragraph{$\sigma = \frac{1}{\beta}$ (CRRA).} Letting $\sigma = \frac{1}{\beta}$ in \eqref{eq:prop4_subutility_general_case}, we obtain the result.

\paragraph{Case B. $\beta \to 0^+$.}
Recall that when $\beta$ goes to $0$ from above, the utility function is given by \eqref{eq:prop4_subutility_HARA} and requires $\alpha>0$. There are three subcases.

\vspace{-.3cm}

\paragraph{$\sigma = 0$ (CARA).}
Noting that $ \lim_{\sigma \to 0}(1+\alpha \sigma q)^\frac{\sigma-1}{\sigma} ={\rm e}^{-\alpha q}$, we obtain
$\lim_{\sigma \to 0} u(q) = \linebreak
\lim_{\sigma \to 0} \frac{K}{\alpha (\sigma-1)} \bigl[ (1+\alpha \sigma q)^\frac{\sigma-1}{\sigma} -1  \bigr] +C =
-\frac{K}{\alpha} ({\rm e}^{-\alpha q}-1)+C = \frac{K}{\alpha} (1- {\rm e}^{-\alpha q})+C$.

\vspace{-.3cm}

\paragraph{$\sigma = -1$ (quadratic).}
Letting $\sigma = -1$ in \eqref{eq:prop4_subutility_HARA}, we obtain
$u(q)=-\frac{K}{2\alpha} [ (1 - \alpha q)^2 -1  ] +C
=-\frac{K}{2\alpha} [ 1 - 2 \alpha q + ( \alpha q)^2 -1 ] +C 
=-\frac{K}{2\alpha} [  ( \alpha q)^2 - 2 \alpha q  ] +C
=-\frac{K}{2} (  \alpha q^2 - 2 q  ) +C$.

\vspace{-.3cm}

\paragraph{$\sigma = 1$ (translated log).}
Noting that $ \lim_{\sigma \to 1}  \frac{(1+\alpha \sigma q)^\frac{\sigma-1}{\sigma}  -1}{\sigma-1}=\ln (1+ \alpha q)$, we obtain
$\lim_{\sigma \to 1} u(q) = \lim_{\sigma \to 1} \frac{K}{\alpha (\sigma-1)} \bigl[ (1+\alpha \sigma q)^\frac{\sigma-1}{\sigma} -1  \bigr] +C =
\frac{K}{\alpha} \ln (1+ \alpha q)+C$.

\paragraph{Case C. $\beta \to 1^-$.}
Recall that when $\beta$ goes to $1$ from below, the utility function is given by \eqref{eq:prop4_subutility_hypergeo}. There are two subcases.

\vspace{-.3cm}

\paragraph{$\sigma =  0$ (exponential integral).}
In this case, equation \eqref{eq:LFRRA_general} reduces to $-\frac{qu''(q)}{u'(q)} =\alpha q+1$, which can be rewritten as $q u''(q) + (\alpha q+1) u'(q)=0$. Let $z= -\alpha q$ and $v(z)= u(q)$, so that $u'(q)= -\alpha v'(z)$ and $u''(q)= (-\alpha)^2 v''(z)$. Thus, we have $- \frac{z}{\alpha}  (-\alpha)^2 v''(z) + (-z+1) (-\alpha) v'(z) =0$. This implies
\begin{equation*}
z  v''(z) + (1-z)  v'(z) =0,
\end{equation*}
which is the confluent hypergeometric differential equation $z  v''(z) + (c-z)  v'(z) - a v(z)=0$ with coefficients $a$ and $c$ satisfying $a = 0$ and $c =1$. This can be rewritten as $\frac{z  v''(z) + v'(z)}{z v'(z)} =1$, so that $\frac{{\rm d}\ln(z v'(z))}{{\rm d}z} = 1$. The differential equation yields $\ln (z v'(z) ) =  z + c_1$, and thus $v'(z) = {\rm e}^{c_1}  \frac{{\rm e}^{z}}{z}$, where $c_1$ is a constant of integration.
Integrating the foregoing equation yields
\linebreak
\vspace{-.5cm}
\begin{equation*}
\textstyle \int_{c_2}^{z} v'(t){\rm d}t  = {\rm e}^{c_1} \int_{c_2}^{z} \frac{{\rm e}^t}{t}{\rm d}t,
\end{equation*}
where $c_2$ is a constant.
Let $s = -t$ and $w(s)=v(t)$. We then have ${\rm d}t = - {\rm d}s$ and $v'(t)=-w'(s)$, so that
\begin{equation*}
\textstyle
\int_{c_2}^{z} v'(t){\rm d}t
= \int_{-c_2}^{-z} w'(s){\rm d}s 
= - {\rm e}^{c_1} \int_{-c_2}^{-z} \frac{{\rm e}^{-s}}{-s}{\rm d}s
= {\rm e}^{c_1} \int_{-z}^{-c_2} \frac{{\rm e}^{-s}}{-s}{\rm d}s
=  {\rm e}^{c_1} \left( \int_{-z}^\infty \frac{{\rm e}^{-s}}{-s}{\rm d}s-  \int_{-c_2}^\infty \frac{{\rm e}^{-s}}{-s}{\rm d}s  \right).
\end{equation*}
Let ${\rm Ei} (z) = - \int_{-z}^\infty \frac{{\rm e}^{-s}}{s}{\rm d}s$ denote the exponential integral. We then have $w(-z)- w(-c_2)  = v(z)-v(c_2)  =  {\rm e}^{c_1} [{\rm Ei}(z)- {\rm Ei}(c_2) ]$, which can be rewritten as $v(z)  = {\rm e}^{c_1} {\rm Ei}(z)+   v(c_2) - {\rm e}^{c_1} {\rm Ei}(c_2)$. Setting $C =  v(c_2) - {\rm e}^{c_1} {\rm Ei}(c_2) $ and $K = {\rm e}^{c_1}$, and recalling that $v(z) = u(q)$ and $z = - \alpha q$, we obtain $u(q)  = K\,  {\rm Ei}(-\alpha q) + C.$

\vspace{-.3cm}

\paragraph{$\sigma = 1$ (log).}

In this case, equation \eqref{eq:LFRRA_general}
reduces to $-\frac{qu''(q)}{u'(q)} =1$, which can be rewritten as $ \frac{u''(q)}{u'(q)}=-\frac{1}{q}$. Integrating both sides yields $\ln u'(q) = c_1 - \ln q$, so that $u'(q) = {\rm e}^{c_1}\frac{1}{q}$, where $c_1$ is a constant of integration. Thus, $u(q) = C +{\rm e}^{c_1}\ln q$, where $C$ is a constant of integration. Hence, letting $K = {\rm e}^{c_1}$, we obtain
$u(q) =  K\ln q+C$.
\hfill$\square$

\subsection{Proof of Corollary~\ref{corollaryMNP}}
\label{app:corollary2}
\vspace{-.1cm}
Let $\alpha = - \frac{1}{\gamma \sigma }$, $K=- \widetilde \beta \gamma(-\frac{1}{\gamma})^\frac{1}{\sigma}$, and $C = \kappa$ in \eqref{eq:prop4_subutility_hypergeo}. We then have
\begin{eqnarray*}
\lim_{\beta \to 1^-} u(q) &=& \textstyle \frac{K \sigma}{\sigma-1} \frac{(1+\alpha \sigma q)^{1-\frac{1}{\sigma}} } {q} \bigl[ q
-\frac{\sigma-1}{\alpha \sigma}
{}_2 F_{1} \bigl(1, 1 ; 1+\frac{1}{\sigma}; - \frac{1}{\alpha \sigma q}   \bigr)  \bigr] +  C \\
&=& \textstyle \frac{- \widetilde \beta \gamma(-\frac{1}{\gamma})^\frac{1}{\sigma} \sigma}{\sigma-1} \frac{(1- \frac{q}{\gamma})^{1-\frac{1}{\sigma}} } {q} \bigl[ q
+\gamma(\sigma-1)
{}_2 F_{1} \bigl(1, 1 ; 1+\frac{1}{\sigma}; \frac{\gamma}{q}   \bigr)  \bigr] + \kappa \\
&=& \textstyle \frac{- \widetilde \beta \gamma(-\frac{1}{\gamma})^\frac{1}{\sigma} \sigma}{\sigma-1} \frac{[(-\frac{1}{\gamma})(q-\gamma)]^{1-\frac{1}{\sigma}} } {q} \bigl[ q
+\gamma(\sigma-1)
{}_2 F_{1} \bigl(1, 1 ; 1+\frac{1}{\sigma}; \frac{\gamma}{q}   \bigr)  \bigr] + \kappa \\
&=& \textstyle \frac{- \widetilde \beta \gamma \sigma}{\sigma-1} \frac{(-\frac{1}{\gamma})(q-\gamma)^{1-\frac{1}{\sigma}} } {q} \bigl[ q
+\gamma(\sigma-1)
{}_2 F_{1} \bigl(1, 1 ; 1+\frac{1}{\sigma}; \frac{\gamma}{q}   \bigr)  \bigr] + \kappa \\
&=& \textstyle \frac{\widetilde \beta \sigma}{\sigma-1} \frac{(q-\gamma)^{1-\frac{1}{\sigma}} } {q} \bigl[ q
+\gamma(\sigma-1)
{}_2 F_{1} \bigl(1, 1 ; 1+\frac{1}{\sigma}; \frac{\gamma}{q}   \bigr)  \bigr] + \kappa.
\end{eqnarray*}
We thus obtain \eqref{eq:LFRRA_utility_C2}.

Equation \eqref{eq:LFRRA_utility_C2_int} can be obtained as follows.
Recall that ${}_2 F_{1} \bigl(1, 1 ; 1+\frac{1}{\sigma}; \frac{\gamma}{q}   \bigr)$ in \eqref{eq:LFRRA_utility_C2} is defined for $\lvert \frac{\gamma}{q}  \rvert <1$. Thus, there are two cases depending on the sign of $\gamma$.\footnote{Strictly speaking, there is another case with $\gamma=0$. Since ${}_2 F_{1} \bigl(1, 1 ; 1+\frac{1}{\sigma}; \frac{\gamma}{q}   \bigr)=1$ for $\gamma=0$,
we do not have to replace ${}_2 F_{1} \bigl(1, 1 ; 1+\frac{1}{\sigma}; \frac{\gamma}{q}   \bigr)$ in \eqref{eq:LFRRA_utility_C2} with the integral in \eqref{eq:LFRRA_utility_C2_int}. However, we can let $\gamma \to 0$ in \eqref{eq:LFRRA_utility_C2_int} to obtain $u(q) = \kappa + \frac{\widetilde \beta \sigma}{\sigma-1} q^\frac{\sigma-1}{\sigma}$ for all $q \in [0,\infty)$, which is
equivalent to \eqref{eq:LFRRA_utility_C2} evaluated at $\gamma=0$.}
First, when $\gamma >0$, $\lvert \frac{\gamma}{q} \rvert <1$ implies $q > \gamma$, which is not restrictive because $u'(q)= \widetilde \beta (q-\gamma)^\frac{\sigma-1}{\sigma} q^{-1}>0$ and $\widetilde \beta>0$ imply $q>\gamma$.
Second, when $\gamma< 0$, $\lvert \frac{\gamma}{q} \rvert <1$ implies $q> -\gamma$, which appears to be restrictive since
\eqref{eq:LFRRA_utility_C2} is not defined for $q \in  (0,  -\gamma]$.
However, since $1+\frac{1}{\sigma}>1>0$ holds in Corollary~\ref{corollaryMNP}, we can extend ${}_2 F_{1} \bigl(1, 1 ; 1+\frac{1}{\sigma}; \frac{\gamma}{q}   \bigr)$ in \eqref{eq:LFRRA_utility_C2} to $q \in  (0,  -\gamma]$
by analytic continuation (see, e.g.,~(9.1.6) in Lebedev, 1965, where $\lvert {\rm arg} (1-z)\rvert =\lvert {\rm arg} (1 -\frac{\gamma}{q}) \rvert =0$ since $1 -\frac{\gamma}{q} > 0$).
Thus, in both cases, we can use the integral representation ${}_2 F_{1} \bigl(1, 1 ; 1+\frac{1}{\sigma}; \frac{\gamma}{q}   \bigr) = \frac{\Gamma(1+\frac{1}{\sigma})}{\Gamma(1)\Gamma(1+\frac{1}{\sigma}-1)}  \int_0^1 (1-t)^{\frac{1}{\sigma}-1} \bigl(1-\frac{t \gamma}{q} \bigr)^{-1}{\rm d}t$, where $\Gamma(\cdot)$ is the gamma function. Using the recurrence relation $\Gamma(1+\frac{1}{\sigma})=\frac{1}{\sigma} \Gamma(\frac{1}{\sigma})$ and $\Gamma (1)=1$, the foregoing expression reduces to ${}_2 F_{1} \bigl(1, 1 ; 1+\frac{1}{\sigma}; \frac{\gamma}{q}   \bigr)  = \frac{1}{\sigma} \int_0^1 (1-t)^{\frac{1}{\sigma}-1} \bigl(1-\frac{t \gamma}{q} \bigr)^{-1}{\rm d}t$. Hence,
we can rewrite \eqref{eq:LFRRA_utility_C2} as \eqref{eq:LFRRA_utility_C2_int}.
\hfill $\square$

\newpage

\numberwithin{equation}{section}
\renewcommand{\thesection}{Supplemental Appendix \Alph{section}}
\renewcommand\thefigure{\Alph{section}.\arabic{figure}}
\renewcommand\thetable{\Alph{section}.\arabic{table}}
\renewcommand\theequation{\Alph{section}.\arabic{equation}}

\section{Proofs}
\label{app:proofs}

\vspace{-.25cm}

\setcounter{table}{0}

\subsection{Proof of Proposition~\ref{prop:markup_bounds}}
\label{app:proofs3}

In the main text, we have already analyzed the case where either $\alpha =0$ or $\beta = \frac{1}{\sigma}$ holds. Thus, in this proof, we focus on the case where $\alpha \neq 0$ and $\beta \neq \frac{1}{\sigma}$.

To simplify the exposition, we split the discussion depending on the sign of $\alpha$ (Cases 1 and 2), which together with $q \ge 0$  determines whether $1-\beta -\mu$ and $1 -  \sigma (1-\mu)$ have the same sign.\footnote{For simplicity, we say that $1-\beta -\mu$ and $1 -  \sigma (1-\mu)$ have the same sign if $1-\beta -\mu \ge 0$ and  $1 -  \sigma (1-\mu)>0$ or $1-\beta -\mu \le 0$ and  $1 -  \sigma (1-\mu)<0$. We also say that $1-\beta -\mu$ and $1 -  \sigma (1-\mu)$ have the opposite sign if $1-\beta -\mu \ge 0$ and  $1 -  \sigma (1-\mu)<0$ or $1-\beta -\mu \le 0$ and  $1 -  \sigma (1-\mu)>0$.}
Each of these two cases is further divided into two depending on the sign of $1-\beta\sigma$ (Cases a and b), which must have the same sign as $1 -  \sigma (1-\mu)$ because $u' > 0$. We will make use of those four different cases throughout the paper.

In Step 1, we derive the ranges of markups and quantities that satisfy the four inequality conditions, $q \ge 0$, $u'> 0$, $u'' < 0$, and $0<\mu \le 1$. In Step 2, we refine the ranges of markups and quantities by imposing the fifth inequality condition, the second-order condition \eqref{eq:SOC_LFRRA}.

\paragraph{Step 1.} We first derive the ranges of markups and quantities that satisfy the four inequality conditions, $q \ge 0$, $u' > 0$, $u'' < 0$, and $0<\mu \le 1$. Table~\ref{tab:tab1} summarizes these ranges.

\begin{table}
\caption{The ranges of markups and quantities satisfying $q \ge 0$, $u' > 0$, $u'' < 0$, and $0<\mu \le 1$.}
\label{tab:tab1}
\hskip -.15cm
\scalebox{0.85}{
\begin{tabular}{l|c|c|c|cc}
\multicolumn{5}{c}{Case 1: $\alpha >0$.} \\
\hline
 & \multicolumn{2}{c|}{Markups} & \multicolumn{2}{c}{Quantities} \\ \cline{2-5}
 & $\sigma \le 1$  & $\sigma > 1 $ & $\sigma \le 1$  & $\sigma > 1 $    \\ \hline
\multirow{2}{*}{Case 1a. $1-\beta \sigma>0$ and $\beta \neq 1$} &
\multirow{2}{*}{$\frac{1}{1-\beta} \le \frac{1}{\mu} < \infty$} &
\multirow{2}{*}{$\frac{1}{1-\beta} \le \frac{1}{\mu} < \frac{\sigma}{\sigma-1}$} &
\multirow{2}{*}{$0 \le q < -\frac{1-\beta}{\alpha (\sigma-1)}$$^\dagger$} &
\multirow{2}{*}{$0 \le q < \infty$} \\
&
&
&
&
\\ \hline
&
&
&
&
\\[-3mm]
\multirow{2}{*}{Case 1b.} \hskip .05cm $1-\beta \sigma<0$ and $\beta\in(0,1)$ &
\multirow{2}{*}{n.a.} &
$\frac{\sigma}{\sigma-1} < \frac{1}{\mu }  \le  \frac{1}{1-\beta}$ &
\multirow{2}{*}{n.a.} & $0 \le q < \infty$  \\[2mm]  
\hskip 1.73cm
$1-\beta \sigma<0$ and $\beta=1$
&
& $\frac{\sigma}{\sigma-1} < \frac{1}{\mu} < \infty$
&
& $0 < q < \infty$
\\[1.5mm] \hline 
\multicolumn{5}{c}{}   \\
 \multicolumn{5}{c}{Case 2: $\alpha <0$.} \\ \hline
 & \multicolumn{2}{c|}{Markups} & \multicolumn{2}{c}{Quantities} \\ \cline{2-5}
 & $\sigma \le 1$  & $\sigma > 1 $ & $\sigma \le 1$  & $\sigma > 1 $    \\ \hline 
 &
&
&
&
\\[-3mm]
\multirow{2}{*}{Case 2a.} \hskip .05cm $1-\beta \sigma>0$ and $\beta\in(0,1)$ &
$1 < \frac{1}{\mu }  \le  \frac{1}{1-\beta}$ &
$1 < \frac{1}{\mu }  \le  \frac{1}{1-\beta}$ &
$0 \le q  < -\frac{\beta}{\alpha} $ &
$0 \le q  < -\frac{\beta}{\alpha} $  \\[2mm]  
\hskip 1.73cm
$1-\beta \sigma>0$ and $\beta=1$
& $1 < \frac{1}{\mu } <\infty^{\ddagger}$
& n.a.
& $0 < q < {-\frac{1}{\alpha}}^{\ddagger}$
& n.a.
\\[1.5mm] \hline 
\multirow{2}{*}{Case 2b. $1-\beta \sigma<0$ and $\beta \in (0,1)$} &
\multirow{2}{*}{n.a.} &
\multirow{2}{*}{$\frac{1}{1-\beta}  \le  \frac{1}{\mu }  < \infty$} &
\multirow{2}{*}{n.a.} &
\multirow{2}{*}{$0  \le q < -\frac{1-\beta}{\alpha (\sigma-1)}$} \\
&
&
&
&
\\ \hline
\multicolumn{5}{p{18.8cm}}{\scriptsize {\it Notes}:
$^{\dagger}$ means that when $\sigma=1$, $0 \le q < -\frac{1-\beta}{\alpha (\sigma-1)}$ must be replaced with $0 \le q < \infty$.
$^{\ddagger}$ means that $\sigma=1$ must be excluded.
The case with $\alpha=0$ corresponds to the CES and requires that $\beta \in (0,1)$.
The case with $1-\beta \sigma=0$ also corresponds to the CES and requires that  $\sigma > 1$.
If $\alpha=0$ or $\beta =\frac{1}{\sigma}$, then the markup is given by $\frac{1}{\mu}  =\frac{1}{1-\beta} \in  (1, \infty)$ or $\frac{1}{\mu}  =\frac{\sigma}{\sigma-1} \in (1, \infty)$, respectively, and the range of quantities is given by $0\le q<\infty$.
Since these two cases are well known, we omit them from the table.}
\end{tabular}}
\end{table}

\vspace{-.25cm}

\paragraph{Case 1: $\alpha >0$.}  \hfill

\vspace{-.4cm}

\paragraph{Case 1a: $1-\beta -\mu \ge 0$, $1 -  \sigma (1-\mu)>0$, and $1 -\beta \sigma>0$.}
In this case, $\beta = 1$ is not admissible since $\mu>0$ must hold. There are three subcases depending on the value of $\sigma$.

\vspace{-.4cm}

\paragraph{Markups $\frac{1}{\mu}$.}
First, when $\sigma>0$, we have $\frac{1}{\sigma}  >  1-\mu$, so that $\mu   >  \frac{\sigma-1}{\sigma}$. Hence, we obtain $\max\{0, \frac{\sigma-1}{\sigma} \} < \mu   \le 1-\beta$. For this inequality to be satisfied, we require $\max\{0, \frac{\sigma-1}{\sigma} \}  < 1-\beta$. If $\frac{\sigma-1}{\sigma} \le 0$, i.e., if $0< \sigma \le 1$, it is always satisfied.
If $\frac{\sigma-1}{\sigma}>0$, i.e., if $\sigma > 1$, then it reduces to $\frac{\sigma-1}{\sigma} < 1-\beta$, which holds because $1 -\beta \sigma>0$. Thus, when $0<\sigma \le 1$, we have $ \frac{1}{1-\beta} \le \frac{1}{\mu} < \infty$ and when $\sigma>1$, we have $\frac{1}{1-\beta} \le \frac{1}{\mu} < \frac{\sigma}{\sigma-1}$.

Second, when $\sigma<0$, we have $\frac{1}{\sigma}  <  1-\mu$, so that $\mu   <  \frac{\sigma-1}{\sigma}$.
Since $\sigma<0$ and $1 -\beta \sigma>0$ hold, we have $1-\beta - \frac{\sigma-1}{\sigma}=
\frac{1}{\sigma}[(1-\beta)\sigma  - (\sigma-1)] =\frac{1}{\sigma} (1-\beta \sigma)<0$, so that  $0 < \mu  \le 1-\beta < \frac{\sigma-1}{\sigma}$.
Thus, we have $\frac{1}{1-\beta}  \le \frac{1}{\mu }  < \infty$.

Last, when $\sigma =0$, we have $0 < \mu \le 1-\beta$, so that $\frac{1}{1-\beta} \le \frac{1}{\mu}< \infty$.

To summarize, we have: (i) $\frac{1}{1-\beta} \le \frac{1}{\mu}< \infty$ when $\sigma \le 1$; and (ii) $ \frac{1}{1-\beta} \le \frac{1}{\mu }  < \frac{\sigma}{\sigma-1}$ when $\sigma>1$. The former corresponds to $0  < \mu \le  1-\beta$ and the latter to $\frac{\sigma-1}{\sigma} < \mu \le 1-\beta$.

\vspace{-.4cm}

\paragraph{Quantities $q$.}
Solving \eqref{eq:FOCprof} for $\mu$ yields $\mu = 1-\frac{\alpha q+\beta}{\alpha\sigma q + 1}$. Plugging it into the bounds for $\mu$ that we have obtained above, we can derive the relevant ranges of quantities as follows: (i) $0 < \mu \le  1-\beta$ implies $0 \le q < -\frac{1-\beta}{\alpha(\sigma-1)}$ when $\sigma < 1$ and $0\le q <\infty$ when $\sigma=1$;
and (ii) $\frac{\sigma-1}{\sigma} < \mu \le 1-\beta$ implies $0 \le q<\infty$ when $\sigma>1$.

\paragraph{Case 1b: $1-\beta -\mu \le 0$, $1 -  \sigma (1-\mu)<0$, and $ 1 -\beta \sigma<0$.}
In this case, $\beta=0$ is not admissible and $\sigma>1$ must hold. 

\vspace{-.4cm}

\paragraph{Markups $\frac{1}{\mu}$.}
We have $\frac{1}{\sigma}  <  1-\mu$, so that $\mu   <  \frac{\sigma-1}{\sigma}$. Hence, we obtain $1-\beta \le \mu   < \frac{\sigma-1}{\sigma}$ for $\beta \in (0,1)$ and $0 < \mu   < \frac{\sigma-1}{\sigma}$ for $\beta=1$ since $\mu>0$ must hold. For these inequalities to be satisfied, we require $1-\beta  < \frac{\sigma-1}{\sigma}$, which is always satisfied because $\sigma>1$ and $1 -\beta \sigma<0$ hold. Thus, when $\sigma>1$, we have $\frac{\sigma}{\sigma-1} < \frac{1}{\mu }  \le \frac{1}{1-\beta}$ for $\beta \in (0,1)$ and $\frac{\sigma}{\sigma-1} < \frac{1}{\mu }  < \infty$ for $\beta=1$.

\vspace{-.4cm}

\paragraph{Quantities $q$.}
Solving \eqref{eq:FOCprof} for $\mu$ yields  $\mu= 1-\frac{\alpha q+\beta}{\alpha\sigma q + 1}$. Plugging it into the bounds for $\mu$, $1-\beta \le \mu < \frac{\sigma-1}{\sigma}$ for $\beta \in (0,1)$ and $0 < \mu   < \frac{\sigma-1}{\sigma}$ for $\beta=1$, we obtain the relevant ranges of quantities as follows: $0 \le q<\infty$ for $\beta \in (0,1)$ and $0 < q<\infty$ for $\beta=1$, when $\sigma>1$.

\paragraph{Case 2: $\alpha <0$.} \hfill

\vspace{-.4cm}

\paragraph{Case 2a: $1-\beta -\mu \le 0$, $1 -  \sigma (1-\mu)>0$, and $ 1 -\beta \sigma>0$.}
In this case, we exclude $\beta=0$
because the condition, $u''<0$, i.e., $\alpha>0$ for $\beta=0$, is violated (see Assumption~\ref{as2}).
Similarly, the case with $\mu=1$ must be excluded for $\beta \in (0,1]$ because \eqref{eq:FOCprof} implies $q=-\frac{\beta}{\alpha}$, thus violating $u''<0$, i.e., $\alpha q+\beta>0$ for $\beta \in (0,1]$.
We also exclude the case with $\beta=1$ and $\sigma \ge 1$ from Case 2a as it violates $1-\beta \sigma>0$.

\vspace{-.4cm}

\paragraph{Markups $\frac{1}{\mu}$.}
First, when $\sigma>0$, we have $\frac{1}{\sigma}  >  1-\mu$, so that $\mu   >  \frac{\sigma-1}{\sigma}$.
Since $\sigma>0$ and $ 1 -\beta \sigma>0$ hold, we have $1-\beta - \frac{\sigma-1}{\sigma}=
\frac{1}{\sigma}[(1-\beta)\sigma - (\sigma-1)] =\frac{1}{\sigma} (1-\beta \sigma)>0$, so that  $\frac{\sigma-1}{\sigma} <1-\beta \le \mu < 1$ for $\beta \in (0,1)$ and $\frac{\sigma-1}{\sigma} < 0 < \mu < 1$ for $\beta=1$.
Thus, we have $1 < \frac{1}{\mu }  \le \frac{1}{1-\beta}$ for $\beta \in (0,1)$ and $1 < \frac{1}{\mu }  < \infty$ for $\beta=1$.
Second, when $\sigma<0$, we have $\frac{1}{\sigma}  <  1-\mu$, so that $\mu   <  \frac{\sigma-1}{\sigma}$, where $\frac{\sigma-1}{\sigma} >1$. Hence, we obtain $1-\beta \le  \mu <  1$ for $\beta \in (0,1)$ and $0 <  \mu <  1$ for $\beta=1$, which implies $1 < \frac{1}{\mu }  \le \frac{1}{1-\beta}$ for $\beta \in (0,1)$ and  $1 < \frac{1}{\mu }  <\infty$ for $\beta=1$.
Last, when $\sigma =0$, we have $1-\beta \le \mu < 1$ for $\beta \in (0,1)$ and $0 < \mu < 1$ for $\beta=1$, so that $1< \frac{1}{\mu} \le \frac{1}{1-\beta}$ for $\beta \in (0,1)$ and  $1 < \frac{1}{\mu }  <\infty$ for $\beta=1$.
To summarize,
we have $1 <  \frac{1}{\mu} \le \frac{1}{1-\beta}$ for $\beta \in (0,1)$ and  $1< \frac{1}{\mu }  <\infty$ for $\beta=1$ and $\sigma < 1$.

\vspace{-.4cm}

\paragraph{Quantities $q$.}
Solving \eqref{eq:FOCprof} for $\mu$ yields $\mu= 1-\frac{\alpha q+\beta}{\alpha\sigma q + 1}$. Plugging it into the bounds for $\mu$, $1-\beta  \le \mu <1$ for $\beta \in (0,1)$ and $0 < \mu < 1$ for $\beta=1$, we obtain the relevant ranges
of quantities
as follows: $0 \le q < -\frac{\beta}{\alpha}$ for $\beta \in (0,1)$ and $0 < q  < -\frac{1}{\alpha} $ for $\beta=1$ and $\sigma < 1$.

\paragraph{Case 2b: $1-\beta -\mu \ge 0$, $1 -  \sigma (1-\mu)<0$, and $ 1 -\beta \sigma<0$.}
In this case, $\beta=0$ and $\beta =1$ are not admissible, whereas $\sigma>1$ must hold.

\vspace{-.4cm}

\paragraph{Markups $\frac{1}{\mu}$.}
We have $\frac{1}{\sigma}  <  1-\mu$, so that $\mu   <  \frac{\sigma-1}{\sigma}$.
Since $\sigma>1$ and $ 1 -\beta \sigma<0$ hold, we have $1-\beta - \frac{\sigma-1}{\sigma}=
\frac{1}{\sigma}[(1-\beta)\sigma - (\sigma-1)] =\frac{1}{\sigma} (1-\beta \sigma)<0$, so that
$0  < \mu   \le 1-\beta <\frac{\sigma-1}{\sigma}$.
Thus, we have $\frac{1}{1-\beta}  \le \frac{1}{\mu} < \infty$.

\vspace{-.4cm}

\paragraph{Quantities $q$.}
Solving \eqref{eq:FOCprof} for $\mu$ yields $\mu= 1-\frac{\alpha q+\beta}{\alpha\sigma q + 1}$. Plugging it into the bounds for $\mu$, $0< \mu   \le 1-\beta$, we obtain the relevant ranges for quantities as follows: $0 \le q <  -\frac{1-\beta}{\alpha(\sigma-1)}$ when $\sigma>1$.

\paragraph{Step 2.}
We now impose the fifth inequality condition, namely the second-order condition \eqref{eq:SOC_LFRRA}.
Recall that $u'>0$ and $u'' < 0$ hold,
where the latter inequality requires that $\alpha q+\beta >0$ for $\beta \in (0,1]$ and $\alpha>0$ for $\beta=0$.

Thus, there are two cases. First, except for the case where $\beta=0$ and $q=0$ simultaneously hold, we have $(\alpha q+\beta)(\alpha \sigma q+1) > 0$.
In this case,
\eqref{eq:SOC_LFRRA}
can be
rewritten as $\frac{\Phi(q)}{(\alpha q+\beta)(\alpha \sigma q+1)}
> 0$, where
\begin{equation}
\Phi(q) \equiv
\left\{
\begin{array}{ll}
\alpha^2 (\sigma -1) \bigl[  q+ \frac{1-\beta}{\alpha (\sigma-1)} \bigr]^2 -\frac{(1-\beta) (1-\beta \sigma)}{\sigma-1} & \mbox{for}  \ \sigma \neq 1 \\
(1-\beta) (2 \alpha q +\beta) &\mbox{for}  \  \sigma=1 
\end{array}
\right..
\label{eq:SOC_LFRRA_modified}
\end{equation}
Observe that $\Phi (0) =\beta (1-\beta) \ge 0$, and that $\Phi$ is concave for $\sigma < 1$, convex for $\sigma >1$, and linear for $\sigma=1$.

Second, if $\beta=0$ and $q=0$ hold, we have $(\alpha q+\beta)(\alpha \sigma q+1) = 0$ and $\Phi(0)=0$.
In this case, we can verify that the second-order condition \eqref{eq:SOC_LFRRA} holds by setting $\beta=0$ and taking a limit as $q \to 0^+$, which yields $\lim_{q \to 0^+} 1+ \frac{1-\alpha q}{\alpha \sigma q+1} =2>0$.

Thus, in what follows, we consider the first case where $\beta=0$ and $q=0$ do not simultaneously hold. In this case,
the SOC holds if
$\Phi(q)
>0$.
We now solve $\Phi(q) =0$ for $q$ and use the solution(s)
to see when $\Phi(q) >0$ holds.

\vspace{-.25cm}

\paragraph{Case with $\sigma \neq 1$.} We first assume that $1 - \beta \sigma>0$.
The equation $\Phi(q)=0$ yields $q = -\frac{1 - \beta}{\alpha (\sigma-1)} \pm \sqrt{ \frac{(1 - \beta) (1 - \beta
\sigma)}{\alpha^2 (\sigma-1)^2}}$. Noting $\sqrt{\alpha^2 (\sigma-1)^2} = |\alpha (\sigma-1)|$, the solutions can be defined~as
\begin{equation} 
\textstyle
q^\pm
= -\frac{1 - \beta \pm \Delta}{\alpha (\sigma-1)}, \quad \mbox{where} \quad \Delta \equiv \sqrt{ (1 - \beta) (1 - \beta \sigma)}.
\label{eq:SOC_roots}
\end{equation}
Second, assume that $1 - \beta \sigma < 0$. If $\beta \neq 1$, the equation $\Phi(q)=0$ has no real roots, whereas
if $\beta=1$, $\Phi(q)=0$ yields $q=0$.

\vspace{-.25cm}

\paragraph{Case with $\sigma = 1$.} If $\beta \neq 1$, the equation $\Phi (q) =0$ yields $q  = -\frac{\beta}{2 \alpha}$, whereas if $\beta = 1$, $\Phi(q)=0$ for all $q$. Since the second-order condition does not hold in the latter case, we exclude it from the analysis.

\bigbreak
\noindent
In what follows, we use those solutions to show under what condition the SOCs hold and to refine the ranges of markups and quantities.

\paragraph{Case 1: $\alpha >0$.} \hfill

\vspace{-.4cm}

\paragraph{Case 1a: $1-\beta -\mu \ge 0$, $1 -  \sigma (1-\mu)>0$, and $1 -\beta \sigma>0$.}
In this case, $\beta = 1$ is not admissible since $\mu>0$ must hold.

First, assume that $\sigma < 1$. Since $1-\beta\sigma>0$ and $\beta \neq 1$ hold, $\Phi(q)=0$ has two real roots \eqref{eq:SOC_roots}. Consider the subcase with $\beta \in (0,1)$. In that case, one of them is positive since $1 - \beta +
\Delta >0$, whereas the other is negative since $1 - \beta -
\Delta < 0$. Furthermore, $\Phi$ is concave, so that it is positive for $0 \le q < -\frac{1 - \beta +
\Delta }{\alpha (\sigma - 1)}$. We also know from Table~\ref{tab:tab1} that $q < -\frac{1 - \beta}{\alpha (\sigma - 1)}$ must hold. Since $-\frac{1 - \beta}{\alpha (\sigma - 1)} <  -\frac{1 - \beta +
\Delta }{\alpha (\sigma - 1)}$ holds for $\sigma < 1$, the SOC holds for all $0 \le q <  -\frac{1 - \beta}{\alpha (\sigma - 1)}$ when $\beta \in (0,1)$. Consider the subcase with $\beta =0$. Then, the two real roots in \eqref{eq:SOC_roots} become $-\frac{2}{\alpha (\sigma-1)}>0$ and $0$. Since $\Phi$ is concave and $-\frac{2}{\alpha (\sigma-1)}$ is greater than $-\frac{1-\beta}{\alpha (\sigma-1)}$ in Table~\ref{tab:tab1} evaluated at $\beta=0$, the SOC holds for all $0< q< -\frac{1}{\alpha (\sigma-1)}$. As $q$ goes to zero, we know that $\frac{\Phi(q)}{(\alpha q+\beta) (\alpha \sigma q+1)}$ evaluated at $\beta=0$ goes to 2.
Hence, the SOC holds for all $0\le q <  -\frac{1-\beta}{\alpha (\sigma-1)}$ when $\beta \in [0,1)$.

Second, assume that $\sigma=1$. Then, the SOC holds for all $q \ge 0$ since both $1- \beta $ and $2\alpha q+\beta$ in \eqref{eq:SOC_LFRRA_modified} are positive when $\beta \in ( 0,1)$ and since $\lim_{\sigma \to 1} \frac{\Phi(q)}{(\alpha q+\beta) (\alpha \sigma q+1)} = \frac{2}{\alpha q+1}>0$ when $\beta=0$. Hence, when $\sigma=1$ and $\beta \in [0,1)$, the SOC holds for all
$0\le q < \infty$ given in Table~\ref{tab:tab1} (recall that when $\sigma=1$, $0 \le q < -\frac{1-\beta}{\alpha (\sigma-1)}$ must be replaced with $0 \le q < \infty$).

Last, assume that $\sigma>1$. Since $1-\beta\sigma>0$ and $\beta \neq 1$ hold, $\Phi(q)=0$ has two real roots \eqref{eq:SOC_roots}. Consider the subcase with $\beta \in (0,1)$. In that case, both are negative since $1 - \beta +
\Delta >0$ and $1 - \beta -
\Delta >0$. Furthermore, $\Phi$ is convex, so that it is positive for all $q \ge 0$. We also know from Table~\ref{tab:tab1} that there is no additional restriction on $q$, so that the SOC holds for all $0 \le q <  \infty$ when $\beta \in (0,1)$. Consider the subcase with $\beta =0$. Then, the two real roots in \eqref{eq:SOC_roots} become $-\frac{2}{\alpha (\sigma-1)}<0$ and $0$. Since $\Phi$ is convex, the SOC holds for all $q>0$. As $q$ goes to zero, we know that $\frac{\Phi(q)}{(\alpha q+\beta) (\alpha \sigma q+1)}$ evaluated at $\beta=0$ goes to 2.
Hence, the SOC holds for all $0\le q <  \infty$ when $\beta \in [0,1)$.

Since the range of $q$ in Table~\ref{tab:tab1} is unaffected by the SOC, that of $\frac{1}{\mu}$ in Table~\ref{tab:tab1} is also unaffected because $\mu =1- \frac{\alpha q +\beta}{\alpha \sigma q+1}$ is monotone with respect to $q$. Hence, the ranges for markups and quantities  in Table~\ref{tab:tab1} remain valid in Table~\ref{tab:tab2}.

\paragraph{Case 1b: $1-\beta -\mu \le 0$, $1 -  \sigma (1-\mu)<0$, and $ 1 -\beta \sigma<0$.}
In this case, $\beta=0$ is not admissible and $\sigma>1$ must hold.

If $\beta \in (0,1)$, then $\Phi(q)=0$ has no real roots since $1-\beta\sigma<0$. Since $\sigma>1$ and $\beta \not\in \{0, 1\}$, $\Phi$ is convex and $\Phi(0) = \beta (1-\beta)>0$, which implies that $\Phi$ is positive for all $q \ge 0$. Hence, the SOC holds for all $q \ge  0$. Since the range of $q$ in Table~\ref{tab:tab1} is unaffected by the SOC, that of $\frac{1}{\mu}$ in Table~\ref{tab:tab1} is also unaffected because $\mu =1- \frac{\alpha q +\beta}{\alpha \sigma q+1}$ is monotone with respect to $q$.

If $\beta =1$, then $\Phi(q)=0$ yields $q=0$, which implies $\Phi(0)=0$. Since $\sigma>1$, $\Phi$ is convex, which implies that $\Phi$ is positive for all $q >0$. Hence, the SOC holds for all $q > 0$.
Since the range of $q$ in Table~\ref{tab:tab1} is unaffected by the SOC, that of $\frac{1}{\mu}$ in Table~\ref{tab:tab1} is also unaffected because $\mu =1- \frac{\alpha q +\beta}{\alpha \sigma q+1}$ is monotone with respect to $q$.

\paragraph{Case 2: $\alpha <0$.} \hfill
\vspace{-.4cm}

\paragraph{Case 2a: $1-\beta -\mu \le0$, $1 -  \sigma (1-\mu)>0$, and $ 1 -\beta \sigma>0$.}
In this case, $\beta =0$ is excluded as mentioned in Step 1.

First, assume that $\sigma <1$. Consider the subcase with $\beta \in (0,1)$. In that case, $\Phi(q)=0$ has two real roots \eqref{eq:SOC_roots}. One of them is negative since $1 - \beta +
\Delta > 0$, whereas the other root is positive since  $1 - \beta -
\Delta < 0$. Furthermore, $\Phi$ is concave, so that it is positive for $0 \le  q < -\frac{1 - \beta -
\Delta }{\alpha (\sigma-1)}$. We also know from Table~\ref{tab:tab1} that $0\le q < - \frac{\beta}{\alpha}$ must hold. Since $-\frac{1 - \beta -
\Delta }{\alpha (\sigma-1)} < - \frac{\beta}{\alpha}$ holds, the second-order condition holds for all $0 \le q < - \frac{1 - \beta -
\Delta }{\alpha (\sigma-1)}$ but it does not hold for $-\frac{1 - \beta -
\Delta }{\alpha (\sigma-1)} \le  q < -\frac{\beta}{\alpha} $. Since the bounds for $q$ are modified, we need to modify those for $\mu$. Noting that $\frac{\partial \mu}{\partial q} = \frac{\partial}{\partial q} \bigl( 1-\frac{\alpha q+\beta}{\alpha\sigma q + 1}  \bigr)=  -\frac{\alpha (1-\beta \sigma)}{(\alpha \sigma q+1)^2}>0$, we evaluate $\mu= 1-\frac{\alpha q+\beta}{\alpha\sigma q + 1}$ at $q=0$ and $q=- \frac{1 - \beta -
\Delta }{\alpha (\sigma-1)} < -\frac{\beta}{\alpha}$ to obtain $1-\beta \le  \mu <  \frac{ (1 - \beta)\sigma +
\Delta }{1+ (1-\beta) \sigma} < 1$.\footnote{The last inequality can be obtained as follows. First, for $\frac{ (1 - \beta)\sigma +
\Delta }{1+ (1-\beta) \sigma} < 1$ to hold, we require $\frac{ -1+
\Delta }{1+ (1-\beta) \sigma} < 0$. Second, the latter inequality requires that the numerator and the denominator must have opposite signs, which is indeed the case.}
Hence, we obtain the bounds for markups $1 <  \frac{1+ (1-\beta) \sigma}{ (1 - \beta)\sigma +
\Delta } <\frac{1}{\mu}  \le \frac{1}{1-\beta}$ for $\sigma < 1$. Consider the subcase with $\beta =1$. Then, $\Phi$ attains the maximum  $\Phi(q) =0$ at  $q=0$. Since $\sigma < 1$, $\Phi$ is concave, so that the second-order condition does not hold for $q \ge 0$, thus excluding the second line in Case 2a in Table~\ref{tab:tab1}.

Second, assume that $\sigma=1$. In this case, $\beta \neq 1$ must hold because of the assumption that $1-\beta \sigma>0$. Thus, $\Phi (q)>0$ implies $q<-\frac{\beta}{2\alpha}$. We also know from Table~\ref{tab:tab1} that $0\le q < - \frac{\beta}{\alpha}$ must hold. Since $-\frac{\beta}{2\alpha}< - \frac{\beta}{\alpha}$ holds, the second-order condition holds for all $0 \le q < -\frac{\beta}{2\alpha}$ but it does not hold for $-\frac{\beta}{2\alpha} \le  q < -\frac{\beta}{\alpha}$.
Since the range of $q$ is modified, we need to modify that for $\mu$. Noting that $\frac{\partial \mu}{\partial q} = \frac{\partial}{\partial q} \bigl( 1-\frac{\alpha q+\beta}{\alpha\sigma q + 1}  \bigr)=  -\frac{\alpha (1-\beta \sigma)}{(\alpha \sigma q+1)^2}>0$, we evaluate $\mu= 1-\frac{\alpha q+\beta}{\alpha\sigma q + 1}$ at $q=0$ and $q= -\frac{\beta}{2\alpha}$ to obtain $1-\beta \le \mu < \frac{2(1-\beta)}{2-\beta} < 1$. Hence, we obtain the range of markups $1 < 
 \frac{2-\beta}{2(1-\beta)} <\frac{1}{\mu}  \le \frac{1}{1-\beta}$ for $\sigma = 1$.

Last, assume that $\sigma> 1$.  In this case, $\beta \neq 1$ must hold because of the assumption that $1-\beta \sigma>0$. Thus, $\Phi(q)=0$ has two real roots \eqref{eq:SOC_roots}. Both roots are positive since $1 - \beta +
\Delta >0$ and $1 - \beta -
\Delta >0$. Since $\Phi$ is convex, it is negative between its roots. Since $ - \frac{1 - \beta -
\Delta }{\alpha (\sigma-1)} < -\frac{\beta}{\alpha} < - \frac{1 - \beta +
\Delta }{\alpha (\sigma-1)}$, the second-order condition holds for all $0 \le q < - \frac{1 - \beta -
\Delta }{\alpha (\sigma-1)}$ but does not hold for $-\frac{1 - \beta -
\Delta }{\alpha (\sigma-1)} \le  q < -\frac{\beta}{\alpha} $.
Since the bounds for $q$ are modified, we need to modify those for $\mu$. Noting that
$\frac{\partial \mu}{\partial q} = \frac{\partial}{\partial q} \bigl( 1-\frac{\alpha q+\beta}{\alpha\sigma q + 1}  \bigr)=  -\frac{\alpha (1-\beta \sigma)}{(\alpha \sigma q+1)^2}>0$, we evaluate $\mu= 1-\frac{\alpha q+\beta}{\alpha\sigma q + 1}$ at $q=0$ and $q=- \frac{1 - \beta -
\Delta }{\alpha (\sigma-1)} < -\frac{\beta}{\alpha}$ to obtain $1-\beta \le \mu <  \frac{ (1 - \beta)\sigma +
\Delta }{1+ (1-\beta) \sigma} <1$. Hence, we obtain the bounds for markups $1 <  \frac{1+ (1-\beta) \sigma}{ (1 - \beta)\sigma +
\Delta } <\frac{1}{\mu} \le \frac{1}{1-\beta} $ for $\sigma > 1$.

\paragraph{Case 2b: $1-\beta -\mu \ge 0$, $1 -  \sigma (1-\mu)<0$, and $ 1 -\beta \sigma<0$.}
In this case, $\beta=0$ and $\beta =1$ are not admissible, whereas $\sigma>1$ must hold.

Since $\beta \neq 1$ and $1-\beta\sigma<0$, $\Phi(q)=0$ has no real roots. Since $\sigma>1$ and $\beta \not\in \{0, 1\}$, $\Phi$ is convex and $\Phi(0) = \beta (1-\beta)>0$, which implies that $\Phi$ is positive for all $q \ge 0$. Hence, the SOC holds for all $q \ge  0$. Therefore, the SOC holds for $0 \le q <  -\frac{1-\beta}{\alpha (\sigma-1)}$ and thus for $\frac{1}{1-\beta} \le \frac{1}{\mu} <\infty$ given in Table~\ref{tab:tab1} when $\beta \in (0,1)$.

\medbreak
Finally, observe from Table~\ref{tab:tab2} that marginal cost pricing $\frac{1}{\mu}=1$ is admissible only when $\beta=0$ in Case 1a. Note that $\frac{\sigma}{\sigma-1} > 1$ holds in Case 1b, $\frac{(1-\beta)\sigma+1}{(1-\beta)\sigma+\Delta} >1$ holds in Case 2a, and that $\beta=0$ is not admissible in Case 2b. \hfill$\square$

\subsection{Proof of Proposition~\ref{prop:propEU}}
\label{app:proofs4}

First, assume that $\alpha \neq 0$, and that $\beta \neq \frac{1}{\sigma}$. Using $q=\frac{1}{\alpha} \frac{1-\beta -\mu}{1- \sigma (1-\mu) } \ge 0$ from \eqref{eq:FOCprof}, we have $1+ \alpha \sigma q = \frac{1 -\beta \sigma}{1- \sigma (1-\mu)} > 0$ because $u' > 0$.
Recall that $q>0$ must hold for $\beta \in (0,1]$, whereas $q=0$ may hold for $\beta=0$ since $\lim_{q\to 0^+} u'(q)=K(1+\alpha \sigma q)^{\beta-\frac{1}{\sigma}} q^{-\beta} =\infty$ for $\beta \in (0,1]$ and the limit is finite for $\beta=0$. We can then
rewrite \eqref{eq:FOCutil} as follows:
\begin{equation}
\mu \left[ \frac{1 -\beta \sigma}{1- \sigma (1-\mu) }  \right]^{\beta- \frac{1}{\sigma}} \left[ \frac{1}{\alpha} \frac{1-\beta -\mu}{1- \sigma (1-\mu) } \right]^{-\beta} = x.
\label{eq:appA1b}
\end{equation}
Consider the subcase with $\beta \in (0,1]$. To obtain the bottom-right expression of Table~\ref{tab:FOC} from \eqref{eq:appA1b}, we first assume that $\sigma\neq 0$. In that case, we can rewrite \eqref{eq:appA1b} as follows:
\begin{equation*}
\left[ \frac{1 -\beta \sigma}{1- \sigma (1-\mu) }  \right]^{\beta \sigma- 1} \left[ \frac{1}{\alpha} \frac{1-\beta -\mu}{1- \sigma (1-\mu) } \right]^{-\beta \sigma} = \left( \frac{x}{\mu} \right)^\sigma.
\end{equation*}
Since both $\frac{1 -\beta \sigma}{1- \sigma (1-\mu) }$ and $\frac{1}{\alpha} \frac{1-\beta -\mu}{1- \sigma (1-\mu) }$ must be positive, we can rewrite the forgoing expression~as
$\bigl[ \frac{1 -\beta \sigma}{1- \sigma (1-\mu) }  \bigr]^{\beta \sigma} \bigl[ \frac{1}{\alpha} \frac{1-\beta -\mu}{1- \sigma (1-\mu) } \bigr]^{-\beta \sigma}
= \bigl( \frac{x}{\mu} \bigr)^\sigma \frac{1 -\beta \sigma}{1- \sigma (1-\mu) }$, which implies
$\bigl[  \frac{\alpha ( 1 -\beta \sigma) }{1-\beta -\mu}  \bigr]^{\beta \sigma} 
= \bigl( \frac{x}{\mu} \bigr)^\sigma \frac{1 -\beta \sigma}{1- \sigma (1-\mu) }$, so that $1- \sigma (1-\mu) 
= \bigl( \frac{\mu}{x} \bigr)^{-\sigma} (1 -\beta \sigma) \bigl[  \frac{1-\beta -\mu}{\alpha ( 1 -\beta \sigma) }  \bigr]^{\beta \sigma}$.
Hence, we obtain 
\begin{equation*}
1-\mu
= \frac{1- \left( \frac{\mu}{x} \right)^{-\sigma} (1 -\beta \sigma) \left[  \frac{1-\beta -\mu}{\alpha ( 1 -\beta \sigma) }  \right]^{\beta \sigma}}{\sigma}.
\end{equation*}
When $\sigma$ goes to $0$, equation \eqref{eq:appA1b}
 reduces to
$\mu {\rm e}^{-(1-\beta-\mu)}\left(\frac{1-\beta-\mu}{\alpha}\right)^{-\beta} = x,$
which can be rewritten as
\begin{equation*}
1-\mu = \ln \left( \frac{\mu}{x} \right) + \beta \left[ 1 -  \ln \left( \frac{1-\beta-\mu}{\alpha} \right) \right].
\end{equation*}

Consider the subcase with $\beta=0$. Then, \eqref{eq:appA1b} implies $\mu [ \frac{1 }{1- \sigma (1-\mu) }  ]^{- \frac{1}{\sigma}}  = x$. We can rewrite it as $ \frac{1 }{1- \sigma (1-\mu) }   = (\frac{x}{\mu})^{-\sigma}$, so that $1- \sigma (1-\mu)   = (\frac{\mu}{x})^{-\sigma}$. Hence, we obtain $1-\mu = \frac{1- ( \frac{\mu}{x} )^{-\sigma} }{\sigma}$. When $\sigma$ goes to $0$, the foregoing equation reduces to $1-\mu = \ln ( \frac{\mu}{x})$.

This establishes the expressions in Table~\ref{tab:FOC}.

We now prove the existence and uniqueness of a markup function.
The right-hand side of expression \eqref{eq:appA1b} is exogenous for each firm, whereas the left-hand side depends on the markup $\mu$ that is endogenously determined.
Differentiating the left-hand side of \eqref{eq:appA1b} with respect to $\mu$ yields
\begin{equation}
\frac{\partial {\rm LHS}}{\partial \mu} = \frac{1-\beta-\mu^2-\sigma(1-\beta)(1-\mu)^2}{(1-\beta-\mu)[1-\sigma(1-\mu)]}  \frac{x}{\mu}.
\label{eq:dLHS}
\end{equation}
We will sign this derivative by using
\eqref{eq:SOC_LFRRA} evaluated at $q= \frac{1}{\alpha} \frac{1-\beta -\mu}{1- \sigma (1-\mu) } $ as follows:
\begin{equation}
\frac{1-\beta-\mu^2-\sigma(1-\beta)(1-\mu)^2}{(1-\mu)(1-\beta\sigma)}>0,
\label{eq:SOCsimplified}
\end{equation}
which holds for all $\mu$ satisfying the range of markups in Table~\ref{tab:tab2}.

As in Proposition~\ref{prop:markup_bounds}, we consider four distinct cases depending on the signs of $\alpha$ and $1-\beta\sigma$.

\paragraph{Case 1: $\alpha >0$.} \hfill

\vspace{-.4cm}

\paragraph{Case 1a: $1-\beta -\mu \ge 0$, $1 -  \sigma (1-\mu)>0$, and $1 -\beta \sigma>0$.}

In this case, $\beta = 1$ is not admissible. Assume that $\sigma \le 1$. The range of markups $\frac{1}{1-\beta} \le \frac{1}{\mu}<\infty$ in Table~\ref{tab:tab2} implies that the range of $\mu$  is given by $0 < \mu \le 1-\beta$.

Consider the subcase with $\beta \in (0,1)$. In this case, $q>0$ must hold as explained before, so that $1-\beta -\mu > 0$. Since $1-\mu > \beta > 0$ and $1 -\beta \sigma>0$ hold, the denominator of \eqref{eq:SOCsimplified} is positive. We thus obtain $1-\beta-\mu^2-\sigma(1-\beta)(1-\mu)^2>0$ from \eqref{eq:SOCsimplified}.

Since $\frac{x}{\mu}>0$, $1-\beta -\mu > 0$, and $1- \sigma (1-\mu)>0$, we know that \eqref{eq:dLHS} is positive, i.e., the LHS of \eqref{eq:appA1b} is strictly increasing in $\mu$. Thus, if a solution $\mu$ to \eqref{eq:appA1b} exists, it is unique. The solution exists when the LHS of \eqref{eq:appA1b} is smaller than the right-hand side, $x$, as $\mu$ goes to $0$ and is greater than $x$ as $\mu$ goes to $1-\beta$. Computing the limits of the LHS, we have
\begin{equation*}
\lim_{\mu\to 0^+} {\rm LHS} = 0 \quad {\rm and} \quad \lim_{\mu\to 1-\beta^-} {\rm LHS} = \infty.
\end{equation*}
Thus, when $\sigma \le 1$ and $0<\beta<1$ hold, there exists a unique solution 
$\mu \in (0, 1-\beta)$  to \eqref{eq:appA1b} regardless of the value of $x$.

Consider the subcase with $\beta =0$. Setting $\beta=0$ and differentiating the LHS of \eqref{eq:appA1b} yield $[1-\sigma(1-\mu)]^{-\frac{\sigma-1}{\sigma}}  [1-\sigma(1-\mu)+\mu]>0$. Computing the limits of the LHS, we have
\begin{equation*}
\lim_{\mu\to 0^+} {\rm LHS} = 0 \quad {\rm and} \quad \lim_{\mu\to 1^-} {\rm LHS} = 1.
\end{equation*}
Thus, when $\sigma \le 1$ and $\beta = 0$ hold, there exists a unique solution $\mu \in (0, 1]$ to \eqref{eq:appA1b} only for $x \le 1$.

Assume that $\sigma > 1$. The range of markups $\frac{1}{1-\beta} \le \frac{1}{\mu}<\frac{\sigma}{\sigma-1}$ in Table~\ref{tab:tab2} implies that the range of $\mu$ is given by $\frac{\sigma-1} {\sigma}< \mu \le 1-\beta$.

Consider the subcase with $\beta \in (0,1)$. In this case, $q>0$ must hold, so that $1-\beta -\mu > 0$. Since $1-\mu > \beta > 0$ and $1 -\beta \sigma>0$ hold, the denominator of \eqref{eq:SOCsimplified} is positive. We thus obtain $1-\beta-\mu^2-\sigma(1-\beta)(1-\mu)^2>0$ from \eqref{eq:SOCsimplified}.

Since $\frac{x}{\mu}>0$, $1-\beta -\mu>0$, and $1- \sigma (1-\mu)>0$, we know that \eqref{eq:dLHS} is positive, i.e., the LHS of \eqref{eq:appA1b} is strictly increasing in $\mu$. Thus, if a solution $\mu$ to \eqref{eq:appA1b} exists, it is unique. The solution exists when the LHS of \eqref{eq:appA1b} is smaller than the right-hand side, $x$, as $\mu$ goes to $\frac{\sigma-1} {\sigma}$ and is greater than $x$ as $\mu$ goes to $1-\beta$. Computing the limits of the LHS, we have
\begin{equation*}
\lim_{\mu\to \frac{\sigma-1} {\sigma}^+} {\rm LHS} = 0 \quad {\rm and} \quad \lim_{\mu\to 1-\beta^-} {\rm LHS} = \infty.
\end{equation*}
Thus, when $\sigma > 1$ and $0<\beta<1$ hold, there exists a unique solution 
$\mu \in \bigl( \frac{\sigma-1}{\sigma}, 1-\beta \bigr)$  to \eqref{eq:appA1b} regardless of the value of $x$.

Consider the subcase with $\beta =0$. Setting $\beta=0$ and differentiating the LHS of \eqref{eq:appA1b} yield $[1-\sigma(1-\mu)]^{-\frac{\sigma-1}{\sigma}}  [1-\sigma(1-\mu)+\mu]>0$. Computing the limits of the LHS, we have
\begin{equation*}
\lim_{\mu\to \frac{\sigma-1} {\sigma}^+} {\rm LHS} = 0 \quad {\rm and} \quad \lim_{\mu\to 1^-} {\rm LHS} = 1.
\end{equation*}
Thus, when $\sigma > 1$ and $\beta = 0$ hold,
there exists a unique solution $\mu \in \bigl( \frac{\sigma-1}{\sigma}, 1 \bigr]$ only for $x \le 1$.

\paragraph{Case 1b: $1-\beta -\mu \le 0$, $1 -  \sigma (1-\mu)<0$, and $ 1 -\beta \sigma<0$.}
In this case,  $\beta = 0$ is not admissible, whereas $\sigma>1$ must hold.

Consider the subcase with $\beta \in (0,1)$. The range of markups $\frac{\sigma}{\sigma-1}<\frac{1}{\mu} \le \frac{1}{1-\beta}$ in Table~\ref{tab:tab2} implies that the range of $\mu$  is given by $1-\beta  \le \mu < \frac{\sigma-1}{\sigma}$. Since $ \mu < \frac{\sigma-1}{\sigma} < 1$ and $1 -\beta \sigma<0$ hold, the denominator of \eqref{eq:SOCsimplified} is negative. We thus obtain $1-\beta-\mu^2-\sigma(1-\beta)(1-\mu)^2<0$ from \eqref{eq:SOCsimplified}.

Recall that when $\beta \in (0,1)$, $q>0$ must hold, so that $1-\beta -\mu < 0$. Since $\frac{x}{\mu}>0$, $1-\beta -\mu<0$, and $1- \sigma (1-\mu)<0$, we know that \eqref{eq:dLHS} is negative, i.e., the LHS of \eqref{eq:appA1b} is strictly decreasing in $\mu$. Thus, if a solution $\mu$ to \eqref{eq:appA1b} exists, it is unique. The solution exists when the LHS of \eqref{eq:appA1b} is greater than the right-hand side, $x$, as $\mu$ goes to $1-\beta$ and is smaller than $x$ as $\mu$ goes to $\frac{\sigma-1}{\sigma}$. Computing the limits of the LHS, we have
\begin{equation*}
\lim_{\mu\to 1-\beta^+} {\rm LHS} = \infty \quad {\rm and} \quad \lim_{\mu\to \frac{\sigma-1}{\sigma}^-} {\rm LHS} = 0.
\end{equation*}
Thus, when $0<\beta<1$, there exists a unique solution 
$\mu \in \bigl( 1-\beta, \frac{\sigma-1}{\sigma} \bigr)$  to \eqref{eq:appA1b} regardless of the value of $x$.

Consider the subcase with $\beta =1$. The range of markups
$\frac{\sigma}{\sigma-1}<\frac{1}{\mu}<\infty$ in Table~\ref{tab:tab2} implies that the range of $\mu$  is given by
$0  < \mu < \frac{\sigma-1}{\sigma}$.
Since $ \mu < \frac{\sigma-1}{\sigma} < 1$ and $1 -\beta \sigma<0$ hold, the denominator of \eqref{eq:SOCsimplified} is negative. We thus obtain $1-\beta-\mu^2-\sigma(1-\beta)(1-\mu)^2<0$ from \eqref{eq:SOCsimplified}.

Recall that when $\beta =1$, $q>0$ must hold, so that $1-\beta -\mu < 0$. Since $\frac{x}{\mu}>0$, $1-\beta -\mu<0$, and $1- \sigma (1-\mu)<0$, we know that \eqref{eq:dLHS} is negative, i.e., the LHS of \eqref{eq:appA1b} is strictly decreasing in $\mu$. Thus, if a solution $\mu$ to \eqref{eq:appA1b} exists, it is unique. The solution exists when the LHS of \eqref{eq:appA1b} is greater than the right-hand side, $x$, as $\mu$ goes to $0$ and is smaller than $x$ as $\mu$ goes to $\frac{\sigma-1}{\sigma}$. Computing the limits of the LHS, we have
\begin{equation*}
\lim_{\mu\to 0^+} {\rm LHS} = \alpha (\sigma-1) \quad {\rm and} \quad \lim_{\mu\to \frac{\sigma-1}{\sigma}^-} {\rm LHS} = 0.
\end{equation*}
Thus,
when $\beta = 1$, there exists a unique solution $\mu \in \bigl( 0, \frac{\sigma-1}{\sigma} \bigr)$ only for $x < \alpha (\sigma-1)$.

\paragraph{Case 2: $\alpha<0$.} \hfill

\vspace{-.4cm}

\paragraph{Case 2a: $1-\beta -\mu \le 0$, $1 -  \sigma (1-\mu)>0$, and $1 -\beta \sigma>0$.}
In this case, $\beta=0$ is excluded by Step 1 of the proof of Proposition~\ref{prop:markup_bounds} and $\beta = 1$ is not admissible by Step 2 of the proof of Proposition~\ref{prop:markup_bounds}. The range of markups $\frac{(1-\beta)\sigma+1}{(1-\beta)\sigma + \Delta} <\frac{1}{\mu} \le \frac{1}{1-\beta}$ in Table~\ref{tab:tab2} implies that the range of $\mu$  is given by $1-\beta \le  \mu < \frac{(1-\beta)\sigma + \Delta}{(1-\beta)\sigma+1}$. Since $ \mu < \frac{(1-\beta)\sigma + \Delta}{(1-\beta)\sigma+1} < 1$ and $1 -\beta \sigma>0$ hold, the denominator of \eqref{eq:SOCsimplified} is positive. We thus obtain $1-\beta-\mu^2-\sigma(1-\beta)(1-\mu)^2>0$ from \eqref{eq:SOCsimplified}.

Recall that when $\beta \in (0,1)$, $q>0$ must hold, so that $1-\beta -\mu < 0$. Since $\frac{x}{\mu}>0$, $1-\beta -\mu < 0$, and $1- \sigma (1-\mu)>0$, we know that \eqref{eq:dLHS} is negative, i.e., the LHS of \eqref{eq:appA1b} is strictly decreasing in $\mu$. Thus, if a solution $\mu$ to \eqref{eq:appA1b} exists, it is unique. The solution exists when the LHS of \eqref{eq:appA1b} is greater than the right-hand side, $x$, as $\mu$ goes to $1-\beta$ and is smaller than $x$ as $\mu$ goes to $\overline \mu \equiv \frac{(1-\beta)\sigma + \Delta}{(1-\beta)\sigma+1}$. Computing the limits of the LHS, we have
\begin{equation*}
\lim_{\mu\to 1-\beta^+} {\rm LHS} = \infty
\quad \mbox{and} \quad
\lim_{{\mu\to \overline \mu}^-}
\textstyle {\rm LHS} =
\textstyle \overline \mu
\left[ \frac{1 -\beta \sigma}{1- \sigma  (1- \overline \mu ) }  \right]^{\beta- \frac{1}{\sigma}} \left[ \frac{1}{\alpha} \frac{1-\beta -\overline \mu}{1- \sigma (1-\overline \mu ) } \right]^{-\beta}  \equiv \underline x.
\end{equation*}
Thus, there exists a unique solution  $\mu \in ( 1-\beta, \overline \mu )$  to \eqref{eq:appA1b} for $x > \underline x$.

\paragraph{Case 2b: $1-\beta -\mu \ge 0$, $1 -  \sigma (1-\mu)<0$,  and $ 1 -\beta \sigma<0$.}

In this case, $\beta=0$ and $\beta =1$ are not admissible, whereas $\sigma>1$ must hold. The range of markups $\frac{1}{1-\beta} \le \frac{1}{\mu}<\infty$ in Table~\ref{tab:tab2} implies that the range of $\mu$  is given by $0  < \mu \le  1-\beta$. Since $ \mu \le 1-\beta < 1$ and $1 -\beta \sigma<0$ hold, the denominator of \eqref{eq:SOCsimplified} is negative. We thus obtain $1-\beta-\mu^2-\sigma(1-\beta)(1-\mu)^2<0$ from \eqref{eq:SOCsimplified}.

Recall that when $\beta \in (0,1)$, $q>0$ must hold, so that $1-\beta -\mu > 0$. Since $\frac{x}{\mu}>0$, $1-\beta -\mu>0$, and $1- \sigma (1-\mu)<0$, we know that \eqref{eq:dLHS} is positive, i.e., the LHS of \eqref{eq:appA1b} is strictly increasing in $\mu$. Thus, if a solution $\mu$ to \eqref{eq:appA1b} exists, it is unique. The solution exists when the LHS of \eqref{eq:appA1b} is smaller than the right-hand side, $x$, as $\mu$ goes to $0$ and is larger than $x$ as $\mu$ goes to $1-\beta$. Computing the limits of the LHS, we have
\begin{equation*}
\lim_{\mu\to 0^+} {\rm LHS} =0 \quad {\rm and} \quad  \lim_{\mu\to 1-\beta^-} {\rm LHS} =  \infty.
\end{equation*}
Thus,  there always exists a unique solution to \eqref{eq:appA1b}.

\medbreak

Next, assume that $\alpha=0$, and that $\beta \in (0,1)$. Then, although the expressions in the right column of Table~\ref{tab:FOC} cannot be used, we can combine \eqref{eq:LFRRA_general} and \eqref{eq:generalFOC} to obtain the first-order condition $1-\mu = \beta$.
Hence, there exists a unique markup function $\frac{1}{\mu} = \frac{1}{{\cal W} (x, 0, \beta,\sigma) } =\frac{1}{1-\beta} >1$ for all $x \in (0,\infty)$. Evaluating the second-order condition \eqref{eq:SOC_LFRRA} at $\alpha=0$, we know that it holds since $\frac{\alpha q}{\alpha q+\beta} +\frac{1-\alpha q -\beta}{\alpha \sigma q+1} = 1-\beta>0$ for $\beta \in (0,1)$.

Finally, assume that $\beta = \frac{1}{\sigma}$, and that $\sigma>1$. Then, although the bottom-right expression of Table~\ref{tab:FOC} cannot be used, we can combine \eqref{eq:LFRRA_general} and \eqref{eq:generalFOC} to obtain the first-order condition $1-\mu = \frac{1}{\sigma}$. Hence, there exists a unique markup function $\frac{1}{\mu} = \frac{1}{{\cal W} (x,\alpha, \frac{1}{\sigma},\sigma) } =\frac{\sigma}{\sigma-1} >1$ for all $x \in (0,\infty)$. Evaluating the second-order condition \eqref{eq:SOC_LFRRA} at $\beta = \frac{1}{\sigma}$, we know that it holds since $\frac{\alpha q}{\alpha q+\beta} +\frac{1-\alpha q -\beta}{\alpha \sigma q+1} = \frac{\sigma-1}{\sigma}>0$ for $\sigma>1$.
\hfill $\square$

\subsection{Proof of Proposition~\ref{prop:compstat}}
\label{app:proofs5}

Assume that $\alpha \neq 0$, and that $\beta \neq \frac{1}{\sigma}$.
Let us start with the markups. First, we differentiate the bottom-right expression of Table~\ref{tab:FOC} to obtain
\begin{equation*}
\textstyle
- {\rm d}\mu = \frac{(1-\beta) (1-\mu)}{\alpha \mu} \left( \frac{\mu}{x} \right)^{-\sigma} \left[  \frac{1-\beta -\mu}{\alpha ( 1 -\beta \sigma) }  \right]^{\beta \sigma-1} {\rm d}\mu
-  \frac{1-\beta \sigma}{x} \left( \frac{\mu}{x} \right)^{-\sigma}
\left[  \frac{1-\beta -\mu}{\alpha ( 1 -\beta \sigma) }  \right]^{\beta \sigma} {\rm d}x,
\end{equation*}
which holds
both for
$\sigma \neq 0$ and
for $\sigma\to 0$.
We thus have
\begin{equation}
\frac{{\rm d} \mu}{{\rm d} x} = 
\frac{ \frac{1-\beta \sigma}{x} \left( \frac{\mu}{x} \right)^{-\sigma}
\left[  \frac{1-\beta -\mu}{\alpha ( 1 -\beta \sigma) }  \right]^{\beta \sigma} }{1+\frac{(1-\beta) (1-\mu)}{\alpha \mu} \left( \frac{\mu}{x} \right)^{-\sigma} \left[  \frac{1-\beta -\mu}{\alpha ( 1 -\beta \sigma) }  \right]^{\beta \sigma-1}}.
\label{eq:dmu_dx_pre}
\end{equation}
Recalling that $1- \sigma (1-\mu) 
= \left( \frac{\mu}{x} \right)^{-\sigma} (1 -\beta \sigma) \left[  \frac{1-\beta -\mu}{\alpha ( 1 -\beta \sigma) } \right]^{\beta \sigma}$, from the bottom-right expression of Table~\ref{tab:FOC}, we can rewrite the foregoing derivative as follows:
\begin{equation}
\frac{{\rm d} \mu}{{\rm d} x} = 
\frac{ \frac{1- \sigma (1-\mu)}{x} }{1+\frac{(1-\beta) (1-\mu)}{\alpha \mu} \frac{1- \sigma (1-\mu)}{1-\beta \sigma} 
\frac{\alpha ( 1 -\beta \sigma)}{1-\beta -\mu }}
=
\frac{ \frac{1- \sigma (1-\mu)}{x} }{1+\frac{(1-\beta) (1-\mu)}{ \mu} \frac{1- \sigma (1-\mu)}{1-\beta -\mu} }.
\label{eq:dmu_dx}
\end{equation}

Recall also that $q>0$ must hold for $\beta \in (0,1]$, whereas $q=0$ may hold for $\beta=0$ since $\lim_{q\to 0^+} u'(q)=K(1+\alpha \sigma q)^{\beta-\frac{1}{\sigma}} q^{-\beta} =\infty$ for $\beta \in (0,1]$ and the limit is finite for $\beta=0$. We use these results throughout the proof.

\paragraph{Case 1: $\alpha>0$.} \hfill

\vspace{-.25cm}

\paragraph{Case 1a: $1-\beta -\mu \ge 0$, $1 -  \sigma (1-\mu)>0$, and $1 -\beta \sigma>0$.}

In this case, $\beta = 1$ is not admissible. Consider the subcase with $\beta \in (0,1)$. In this case, $q > 0$ must hold, so that ${1-\beta -\mu}>0$. This, together with $1- \sigma (1-\mu)>0$, implies that the denominator of \eqref{eq:dmu_dx} is positive. Since $x>0$, we know that the numerator of \eqref{eq:dmu_dx} is positive. We thus obtain $\frac{{\rm d} \mu}{{\rm d} x} >0$. Since $x=\frac{{\lambda} m}{K}$, we may conclude that $\frac{{\rm d}  \mu}{{\rm d}  m}>0$. Consider the subcase with $\beta =0$. In this case, taking the limit of \eqref{eq:dmu_dx_pre} as $\beta \to 0^+$ yields $\frac{{\rm d} \mu}{{\rm d} x}  = \frac{1}{x} \bigl[ \frac{1}{\mu} + \bigl(\frac{\mu}{x} \bigr)^\sigma \bigr]^{-1}>0$, which implies $\frac{{\rm d}  \mu}{{\rm d}  m}>0$.

\paragraph{Case 1b: $1-\beta -\mu \le 0$, $1 -  \sigma (1-\mu)<0$, and $ 1 -\beta \sigma<0$.}
In this case, $\beta = 0$ is not admissible, whereas $\sigma>1$ must hold. Consider the subcase with $\beta \in (0,1)$. In this case, $q > 0$ must hold, so that ${1-\beta -\mu}<0$. This, together with $1- \sigma (1-\mu)<0$, implies that the denominator of \eqref{eq:dmu_dx} is positive. Since $x>0$, we know that the numerator of \eqref{eq:dmu_dx} is negative. We thus obtain $\frac{{\rm d} \mu}{{\rm d} x} <0$. Since $x=\frac{{\lambda} m}{K}$, we may conclude that $\frac{{\rm d}  \mu}{{\rm d}  m}<0$. Consider the subcase with $\beta =1$. In this case, taking the limit of \eqref{eq:dmu_dx_pre} as $\beta \to 1^-$ yields $\frac{{\rm d} \mu}{{\rm d} x}  = - \frac{\sigma-1}{x} \bigl(\frac{\mu}{x} \bigr)^{-\sigma}  \bigl[\frac{\mu}{\alpha (\sigma-1)} \bigr]^\sigma <0$, which implies $\frac{{\rm d}  \mu}{{\rm d}  m}<0$.

\paragraph{Case 2: $\alpha<0$.} \hfill

\vspace{-.25cm}

\paragraph{Case 2a: $1-\beta -\mu \le 0$, $1 -  \sigma (1-\mu)>0$, and $1 -\beta \sigma>0$.}
In this case, $\beta=0$ is excluded by Step 1 of the proof of Proposition~\ref{prop:markup_bounds} and $\beta = 1$ is not admissible by Step 2 of the proof of Proposition~\ref{prop:markup_bounds}. Since $\beta \in (0,1)$ in this case, $q > 0$ must hold, so that ${1-\beta -\mu}<0$. By the second-order condition in~\eqref{eq:SOC_LFRRA} we can show that the denominator of \eqref{eq:dmu_dx} is negative.\footnote{Since $\alpha q +\beta > 0$ holds for all quantities such that the second-order condition is satisfied (see Step 2 of the proof of Proposition~\ref{prop:markup_bounds} and Table~\ref{tab:tab2}), the second-order condition for profit-maximization \eqref{eq:SOC_LFRRA} implies that $\alpha q> -(1-\alpha q -\beta) \frac{\alpha q +\beta}{\alpha \sigma q+1}.$ At any price equilibrium, $1-\mu = \frac{\alpha q +\beta}{\alpha \sigma q+1}$ must hold. This implies $\alpha q> -(1-\alpha q -\beta) (1-\mu)$, so that $\bigl[ \frac{1-\beta-\mu}{1-\sigma (1-\mu)} \bigr] \mu > -(1 -\beta) (1-\mu)$, where we use $q=\frac{1}{\alpha} \frac{1-\beta-\mu}{1-\sigma (1-\mu)}$ to get the last inequality. Noting that the left-hand side is negative, we can show that $ 1 < -\frac{(1 -\beta) (1-\mu)}{\mu} \frac{1-\sigma (1-\mu)}{1-\beta-\mu} $, which yields $ 1 + \frac{(1 -\beta) (1-\mu)}{\mu} \frac{1-\sigma (1-\mu)}{1-\beta-\mu} < 0$.\label{fn:25}}
Since $x>0$, we know that the numerator of \eqref{eq:dmu_dx} is positive. We thus obtain $\frac{{\rm d} \mu}{{\rm d} x} <0$. Since $x=\frac{{\lambda} m}{K}$, we may conclude that $\frac{{\rm d}  \mu}{{\rm d}  m}<0$.

\paragraph{Case 2b: $1-\beta -\mu \ge 0$, $1 -  \sigma (1-\mu)<0$,  and $ 1 -\beta \sigma<0$.}

In this case, $\beta=0$ and $\beta =1$ are not admissible, whereas $\sigma>1$ must hold. Since $\beta \in (0,1)$ in this case, $q > 0$ must hold, so that ${1-\beta -\mu}>0$. By the second-order condition in~\eqref{eq:SOC_LFRRA} we can show as before that the denominator of \eqref{eq:dmu_dx} is negative (see footnote~\ref{fn:25}). Since $x>0$, we know that the numerator of \eqref{eq:dmu_dx} is negative. We thus obtain $\frac{{\rm d} \mu}{{\rm d} x} >0$. Since $x=\frac{{\lambda} m}{K}$, we may conclude that $\frac{{\rm d}  \mu}{{\rm d}  m}>0$.

\bigbreak
Turning next to prices, we
differentiate $\frac{m}{\mu (x(m))}$ with respect to $m$ to obtain
\begin{eqnarray*}
\textstyle
\frac{\partial}{\partial m}\left( \frac{m}{\mu (x(m))}\right) &=&
\textstyle
\frac{\mu (x(m)) - m x' (m) \mu' (x(m))}{ [\mu (x(m))]^2 } 
= \frac{1}{\mu (x(m))}   - \frac{m x' (m)}{\mu (x(m))} \frac{\mu' (x(m))}{ \mu (x(m)) } \\
&=&
\textstyle
\frac{1}{\mu (x(m))}   - \frac{m x' (m)}{x(m) \mu (x(m))} \frac{x(m) \mu' (x(m))}{ \mu (x(m)) }
=\frac{1}{\mu (x(m))}  \left[ 1  -  \frac{x(m) \mu' (x(m))}{ \mu (x(m)) } \right],
\end{eqnarray*}
where we use $\frac{m x'(m)}{x(m)}=1$, which follows from $x=\frac{{\lambda} m}{K}$, to get the last equality.
Using \eqref{eq:dmu_dx},
we can rewrite it as follows:
\begin{equation}
\textstyle
\frac{\partial}{\partial m}\left( \frac{m}{\mu (x(m))}\right)
= \textstyle
\frac{1}{\mu} \left[  1  - \frac{x}{ \mu }
\frac{ \frac{1- \sigma (1-\mu)}{x} }{1+\frac{(1-\beta) (1-\mu)}{ \mu} \frac{1- \sigma (1-\mu)}{1-\beta -\mu} }
\right],
\label{eq:price_d}
\end{equation}
where we suppress the argument of $\mu$ and that of $x$ in the right-hand side.
Hence, we have
\begin{eqnarray}
\textstyle
\frac{\partial}{\partial m}\left( \frac{m}{\mu (x(m))}\right) &=&
\textstyle
\frac{1}{\mu} \left[1  -
\frac{ 1- \sigma (1-\mu) }{\mu+(1-\beta) (1-\mu) \frac{1- \sigma (1-\mu)}{1-\beta -\mu} } \right]
= \textstyle
\frac{1}{\mu} \left[
\frac{ \mu+(1-\beta) (1-\mu) \frac{1- \sigma (1-\mu)}{1-\beta -\mu} - 1+ \sigma (1-\mu) }{\mu+(1-\beta) (1-\mu) \frac{1- \sigma (1-\mu)}{1-\beta -\mu} } \right]    \notag \\
&=& \textstyle \frac{1}{\mu} \left[ \frac{ \frac{(1-\beta) (1-\mu)}{\mu} \frac{1- \sigma (1-\mu)}{1-\beta -\mu} - \frac{(1- \sigma) (1-\mu)}{\mu}  }{1+\frac{(1-\beta) (1-\mu)}{\mu} \frac{1- \sigma (1-\mu)}{1-\beta -\mu} } \right]
=
\frac{\frac{1-\mu}{\mu}}{\mu} \left[ \frac{  \frac{(1-\beta) [1- \sigma (1-\mu)]}{1-\beta -\mu} - (1- \sigma)  }{1+\frac{(1-\beta) (1-\mu)}{\mu} \frac{1- \sigma (1-\mu)}{1-\beta -\mu} } \right]   \notag \\
&=& \textstyle
\frac{\frac{1-\mu}{\mu}}{\mu} \left[ \frac{  \frac{(1-\beta) [1- \sigma (1-\mu)]   - (1- \sigma)(1-\beta -\mu)  }{1-\beta -\mu}   }{1+\frac{(1-\beta) (1-\mu)}{\mu} \frac{1- \sigma (1-\mu)}{1-\beta -\mu} } \right]
=
\frac{1-\mu}{\mu} \left[ \frac{ \frac{1-\beta \sigma}{1-\beta -\mu}  }{1+\frac{(1-\beta) (1-\mu)}{\mu} \frac{1- \sigma (1-\mu)}{1-\beta -\mu} } \right].
\label{eq:price_d_rev}
\end{eqnarray}
By Table~\ref{tab:tab2}, we know that
$\frac{1-\mu}{\mu}>0$
except for the special case of Case 1a, where the marginal cost pricing $\frac{1}{\mu}=1$ is admissible when $\beta=0$ (in which case, \eqref{eq:price_d_rev} reduces to $\frac{\partial}{\partial m}\left( \frac{m}{\mu (x(m))}\right)= \frac{1}{\mu + 1-\sigma (1-\mu)}$, so that we have $\frac{\partial}{\partial m}\left( \frac{m}{\mu (x(m))}\right)= \frac{1}{2}>0$ at $\mu=1$).

\paragraph{Case 1: $\alpha>0$.} \hfill

\vspace{-.25cm}

\paragraph{Case 1a: $1-\beta -\mu \ge 0$, $1 -  \sigma (1-\mu)>0$, and $1 -\beta \sigma>0$.}
In this case, $\beta = 1$ is not admissible. Consider the subcase with $\beta \in (0,1)$. In this case, $q > 0$ must hold, so that ${1-\beta -\mu}>0$. Since $\beta \in (0,1)$, we know by Table~\ref{tab:tab2} that $\frac{1-\mu}{\mu}>0$. Thus, we focus on the terms in the
square brackets of \eqref{eq:price_d_rev}.
In this case, both the numerator $\frac{1-\beta \sigma}{1-\beta -\mu}$ and the denominator $1+\frac{(1-\beta) (1-\mu)}{\mu} \frac{1- \sigma (1-\mu)}{1-\beta -\mu}$ are positive. Hence, $\frac{\partial}{\partial m} \bigl(\frac{m}{\mu (x(m))} \bigr)>0$. Consider the subcase with $\beta =0$. Taking the limit of
\eqref{eq:price_d_rev}
as $\beta \to 0^+$ yields
\begin{equation*}
\textstyle
\frac{\partial}{\partial m}\left( \frac{m}{\mu (x(m))}\right)
= \textstyle
\frac{1-\mu}{\mu} \left[ \frac{ \frac{1}{1 -\mu}  }{1+\frac{ 1-\mu}{\mu} \frac{1- \sigma (1-\mu)}{1 -\mu} } \right]
= \frac{ 1 }{\mu +  1- \sigma (1-\mu)}>0.
\end{equation*}

\paragraph{Case 1b: $1-\beta -\mu \le 0$, $1 -  \sigma (1-\mu)<0$, and $ 1 -\beta \sigma<0$.}
In this case,  $\beta = 0$ is not admissible, whereas $\sigma>1$ must hold. Consider the subcase with $\beta \in (0,1)$. In this case, $q > 0$ must hold, so that ${1-\beta -\mu}<0$. Since $\beta \in (0,1)$, we know by Table~\ref{tab:tab2} that $\frac{1-\mu}{\mu}>0$. Thus, we focus on the terms in the
square brackets of \eqref{eq:price_d_rev}. In this case, both the numerator $\frac{1-\beta \sigma}{1-\beta -\mu}$ and the denominator $1+\frac{(1-\beta) (1-\mu)}{\mu} \frac{1- \sigma (1-\mu)}{1-\beta -\mu}$ are positive. Hence, $\frac{\partial}{\partial m} \bigl(\frac{m}{\mu (x(m))} \bigr)>0$. Consider the subcase with $\beta =1$. Taking the limit of
\eqref{eq:price_d_rev}
as $\beta \to 1^-$ yields
\begin{equation*}
\textstyle
\frac{\partial}{\partial m}\left( \frac{m}{\mu (x(m))}\right)
= \textstyle \frac{1-\mu}{\mu} \left[ \frac{1- \sigma}{ -\mu}  \right] = \frac{(1-\mu) (\sigma-1)}{\mu^2}>0.
\end{equation*}

\paragraph{Case 2: $\alpha<0$.} \hfill

\vspace{-.25cm}

\paragraph{Case 2a: $1-\beta -\mu \le 0$, $1 -  \sigma (1-\mu)>0$, and $ 1 -\beta \sigma>0$.}
In this case, $\beta=0$ is excluded by Step 1 of the proof of Proposition~\ref{prop:markup_bounds} and $\beta = 1$ is not admissible by Step 2 of the proof of Proposition~\ref{prop:markup_bounds}. Since $\beta \in (0,1)$ in this case, $q > 0$ must hold, so that ${1-\beta -\mu}<0$. Since $\beta \in (0,1)$, we know by Table~\ref{tab:tab2} that $\frac{1-\mu}{\mu}>0$. Thus, we focus on the terms in the
square brackets of \eqref{eq:price_d_rev}. In this case, the numerator $\frac{1-\beta \sigma}{1-\beta -\mu}$ is negative. As shown in footnote~\ref{fn:25},  the denominator $1+\frac{(1-\beta) (1-\mu)}{\mu} \frac{1- \sigma (1-\mu)}{1-\beta -\mu}$ is negative for Case 2a. Hence, $\frac{\partial}{\partial m} \bigl(\frac{m}{\mu (x(m))} \bigr)>0$.

\paragraph{Case 2b: $1-\beta -\mu \ge 0$, $1 -  \sigma (1-\mu)<0$, and  $ 1 -\beta \sigma<0$.}
In this case, $\beta=0$ and $\beta =1$ are not admissible, whereas $\sigma>1$ must hold. Since $\beta \in (0,1)$ in this case, $q > 0$ must hold, so that ${1-\beta -\mu}>0$. Since $\beta \in (0,1)$, we know by Table~\ref{tab:tab2} that $\frac{1-\mu}{\mu}>0$. Thus, we focus on the terms in the
square brackets of \eqref{eq:price_d_rev}. In this case, the numerator $\frac{1-\beta \sigma}{1-\beta -\mu}$ is negative. As shown in footnote~\ref{fn:25}, the denominator $1+\frac{(1-\beta) (1-\mu)}{\mu} \frac{1- \sigma (1-\mu)}{1-\beta -\mu}$ is also negative for Case 2b. Hence, $\frac{\partial}{\partial m} \bigl(\frac{m}{\mu (x(m))} \bigr)>0$.

\bigbreak
This establishes that $p(m)$ is increasing in $m$ (decreasing in productivity $1/m$) for $\alpha \neq 0$  and $\beta \neq \frac{1}{\sigma}$.

\bigbreak
Turning to quantities, we differentiate
$q = \frac{1}{\alpha} \frac{1-\beta -\mu}{1-\sigma (1-\mu)}$ with respect to $m$
to obtain
$\frac{\partial q}{\partial m} =  -\frac{1}{\alpha} \frac{1-\beta \sigma}{[1-\sigma (1-\mu)]^2 } \frac{\partial \mu}{\partial m}$.
When $\alpha >0,$ $1-\beta \sigma$ and $\frac{\partial \mu}{\partial m}$ have the same sign, so that $\frac{\partial q}{\partial m}<0$. When $\alpha <0,$ $1-\beta \sigma$ and $\frac{\partial \mu}{\partial m}$ have the opposite sign, so that $\frac{\partial q}{\partial m}<0$.
This establishes that $q(m)$ is decreasing in $m$ (increasing in productivity $1/m$) in all cases.
This completes the proof for $\alpha \neq 0$  and $\beta \neq \frac{1}{\sigma}$.

\medbreak
Finally, assume that either $\alpha=0$ or $\beta =\frac{1}{\sigma}$ holds. In the former case, we have $p= \frac{m}{1-\beta}$, whereas in the latter case, we have $p=\frac{\sigma m}{\sigma-1}$. Hence, the markups are constant in $m$ and the prices are increasing in $m$ regardless of parameter values. Turning to quantities, we differentiate the first-order condition $u'(q) ={\lambda} p$ with respect to $m$ to obtain $u''(q) \frac{\partial q}{\partial m} ={\lambda} \frac{\partial p}{\partial m}$. Since $u''(q)<0$ and ${\lambda}>0$, $\frac{\partial q}{\partial m}$ and $\frac{\partial p}{\partial m}$ must have opposite signs, thus implying that the quantities are decreasing in $m$ regardless of parameter values.
This completes the proof.
\hfill$\square$

\section{Numerical
implementation}
\label{app:empirics}

\paragraph{Numerical procedure.} We solve the minimization problem \eqref{eq:estimation}
subject to the theoretical constraints $\alpha  q (\omega)+\beta > 0$, $\alpha \sigma q(\omega)+1>0$, and $0 \le \frac{\alpha q  (\omega)+\beta}{\alpha \sigma q(\omega)+1} <1$. As shown by \eqref{eq:estimation}, conditional on observed markups $\mu(\omega)$ and quantities $ q(\omega)$, only the product $\alpha q(\omega)$ matters for the LFRRA. Without loss of generality, we therefore scale quantities by some multiple $\chi>0$ of the mean quantity $\overline q$ in the sample and use $q(\omega)/(\chi \overline q)$  as our measure of quantities. Doing so scales the estimated magnitude of $\alpha$ by $\chi \overline q$ without changing its sign and improves both the numerical computations (since quantities are extremely skewed) and the legibility of the output tables. Hence, in what follows, we consider the sign of $\alpha$ but do not interpret the magnitude of $\alpha$ as it depends on the scaling.

Following Arkolakis et al. (2019), we split the estimation procedure into an inner loop and an outer loop. In the inner loop, we fix the value of $\alpha$ and minimize the RSS in
\eqref{eq:estimation}
with respect to $\beta$ and $\sigma$, which yields
$\widehat\beta(\alpha)$ and
$\widehat\sigma(\alpha)$.
In the outer loop, we do a grid search for the value $\widehat\alpha$ that minimizes the RSS in \eqref{eq:estimation} conditional on $\widehat\beta(\alpha)$ and $\widehat\sigma(\alpha)$.

To implement the procedure, we have written a recursive Mathematica solver. At iteration $t$, the solver starts with a given domain $[\underline\alpha_t, \overline\alpha_t]$ and solves the system for $i=1,2,\ldots,100$ equally-spaced values $\alpha_t^i$, in that domain.
Let $\widehat\alpha_t$ denote the value of $\alpha_t^i$ such that $\{\widehat\alpha_t, \widehat\beta(\widehat\alpha_t), \widehat\sigma(\widehat\alpha_t)\}$ minimizes the RSS on the domain $[\underline\alpha_t, \overline\alpha_t]$.
If
$\widehat\alpha_t$
is interior to the domain $[\underline\alpha_t, \overline\alpha_t]$, say at $\alpha_t^i$, we bracket that domain by letting $\underline\alpha_{t+1} =\alpha^{i-1}_t$ and $\overline\alpha_{t+1} = \alpha^{i+1}_t$ and solve the system again for $100$ points over the new domain $[\underline\alpha_{t+1}, \overline\alpha_{t+1}]$. If the solution is not interior, i.e., the minimizer hits the upper bound $\overline\alpha_t$ or the lower bound $\underline\alpha_t$ of the domain, we shift and expand the domain in that direction. For example, if we hit the upper bound at iteration $t$, we let $\underline\alpha_{t+1} = \overline\alpha_t$ and $\overline\alpha_{t+1} = \xi \times \overline\alpha_t$, with $\xi>1$ and solve the system again for 100 equally-spaced points over the new domain. We run the procedure until we converge to a solution, where convergence at iteration $t$ requires that $|\widehat\alpha_t - \widehat\alpha_{t-1}| < 10^{-12}$.

Since our grid search (like any grid search) may pick up local minima, we implement our procedure in both ascending and descending orders. In the ascending version, we start with a domain of $\alpha$ between $\underline \alpha_0 = -\frac{1}{q^{\rm max}}$, where $q^{\rm max}$ is the largest scaled quantity in our data, and $\overline \alpha_0 = 450$ and expand the search domain if required.
In the descending version, we start with a domain of $\alpha$ between $\underline \alpha_0  = -\frac{1}{q^{\rm max}}$ and $\overline \alpha_0  = 1.5\times 10^8$ and reduce the search domain if required. If both procedures yield different values, we pick the solution that gives the smaller RSS.

Note that we define $\underline \alpha_0$ as the largest value of $\alpha$ for which there exists no $\beta \in [0,1]$ such that $\alpha q + \beta > 0$ holds for all $q \in [q^{\min} ,q^{\max}]$, where $q^{\min}>0$ is the smallest scaled quantity in our data. Thus, it follows that if $\alpha  \le \underline \alpha_0$ were to hold, then the numerical problem would not be well defined since there would be no value of $\beta \in [0,1]$ such that $\alpha q + \beta > 0$ for all $q \in [q^{\min} ,q^{\max}]$. The foregoing definition allows us to focus on $\alpha >\underline \alpha_0$, in which case there exists a non-empty set of $\beta$ such that $\alpha q + \beta > 0$ for all $q \in [q^{\min} ,q^{\max}]$.

To show that such a definition implies $\underline \alpha_0 = -\frac{1}{q^{\rm max}}$, we consider three cases classified by the sign of $\alpha$. First, we start with $\alpha>0$. In this case, the inequality $\alpha q + \beta > 0$ holds for all $q \in [q^{\min} ,q^{\max}]$ and for all $\beta \in [0,1]$, which implies that $\underline \alpha_0$ cannot be positive. Thus, we move on to the second case with $\alpha=0$. Then, the inequality becomes $\beta > 0$, which holds for all $\beta \neq 0$. This implies that $\underline \alpha_0$ cannot be zero. Hence, in what follows, we focus on the third case with $\alpha < 0$.  In this case, the left-hand side of $\alpha q + \beta > 0$ is decreasing in $q$. To make sure if this inequality holds for all $q \in [q^{\min} ,q^{\max}]$, it is enough to evaluate it at $q^{\max}$. That is, for any given $\beta \in [0,1]$, if $\alpha q^{\max} + \beta > 0$ holds, the inequality is satisfied for all $q \in [q^{\min} ,q^{\max}]$. Finally, since the left-hand side of $\alpha q^{\max} + \beta>0$ is increasing in $\beta$, if $\alpha q^{\max} + \beta= 0$ were to hold for some $\check \beta \in [0,1)$, then $\alpha q + \beta > 0$ would hold for all $\beta \in (\check \beta, 1]$. Hence, letting $\beta=1$ in the left-hand side of $\alpha q^{\max} + \beta > 0$ and setting $\alpha q^{\max} + 1= 0$, we obtain $\underline \alpha_0 = -\frac{1}{q^{\rm max}}$.

\vspace{-.25cm}

\paragraph{Bootstrapping.} 
Turning to the bootstrapping, since our procedure is computationally demanding, as it requires solving a constrained minimization problem at each point of the grid, we use
50 steps per iteration in this case.\footnote{Recall that we use 100 steps at each iteration during the parameter estimation. However, this becomes impractical 
when running the bootstrap replications 200 times for each sector and the aggregate economy.
For instance, even with a 50-step grid and parallel computation with Mathematica 13.1 on a 2021 MacBook Pro with an Apple M1 Pro processor, bootstrapping
the aggregate economy takes more than 18 hours.}

\vspace{-.25cm}

\paragraph{Alternative specifications.} 
For the sake of completeness, we also re-estimate the values of $\alpha$ and $\sigma$ by constraining $\beta$ a priori to be either $0$ or~$1$. This allows us to check that our procedure has `correct' numerical behavior at the corners.
Reassuringly, as should be the case, the RSS
is never smaller than the one we estimate in the unconstrained model.

Next, although by Assumption~\ref{as2} the case with $\alpha \sigma<0$ is not nested in the CREMR specification \eqref{eq:prop4_subutility_hypergeo} in Theorem~\ref{prop:utility_NEW}, we can estimate this case using the specification \eqref{eq:LFRRA_utility_C2} in Corollary~\ref{corollaryMNP}, which requires the restrictions, $\alpha = -\frac{1}{\gamma\sigma}$, $\sigma>0$, $q^{\min} > \gamma$, and $\gamma > 0$. The results are given in Table~\ref{tab:tab_app_est_beta1MNP}. We do not find any case in which the specification \eqref{eq:LFRRA_utility_C2} in Corollary~\ref{corollaryMNP} yields a smaller RSS than the specification  \eqref{eq:prop4_subutility_hypergeo} in Theorem~\ref{prop:utility_NEW} in Table~\ref{tab:tab_est_3pct_comb_rounded}.\footnote{Our exercise differs from that in Mr\'azov\'a et al.~(2021) who estimate preference and productivity distribution parameters for a given pair of utility and productivity distribution functions. In contrast, we estimate three preference parameters in the LFRRA without specifying the productivity distribution function.}

Finally, we consider four non-linear fractional forms introduced in Section~\ref{sec:nonlinear_RRA}, namely, 
$-\frac{q u''(q)}{u'(q)} = \frac{\frac{A}{B} \eta q^{\vartheta-\eta} + \vartheta}{\frac{A}{B} q^{\vartheta-\eta} +1}$ and 
$-\frac{q u''(q)}{u'(q)} =  \frac{\gamma}{q^\frac{k-1}{k} +\gamma}$ as in Mr\'{a}zov\'{a} and Neary (2017),
$-\frac{q u''(q)}{u'(q)} =  \rho+(1-\rho) \alpha q^{1-\rho}$ as in Dhingra and Morrow (2019),
and
$-\frac{\theta \Upsilon''(\theta)}{\Upsilon'(\theta)} = \frac{1}{\overline \sigma \theta^{-\frac{\varepsilon}{\overline \sigma}}}$ with $\theta=\frac{q}{U}$ as in Klenow and Willis (2016) and Edmond et al.~(2023). In the first three cases, we use the scaled quantity for $q$, whereas in the last case we use the scaled quantity for $\theta$ (so that the constant term $U$ can be incorporated into $\chi \overline q$). Comparing Table~\ref{tab:tab_est_3pct_comb_rounded} with Tables~\ref{tab:tab_app_est_bipower}--\ref{tab:tab_app_est_edmond} shows that the RRS for each sector and the aggregate economy is smaller for the LFRRA than for these four non-linear fractional forms. Hence, for the data from De Loecker et al.~(2016), the LFRRA provides a better fit than the four non-linear fractional forms.

\vspace{-.25cm}

\paragraph{Second-order conditions.} 
Last, we verify that the second-order condition evaluated at the fitted values $\{\widehat \alpha, \widehat \beta, \widehat \sigma\}$, i.e.,
\begin{equation*}
\frac{\widehat \alpha q}{\widehat \alpha q + \widehat\beta} + \frac{1-\widehat \alpha q-\widehat\beta}{\widehat \alpha \widehat \sigma q + 1} > 0
\end{equation*}
holds for all quantities $q$ in each sector and the aggregate economy.

\newpage
\clearpage

\setcounter{table}{0}

\bigbreak
\begin{sidewaystable}
\caption{Sectoral and aggregate estimates (CREMR in Mr\'azov\'a et al., 2021).}
\label{tab:tab_app_est_beta1MNP}
\centerline{\scalebox{0.75}{
\hskip -0.2cm
\begin{tabular}{llcccccccc}
\hline\hline
Sector & Name & RSS & $\widehat \alpha$ & $\widehat\beta$ & $\widehat\sigma$ & $\widehat\beta-1/\widehat\sigma$ & $\widehat\alpha(1-\widehat\beta\widehat\sigma)$ & Type & Obs \\ \hline
15 & Food products and beverages & $53.713$ & $-16.281$ & $1.000$ & $2.713$ & $0.631$ & $27.885$ & IRRA  & $1044$ \\
 &  &  & $[-24.675, -11.559]$ &  & $[2.598, 2.812]$ & $[0.615, 0.644]$ & $[19.609, 42.897]$ & &  \\
17 & Textiles, apparel & $22.709$ & $0.756$ & $1.000$ & $4.012$ & $0.751$ & n.a.$^\dagger$ & CRRA & $1534$ \\
 &  &  & $[-1.696, 2.974]$ &  & $[3.903, 4.132]$ & $[0.744, 0.758]$ & & &  \\
21 & Paper and paper products & $6.573$ & $-0.706$ & $1.000$ & $3.211$ & $0.689$ & n.a.$^\dagger$ & CRRA & $546$ \\
 &  &  & $[-6.613, 4.568]$ & & $[3.122, 3.301]$ & $[0.680, 0.697]$ & & &  \\
24 & Chemicals & $37.602$ & $-2.164$ & $1.000$ & $3.361$ & $0.702$ & $5.108$ & IRRA & $1684$ \\
 &  &  & $[-4.460, -1.361]$ &  & $[3.287, 3.433]$ & $[0.696, 0.709]$ & $[3.180, 10.628]$ & &  \\
25 & Rubber and plastic & $11.898$ & $-16.037$ & $1.000$ & $3.287$ & $0.696$ & n.a.$^\dagger$ &CRRA  & $676$ \\
 &  &  & $[-47.205, 75.685]$ &  & $[3.186, 3.406]$ & $[0.686, 0.706]$ &  & &  \\
26 & Nonmetallic mineral products & $24.854$ & $-31.201$ & $1.000$ & $1.832$ & $0.454$ & $25.971$ & IRRA & $744$ \\
 &  &  & $[-90.855, -21.824]$ & & $[1.789, 1.877]$ & $[0.441, 0.467]$ & $[18.077, 74.889]$ & &  \\
27 & Basic metals & $20.934$ & $3.420$ & $1.000$ & $4.535$ & $0.779$ & $-12.088$ & DRRA & $988$ \\
 &  &  & $[1.753, 10.292]$ &  & $[4.362, 4.753]$ & $[0.771, 0.790]$ & $[-35.511, -6.394]$ & &  \\
28 & Fabricated metal products & $7.085$ & $1.918$ & $1.000$ & $3.472$ & $0.712$ & $-4.742$ & DRRA & $445$ \\
 &  &  & $[0.555, 10.421]$ &  & $[3.332, 3.621]$ & $[0.700, 0.724]$ & $[-25.283, -1.391]$ & &  \\
29 & Machinery and equipment & $15.390$ & $37.424$ & $1.000$ & $3.382$ & $0.704$ & n.a.$^\dagger$ & CRRA  & $745$ \\
 &  &  & $[-118.510, 197.800]$ &  & $[3.263, 3.516]$ & $[0.694, 0.716]$ &  & &  \\
31 & Electrical machinery & $8.193$ & $247.543$ & $1.000$ & $3.547$ & $0.718$ & $-630.438$ & DRRA & $495$ \\
 &  \quad and communications  &  & $[141.819, 670.674]$ &  & $[3.381, 3.681]$ & $[0.704, 0.728]$ & $[-1708.863, -353.146]$ & &  \\
34 & Motor vehicles, trailers & $8.903$ & $-65.492$ & $1.000$ & $2.982$ & $0.665$ & n.a.$^\dagger$ & CRRA & $331$ \\
 &  &  & $[-234.988, 476.882]$ &  & $[2.819, 3.136]$ & $[0.645, 0.681]$ &  & &  \\
0 & All sectors pooled & $275.952$ & $233.459$ & $1.000$ & $3.219$ & $0.689$ & $-518.155$ & DRRA  & $9232$ \\
 &  &  & $[96.565, 1475.375]$ & & $[3.186, 3.260]$ & $[0.686, 0.693]$ & $[-3280.142, -217.310]$ & &  \\
\hline \hline
\multicolumn{10}{p{28cm}}{\small {\it Notes:} We use markup and quantity data from De Loecker et al. (2016). We restrict ourselves to single-product firms and products for which the markup exceeds one. Following De Loecker et al. (2016), we trim the top- and bottom-3\% of quantities and markups by sector. We run 200 bootstrap replications with replacement, and the 95\% confidence intervals are provided below the coefficients. RSS denotes the residual sum of squares, i.e., the value of our objective function in \eqref{eq:estimation}. Obs stands for the number of observations. Details on the numerical procedure are provided in \ref{app:empirics}.
$^\dagger$ means that the derivative of RRA is zero when $\widehat \alpha$ or $\widehat \beta - {1}/{\widehat \sigma}$ is not different from zero and is therefore not applicable.
} \\
\end{tabular}
}}
\end{sidewaystable}

\newpage
\clearpage

\bigbreak
\begin{sidewaystable}
\caption{Sectoral and aggregate estimates (Bipower in Mr\'azov\'a and Neary, 2017).}
\label{tab:tab_app_est_bipower}
\centerline{\scalebox{0.7}{
\hskip 0.3cm
\begin{tabular}{llcccccccc}
\hline\hline
Sector & Name & RSS & $\widehat{A/B}$ & $\widehat\eta$ & $\widehat\vartheta $ & $\widehat{\eta}-\widehat{\vartheta}$ & $(\widehat{A/B})(\widehat \eta-\widehat \vartheta)^2$ & Type & Obs \\ \hline
15 & Food products and beverages & 54.482 & 1.878 & 0.318 & 0.404 & $-0.086$ & n.a.$^\dagger$ & CRRA & 1044 \\
 &  &  & $[0.282, 12.972]$ & $[0.266, 0.375]$ & $[0.341, 0.488]$ & $[-0.170, 0.032]$ & &  &  \\
17 & Textiles, apparel & 22.798 & 0.071 & $-0.032$ & 0.259 & $-0.291$ & n.a.$^\dagger$ & CRRA & 1534 \\
 &  &  & $[0.022, 5.87\times 10^6]$ & $[-0.043, 0.898]$ & $[-0.038, 1.566]$ & $[-1.316, 0.788]$ & &  &  \\
21 & Paper and paper products & 6.578 & 0.725 & 0.310 & 0.310 & $0.000$ & n.a.$^\dagger$ & CRRA & 546 \\
 &  &  & $[-0.999, 7.559]$ & $[0.301, 0.660]$ & $[-0.009, 0.394]$ & $[-8.94\times 10^{-8}, 0.348]$ & &  &  \\
24 & Chemicals & 37.830 & 2.705 & 0.294 & 0.294 & $0.000$ & n.a.$^\dagger$ & CRRA & 1684 \\
 &  &  & $[2.728, 3.398]$ & $[0.288, 0.300]$ & $[0.288, 0.300]$ & $[-2.98\times 10^{-8}, 0.000]$ & &  &  \\
25 & Rubber and plastic & 11.899 & 4.259 & 0.304 & 0.304 & $0.000$ & n.a.$^\dagger$ & CRRA & 676 \\
 &  &  & $[0.052, 4.391]$ & $[-0.019, 0.313]$ & $[0.291, 0.319]$ & $[-0.327, 0.000]$ & &  &  \\
26 & Nonmetallic mineral products & 24.611 & $-1.000$ & 0.620 & 0.620 & $1.97 \times 10^{-6}$ & n.a.$^\dagger$ & CRRA & 744 \\
 &  &  & $[-1.000, 1.728]$ & $[0.536, 0.627]$ & $[0.536, 0.626]$ & $[-5.96\times 10^{-8}, 0.000]$ & &  &  \\
27 & Basic metals & 20.737 & 0.108 & 0.418 & 0.164 & 0.254 & 0.007 & DRRA & 988 \\
 &  &  & $[4.41\times 10^{-12}, 0.947]$ & $[0.292, 2.503]$ & $[0.065, 0.226]$ & $[0.186, 2.279]$ & $[2.29\times 10^{-11}, 0.046]$ &  &  \\
28 & Fabricated metal products & 6.930 & 1171.087524 & 0.271 & 0.652 & $-0.381$ & 169.939 & DRRA & 445 \\
 &  &  & $[200.487, 2.65\times 10^4]$ & $[0.252, 0.290]$ & $[0.546, 0.837]$ & $[-0.559, -0.283]$ & $[17.094, 8296.303]$ &  &  \\
29 & Machinery and equipment & 15.342 & 7.705 & 0.266 & 0.394 & $-0.128$ & n.a.$^\dagger$ & CRRA & 745 \\
 &  &  & $[0.598, 7.53\times 10^6]$ & $[0.190, 0.298]$ & $[0.298, 1.268]$ & $[-0.976, 0.000]$ &  &  &  \\
31 & Electrical machinery & 8.467 & 4.007 & 0.289 & 0.289 & $0.000$ & n.a.$^\dagger$ & CRRA & 495 \\
 & \quad and communications &  & $[6.81\times 10^{-8}, 7.61\times 10^6]$ & $[0.275, 0.911]$ & $[0.277, 0.850]$ & $[-0.567, 0.627]$ &  &  &  \\
34 & Motor vehicles, trailers & 8.411 & 1.577 & 0.377 & 0.022 & 0.356 & n.a.$^\dagger$ & CRRA & 331 \\
 &  &  & $[-8.50\times 10^{-6}, 3.394]$ & $[-0.001, 1.383]$ & $[-0.002, 0.395]$ & $[-0.366, 1.057]$ & &  &  \\
0 & All sectors pooled & 276.077 & 3.328 & 0.311 & 0.311 & $-2.98\times 10^{-8}$ & n.a.$^\dagger$ & CRRA & 9232 \\
 &  &  & $[3.241, 3.424]$ & $[0.307, 0.314]$ & $[0.307, 0.314]$ & $[-5.96\times 10^{-8}, 2.98\times 10^{-8}]$ & &  &  \\
\hline \hline
\multicolumn{10}{p{33.5cm}}{\small {\it Notes:} We use markup and quantity data from De Loecker et al. (2016). We restrict ourselves to single-product firms and products for which the markup exceeds one. Following De Loecker et al. (2016), we trim the top- and bottom-3\% of quantities and markups by sector. We run 200 bootstrap replications with replacement, and the 95\% confidence intervals are provided below the coefficients. RSS denotes the residual sum of squares, i.e., the value of our objective function in \eqref{eq:estimation}. Obs stands for the number of observations. We report all values greater than $10^4$ or smaller than $10^{-4}$ using scientific notation. Details on the numerical procedure are provided in \ref{app:empirics}. $^\dagger$ means that the derivative of RRA is zero when $\widehat{A/B}$ or $\widehat \eta - \widehat \vartheta$ is not different from zero and is therefore not applicable.
} \\
\end{tabular}
}}
\end{sidewaystable}

\bigbreak
\begin{table}
\caption{Sectoral and aggregate estimates (CPPT in Mr\'azov\'a and Neary, 2017).}
\label{tab:tab_app_est_cppt}
\centerline{\scalebox{0.7}{
\hskip -0.2cm
\begin{tabular}{llcccccc}
\hline\hline
Sector & Name & RSS & $\widehat \gamma$ & $\widehat{k}$ &
Type & Obs \\ \hline
15 & Food products and beverages & 54.487 & 0.538 & 1.007 & CRRA & 1044 \\
 &  &  & $[0.480, 0.601]$ & $[0.991, 1.021]$ &  & \\
17 & Textiles, apparel & 22.833 & 0.331 & 1.005 & CRRA & 1534 \\
 &  &  & $[0.287, 0.366]$ & $[0.986, 1.023]$ &  &  \\
21 & Paper and paper products & 6.540 & 0.504 & 0.976 & CRRA & 546 \\
 &  &  & $[0.006, 0.576]$ & $[0.276, 1.007]$ &  &  \\
24 & Chemicals & 35.911 & 0.613 & 0.939 & IRRA & 1684 \\
 &  &  & $[0.552, 0.675]$ & $[0.924, 0.954]$ &  &  \\
25 & Rubber and plastic & 11.898 & 0.443 & 0.998 & CRRA & 676 \\
 &  &  & $[0.395, 0.494]$ & $[0.984, 1.014]$ &  &  \\
26 & Nonmetallic mineral products & 24.242 & 1.467 & 0.952 & IRRA & 744 \\
 &  &  & $[1.320, 1.633]$ & $[0.935, 0.973]$ &  &  \\
27 & Basic metals & 20.759 & 0.219 & 1.074 & CRRA & 988 \\
 &  &  & $[1.597\times 10^{-8}, 0.247]$ & $[0.117, 1.103]$ &  &  \\
28 & Fabricated metal products & 7.155 & 0.353 & 1.018 & DRRA & 445 \\
 &  &  & $[0.292, 0.426]$ & $[1.000, 1.038]$ &  &  \\
29 & Machinery and equipment & 15.351 & 0.387 & 1.013 & DRRA & 745 \\
 &  &  & $[0.337, 0.429]$ & $[1.002, 1.031]$ &  &  \\
31 & Electrical machinery & 8.444 & 0.433 & 0.995 & CRRA & 495 \\
 & \quad  and communications &  & $[0.389, 0.482]$ & $[0.986, 1.003]$ &  &  \\
34 & Motor vehicles, trailers  & 8.611 & 0.404 & 1.039 & IRRA & 331 \\
 &  &  & $[0.343, 0.472]$ & $[1.013, 1.061]$ &  &  \\
0 & All sectors pooled & 273.993 & 0.596 & 0.980 & IRRA & 9232 \\
 &  &  & $[7.326\times 10^{-11}, 0.634]$ & $[0.122, 0.985]$ &  &  \\
\hline \hline
\multicolumn{8}{p{18.5cm}}{\small {\it Notes:} We use markup and quantity data from De Loecker et al. (2016). We restrict ourselves to single-product firms and products for which the markup exceeds one. Following De Loecker et al. (2016), we trim the top- and bottom-3\% of quantities and markups by sector. We run 200 bootstrap replications with replacement, and the 95\% confidence intervals are provided below the coefficients. RSS denotes the residual sum of squares, i.e., the value of our objective function in \eqref{eq:estimation}. Obs stands for the number of observations. We report all values smaller than $10^{-4}$ using scientific notation. Details on the numerical procedure are provided in \ref{app:empirics}.} \\
\end{tabular}
}}
\end{table}

\bigbreak
\begin{table}
\caption{Sectoral and aggregate estimates (Expo-power in Dhingra and Morrow, 2019).}
\label{tab:tab_app_est_expo_power}
\centerline{\scalebox{0.7}{
\hskip -0.2cm
\begin{tabular}{llccccccc}
\hline\hline
Sector & Name & RSS & $\widehat\alpha $ & $\widehat\rho$ & Type & Obs \\ \hline
15 & Food products and beverages  & $54.4876$ & $-226.6083$ & $0.9971$ & DRRA & 1044 \\
 &  &  & $[-1012.2740, -87.3480]$ & $[0.9925, 0.9994]$ &  & \\
17 & Textiles, apparel & $22.8341$ & $-419.2237$ & $0.9982$ & DRRA & 1534 \\
 &  &  & $[-1080.4592, -117.9296]$ & $[0.9934, 0.9993]$ &  & \\
21 & Paper and paper products & $6.5958$ & $-418.0499$ & $0.9983$ & DRRA & 546 \\
 &  &  & $[-2092.4607, -80.3307]$ & $[0.9912, .9997]$ &  & \\
24 & Chemicals & $38.6397$ & $-192.6799$ & $0.9963$ & DRRA & 1684 \\
 &  &  & $[-7220.7041, -162.2823]$ & $[0.9956, 0.9999]$ &  & \\
25 & Rubber and plastic & $11.9952$ & $-134.5331$ & $0.9947$ & DRRA & 676 \\
 &  &  & $[-1826.0720, -125.2231]$ & $[0.9943, 0.9996]$ &  &  \\
26 & Nonmetallic mineral products & $25.0078$ & $-2026.0922$ & $0.9998$ & DRRA & 744 \\
 &  &  & $[-6222.4185, -84.7281]$ & $[0.9946, 0.9999]$ &  &  \\
27 & Basic metals & $20.7998$ & $-58.5374$ & $0.9863$ & DRRA & 988 \\
 &  &  & $[-84.2632, -53.9247]$ & $[0.9851, 0.9906]$ &  &  \\
28 & Fabricated metal products & $7.1663$ & $-168.3581$ & $0.9957$ & DRRA & 445 \\
 &  &  & $[-831.2285, -85.1215]$ & $[0.9911, 0.9991]$ &  &  \\
29 & Machinery and equipment  & $15.3646$ & $-133.8742$ & $0.9946$ & DRRA & 745 \\
 &  &  & $[-706.5084, -86.5909]$ & $[0.9914, 0.9990]$ &  &  \\
31 &  Electrical machinery & $8.4838$ & $-1636.2917$ & $0.9996$ & DRRA & 495 \\
 &  \quad and communications  &  & $[-7872.1021, -149.0634]$ & $[0.9950, 0.9999]$ &  &  \\
34 & Motor vehicles, trailers  & $8.5823$ & $-50.6071$ & $0.9861$ & DRRA & 331 \\
 &  &  & $[-130.9806, -35.7461]$ & $[0.9800, 0.9949]$ &  &  \\
0 & All sectors pooled & $276.5444$ & $-1004.0812$ & $0.9993$ & DRRA & 9232 \\
 &  &  & $[-2087.9907, -149.2689]$ & $[0.9951, 0.9997]$ &  &  \\
\hline \hline
\multicolumn{8}{p{19.7cm}}{\small {\it Notes:} We use markup and quantity data from De Loecker et al. (2016). We restrict ourselves to single-product firms and products for which the markup exceeds one. Following De Loecker et al. (2016), we trim the top- and bottom-3\% of quantities and markups by sector. We run 200 bootstrap replications with replacement, and the 95\% confidence intervals are provided below the coefficients. RSS denotes the residual sum of squares, i.e., the value of our objective function in \eqref{eq:estimation}. Obs stands for the number of observations. Details on the numerical procedure are provided in \ref{app:empirics}.} \\
\end{tabular}
}}
\end{table}

\newpage
\clearpage

\bigbreak
\begin{table}
\caption{Sectoral and aggregate estimates (Klenow and Willis, 2016; Edmond et al., 2023).}
\label{tab:tab_app_est_edmond}
\centerline{\scalebox{0.7}{
\hskip -0.2cm
\begin{tabular}{llccclcc}
\hline\hline
Sector & Name & RSS & $\widehat \varepsilon$ & $\widehat{\overline \sigma}$ &
$-{\theta \Upsilon''(\theta)}/{\Upsilon'(\theta)}$
& Obs \\ \hline
15 & Food products and beverages & 54.487 & $-0.012$ & 2.857 & \ \ \ \ constant
& 1044 \\
 &  &  & $[-0.041, 0.014]$ & $[2.663, 3.076]$ &  &  \\
17 & Textiles, apparel & 22.833 & $-0.015$ & 4.021 &
 \ \ \ \ constant
& 1534 \\
 &  &  & $[-0.073, 0.026]$ & $[3.730, 4.335]$ &  &  \\
21 & Paper and paper products & 6.541 & 0.049 & 2.988 &
 \ \ \ \ constant
& 546 \\
 &  &  & $[-0.016, 0.105]$ & $[2.725, 3.297]$ &  &  \\
24 & Chemicals & 35.971 & 0.114 & 2.636 &
 \ \ \ \ increasing
& 1684 \\
 &  &  & $[0.091, 0.133]$ & $[2.494, 2.813]$ &  &  \\
25 & Rubber and plastic & 11.898 & 0.004 & 3.258 &
 \ \ \ \ constant
& 676 \\
 &  &  & $[-0.031, 0.033]$ & $[3.026, 3.511]$ &  &  \\
26 & Nonmetallic mineral products & 24.169 & 0.042 & 1.657 &
 \ \ \ \ increasing
& 744 \\
 &  &  & $[0.025, 0.057]$ & $[1.576, 1.741]$ &  &  \\
27 & Basic metals & 20.753 & $-0.291$ & 5.524 & 
 \ \ \ \ decreasing
& 988 \\
 &  &  & $[-0.427, -0.162]$ & $[5.032, 6.056]$ &  &  \\
28 & Fabricated metal products & 7.150 & $-0.051$ & 3.858 &
 \ \ \ \ decreasing
& 445 \\
 &  &  & $[-0.120, -0.002]$ & $[3.345, 4.401]$ &  &  \\
29 & Machinery and equipment & 15.351 & $-0.033$ & 3.585 & 
 \ \ \ \ decreasing
& 745 \\
 &  &  & $[-0.085, -0.004]$ & $[3.334, 3.952]$ &  &  \\
31 & Electrical machinery & 8.443 & 0.012 & 3.300 &
 \ \ \ \ constant
& 495 \\
 & \quad  and communications &  & $[-0.009, 0.031]$ & $[3.057, 3.580]$ &  &  \\
34 & Motor vehicles, trailers & 8.625 & $-0.080$ & 3.434 &
 \ \ \ \ decreasing
& 331 \\
 &  &  & $[-0.133, -0.026]$ & $[3.100, 3.796]$ &  &  \\
0 & All sectors pooled & 274.005 & 0.037 & 2.669 &
 \ \ \ \ increasing
& 9232 \\
 &  &  & $[0.030, 0.043]$ & $[2.579, 2.795]$ &  &  \\
\hline \hline
\multicolumn{8}{p{19.2cm}}{\small {\it Notes:} We use markup and quantity data from De Loecker et al. (2016). We restrict ourselves to single-product firms and products for which the markup exceeds one. Following De Loecker et al. (2016), we trim the top- and bottom-3\% of quantities and markups by sector. We run 200 bootstrap replications with replacement, and the 95\% confidence intervals are provided below the coefficients. RSS denotes the residual sum of squares, i.e., the value of our objective function in \eqref{eq:estimation}. Obs stands for the number of observations. Details on the numerical procedure are provided in \ref{app:empirics}.
} \\
\end{tabular}
}}
\end{table}

\end{document}